\documentclass[aps,prl,twocolumn,superscriptaddress,floatfix,nofootinbib,showpacs,amsmath,amssymb]{revtex4-1}
\usepackage[colorlinks=true,pdfstartview=FitV,linkcolor=blue,citecolor=blue,urlcolor=blue]{hyperref}
\usepackage{wasysym}
\usepackage[separate-uncertainty,retain-explicit-plus,per-mode=symbol,binary-units]{siunitx}
\usepackage{array,mathtools,amssymb,dcolumn}
\usepackage[below]{placeins}
\usepackage[dvipsnames, table]{xcolor}
\usepackage{tikz}
\usepackage{multirow}
\usepackage{afterpage}
\usepackage{lineno}
\usepackage{paralist}
\usepackage{listings}
\usepackage{array}
\usepackage{cancel}
\usepackage[version=4]{mhchem}
\usepackage{calc}
\newcommand{\instr}[1]{{\color{Black}#1}} 

\newcommand{\etal}{et al.}

\newcommand{\darkmatteredensitypercent}{\SI{27}{\percent}}
\newcommand{\baryonicmatteredensitypercent}{\SI{5}{\percent}}

\newcommand{\Geant}{Geant4}

\newcommand{\SOURCES}{SOURCES-4C}
\newcommand{\neucbot}{NeuCBOT}

\newcommand{\CERNRoot}{ROOT}
\newcommand{\MIDAS}{MIDAS}
\newcommand{\RAT}{RAT}

\newcommand{\E}[1]{\ensuremath{\times 10^{#1}}}
\DeclareSIUnit\c{\mbox{$c$}}
\DeclareSIUnit\week{w}
\DeclareSIUnit\year{yr}
\DeclareSIUnit\yr{yr}
\DeclareSIUnit\yr{yr}
\DeclareSIUnit\standard{std}
\DeclareSIUnit\str{sr}
\DeclareSIUnit\ppm{ppm}
\DeclareSIUnit\ppb{ppb}
\DeclareSIUnit\ppt{ppt}
\DeclareSIUnit\pe{PE}
\DeclareSIUnit\spe{SPE}
\DeclareSIUnit\ev{events}
\DeclareSIUnit\hit{hit}
\DeclareSIUnit\hits{hits}
\DeclareSIUnit\bin{(\mbox{5-PE}~bin)}
\DeclareSIUnit\sgm{\mbox{$\sigma$}}
\DeclareSIUnit\rms{RMS}
\DeclareSIUnit\keVr{\mbox{keV$_{\rm nr}$}}
\DeclareSIUnit\keVee{\mbox{keV$_{\rm ee}$}}
\DeclareSIUnit\ph{photons}
\DeclareSIUnit\pm{PMT}
\DeclareSIUnit\inch{''}
\DeclareSIUnit\bit{bit}
\DeclareSIUnit\sample{S}
\DeclareSIUnit\barn{b}
\DeclareSIUnit\bara{bar}
\DeclareSIUnit\Curie{Ci}
\DeclareSIUnit\psi{psi}
\DeclareSIUnit\mK{\milli\kelvin}
\DeclareSIUnit\micron{\micro\metre}
\DeclareSIUnit\liveday{\mbox{live-days}}
\DeclareSIUnit\tonneday{\mbox{tonne$\cdot$day}}
\DeclareSIUnit\days{\mbox{days}}

\newcommand{\FPrange}[2]{\mbox{\SI{#1}{}$<$\FPrompt\SI{<#2}{}}}

\usepackage[symbol*]{footmisc}
\DefineFNsymbolsTM{otherfnsymbols}{%
  \textdagger    \dagger
  \textasteriskcentered *
  \textbardbl    \|%
  \textparagraph \mathparagraph
  \textbullet \circ
  \textdaggerdbl \ddagger
}%
\setfnsymbol{otherfnsymbols}

\newcommand{\WIMPEscapeVelocity}{\SI{544}{\kilo\metre\per\second}}
\newcommand{\WIMPVzero}{\SI{220}{\kilo\metre\per\second}}
\newcommand{\WIMPDensity}{\SI{0.3}{GeV\per\cubic\cm}}

\newcommand{\CurrentBestWIMPLimit}{\SI{9.0E-47}{\square\cm}}

\newcommand{\PaperTwoWIMPLimitOneHundredGeV}{\SI{3.9E-45}{\square\cm}}
\newcommand{\PaperTwoWIMPLimitOneTeV}{\SI{1.5E-44}{\square\cm}}

\newcommand{\ERLeakage}{\instr{\SI{<0.05}{}}} 
\newcommand{\NeckAlphaLeakage}{\instr{\SI{<0.5}{}}}
\newcommand{\PaperTwoROIAcceptanceLossTopFP}{\instr{\SI{30}{\percent}}}
\newcommand{\PaperTwoTotalPredictionHighFP}{\SI{0.46}{}$^{+0.13}_{-0.18}$} 
\newcommand{\PaperTwoTotalPredictionHighPE}{\SI{1.25}{}$^{+0.26}_{-0.42}$} 
\newcommand{\NeutronCRFPromptRange}{\FPrange{0.6}{0.8}}
\newcommand{\NeutronCRRadius}{\instr{\SI{<800}{\mm}}}
\newcommand{\NeutronTaggingEfficiency}{\instr{\SI{22.5\pm0.5}{\percent}}}  
\newcommand{\PaperTwoTotalCosmoNeutronBkgd}{\instr{\SI{<0.11}{}}}          
\newcommand{\PaperTwoTotalCosmoNeutronBkgdBeforeCuts}{\instr{\SI{<0.2}{}}} 
\newcommand{\PaperTwoTotalRadioNeutronBkgd}{\instr{\SI{0.10}{}$^{+0.10}_{-0.09}$}}  
\newcommand{\PaperTwoTotalRadioNeutronBkgdLL}{\instr{\SI{11}{}$^{+8}_{-9}$}}        
\newcommand{\PaperTwoTotalRadioNeutronBkgdBeforeCutsWIMPPE}{\instr{\SI{6\pm4}{}}}   
\newcommand{\PaperTwoTotalRadioNeutronBkgdBeforeCuts}{\instr{\SI{23}{}$^{+17}_{-14}$}}
\newcommand{\PaperTwoPMTGlassRadioNeutronsNeuCBOTROI}{\instr{\SI{0.016}{}$^{+0.013}_{-0.007}$}}   
\newcommand{\PaperTwoPMTGlassRadioNeutronsNeuCBOTCR}{\instr{\SI{4.1}{}$^{+2.0}_{-1.3}$}}          
\newcommand{\PaperTwoPMTCeramicRadioNeutronsNeuCBOTROI}{\instr{\SI{<0.03}{}}}                     
\newcommand{\PaperTwoPMTCeramicRadioNeutronsNeuCBOTCR}{\instr{\SI{0.36}{}$^{+0.09}_{-0.15}$}}     
\newcommand{\PaperTwoPMTMountRadioNeutronsNeuCBOTROI}{\instr{\SI{0.0004}{}$^{+0.0003}_{-0.0001}$}}
\newcommand{\PaperTwoPMTMountRadioNeutronsNeuCBOTCR}{\instr{\SI{0.10}{}$^{+0.04}_{-0.05}$}}       
\newcommand{\PaperTwoNVPMTRadioNeutronsNeuCBOTROI}{\instr{\SI{<0.02}{}}}                          
\newcommand{\PaperTwoNVPMTRadioNeutronsNeuCBOTCR}{\instr{\SI{0.060}{}$^{+0.036}_{-0.049}$}}       
\newcommand{\PaperTwoFillerBlocksRadioNeutronsNeuCBOTROI}{\instr{\SI{0.048}{}$^{+0.115}_{-0.048}$}}  
\newcommand{\PaperTwoFillerBlocksRadioNeutronsNeuCBOTCR}{\instr{\SI{8.1}{}$^{+9.2}_{-7.7}$}}         
\newcommand{\PaperTwoFillerFoamRadioNeutronsNeuCBOTROI}{\instr{\SI{0.0088}{}$^{+0.0123}_{-0.0067}$}} 
\newcommand{\PaperTwoFillerFoamRadioNeutronsNeuCBOTCR}{\instr{\SI{0.95}{}$^{+0.50}_{-0.47}$}}        
\newcommand{\PaperTwoTotalRadioNeutronsNeuCBOTROI}{\instr{\SI{0.073}{}$^{+0.119}_{-0.048}$}}         
\newcommand{\PaperTwoTotalRadioNeutronsNeuCBOTCR}{\instr{\SI{13.6}{}$^{+9.4}_{-7.8}$}}               
\newcommand{\PaperTwoPMTGlassRadioNeutronsSOURCESROI}{\instr{\SI{0.009}{}$^{+0.008}_{-0.004}$}}      
\newcommand{\PaperTwoPMTGlassRadioNeutronsSOURCESCR}{\instr{\SI{2.4}{}$^{+1.2}_{-0.8}$}}
\newcommand{\PaperTwoPMTCeramicRadioNeutronsSOURCESROI}{\instr{\SI{<0.02}{}}}           
\newcommand{\PaperTwoPMTCeramicRadioNeutronsSOURCESCR}{\instr{\SI{0.22}{}$^{+0.06}_{-0.11}$}}       
\newcommand{\PaperTwoPMTMountRadioNeutronsSOURCESROI}{\instr{\SI{0.0004}{}$^{+0.0002}_{-0.0001}$}}  
\newcommand{\PaperTwoPMTMountRadioNeutronsSOURCESCR}{\instr{\SI{0.095}{}$^{+0.032}_{-0.041}$}}      
\newcommand{\PaperTwoNVPMTRadioNeutronsSOURCESROI}{\instr{\SI{<0.01}{}}}                            
\newcommand{\PaperTwoNVPMTRadioNeutronsSOURCESCR}{\instr{\SI{0.038}{}$^{+0.022}_{-0.032}$}}         
\newcommand{\PaperTwoFillerBlocksRadioNeutronsSOURCESROI}{\instr{\SI{0.042}{}$^{+0.102}_{-0.042}$}}   
\newcommand{\PaperTwoFillerBlocksRadioNeutronsSOURCESCR}{\instr{\SI{7.1}{}$^{+8.2}_{-7.0}$}}          
\newcommand{\PaperTwoFillerFoamRadioNeutronsSOURCESROI}{\instr{\SI{0.0076}{}$^{+0.0107}_{-0.0063}$}}  
\newcommand{\PaperTwoFillerFoamRadioNeutronsSOURCESCR}{\instr{\SI{0.79}{}$^{+0.43}_{-0.41}$}}         
\newcommand{\PaperTwoTotalRadioNeutronsSOURCESROI}{\instr{\SI{0.060}{}$^{+0.104}_{-0.045}$}}          
\newcommand{\PaperTwoTotalRadioNeutronsSOURCESCR}{\instr{\SI{10.6}{}$^{+8.3}_{-7.1}$}}                
\newcommand{\PaperTwoIFGISRate}{\instr{\SI{14.1\pm1.3}{\micro\Hz}}} 
\newcommand{\PaperTwoIFGOSRate}{\instr{\SI{16.8\pm1.4}{\micro\Hz}}} 
\newcommand{\PaperTwoOFGISRate}{\instr{\SI{22.7\pm1.6}{\micro\Hz}}} 
\newcommand{\PaperTwoIFGISAlphaBkgd}{\instr{\SI{0.07}{}$^{+0.13}_{-0.07}$}} 
\newcommand{\PaperTwoIFGOSAlphaBkgd}{\instr{\SI{0.17}{}$^{+0.12}_{-0.14}$}} 
\newcommand{\PaperTwoOFGISAlphaBkgd}{\instr{\SI{0.25}{}$^{+0.21}_{-0.20}$}} 

\newcommand{\PaperTwoIFGISAlphaBkgdBeforeCuts}{\instr{\SI{12}{}$^{+9}_{-7}$}} 
\newcommand{\PaperTwoIFGOSAlphaBkgdBeforeCuts}{\instr{\SI{8}{}$^{+6}_{-5}$}} 
\newcommand{\PaperTwoOFGISAlphaBkgdBeforeCuts}{\instr{\SI{8}{}$^{+7}_{-5}$}} 
\newcommand{\PaperTwoLArTPBSurfaceAlphaLeakage}{\instr{\mbox{\SI{0.8}{}$^{+1.8}_{-0.8}\times10^{-5}$}}} 
\newcommand{\PaperTwoTotalNeckAlphaBkgd}{\instr{\SI{0.49}{}$^{+0.27}_{-0.26}$}}
\newcommand{\PaperTwoTotalNeckAlphaBkgdBeforeCuts}{\instr{\SI{28}{}$^{+13}_{-10}$}}
\newcommand{\PaperTwoTotalNeckAlphaBkgdBeforeCutsLL}{\instr{\SI{28}{}$^{+13}_{-10}$}}

\newcommand{\PaperTwoERBkgdBeforeCuts}{\instr{\SI{2.44e9}{}}}     
\newcommand{\PaperTwoERBkgdAfterCuts}{\instr{\SI{0.03\pm0.01}{}}} 
\newcommand{\PaperTwoERBkgdAfterLLCuts}{\instr{\SI{0.34\pm0.11}{}}}

\newcommand{\PaperTwoCherenkovEventsBeforeCuts}{\instr{$<3.3\times10^{5}$}} 

\newcommand{\PaperTwoCherenkovEventsLGBeforeCuts}{\instr{\SI{<325000}{}}} 
\newcommand{\PaperTwoCherenkovEventsLGAfterCuts}{\instr{$<0.11$}} 
\newcommand{\PaperTwoCherenkovEventsLGLF}{\instr{$<4.62\times10^{-7}$}} 
\newcommand{\PaperTwoCherenkovEventsNeckBeforeCuts}{\instr{\SI{<3890}{}}}
\newcommand{\PaperTwoCherenkovEventsNeckAfterCuts}{\instr{$<0.09$}} 
\newcommand{\PaperTwoCherenkovEventsNeckLF}{\instr{$<6.13\times10^{-5}$}} 

\newcommand{\PaperTwoCherenkovEventsAfterCuts}{\instr{$<0.14$}} 
\newcommand{\PaperTwoCherenkovEventsAfterLLCuts}{\instr{$<3890$}} 
\newcommand{\MeasuredMuonFlux}{\instr{(3--4)$\times10^{-10}$\SI{}{~muons/\square\cm/\s}}}

\newcommand{\AVSurfaceEventsAfterLLCuts}{\instr{$<3000$}}
\newcommand{\AVSurfaceEventsAfterCuts}{\instr{$<0.08$}} 

\newcommand{\TotalBackgroundExpectationAfterCuts}{\instr{0.62$^{+0.31}_{-0.28}$}} 
\newcommand{\TotalBackgroundExpectationAfterLLCuts}{\instr{$<4910$}} 

\newcommand{\WIMPFiducialAcceptanceNeckVeto}{\instr{$92.0^{+1.0}_{-0.1}$}}
\newcommand{\WIMPFiducialAcceptancePIFGAR}{\instr{$45.4^{+1.5}_{-0.1}$}}
\newcommand{\WIMPFiducialAcceptance}{\instr{$35.4^{+2.5}_{-0.1}$}}

\newcommand{\PredEventsROINeckVeto}{\instr{$9.2^{+4.4}_{-3.5}$}}
\newcommand{\PredEventsROIPIFGAR}{\instr{$2.3^{+1.1}_{-0.9}$}}

\newcommand{\EventsROINeckVeto}{\instr{29}}
\newcommand{\EventsROIPIFGAR}{\instr{2}}
\newcommand{\EventsROI}{\instr{0}}

\newcommand{\AveDarkNoisePhysics}{\instr{\SI{1.1\pm0.2}{\pe}}}
\newcommand{\AveDarkNoiseSodium}{\instr{\SI{2.1\pm0.2}{\pe}}} 
\newcommand{\LightYieldNum}{6.1} 

\newcommand{\LightYieldSpread}{\SI{\pm1.3}{\percent}} 
\newcommand{\FpromptSpread}{\instr{\SI{\pm0.7}{\percent}}} 
\newcommand{\LightYieldWithErr}{\instr{\SI{\LightYieldNum\pm0.4}{\PE\per\keVee}}}
\newcommand{\DarkNoiseNum}{1.1} 

\newcommand{\DarkNoiseWithErr}{\instr{\SI{\DarkNoiseNum\pm0.2}{\PE}}} 
\newcommand{\ResFactorNum}{1.4} 

\newcommand{\ResFactorWithErr}{\SI{\ResFactorNum\pm0.1}{\PE}} 
\newcommand{\RelLYVarNum}{0.0004} 

\newcommand{\RelLYVarWithErr}{\instr{\SI{\RelLYVarNum}{}$^{+0.0010}_{-0.0004}$}} 
\newcommand{\ArSpectrumFitChiSqNDF}{\instr{\mbox{542/433}}} 
\newcommand{\ArThreeNineFitRange}{\SIrange{80}{4500}{\pe}} 
\newcommand{\ArThreeNineFitChisqNDF}{\SI{1252/534}{}} 

\newcommand{\ArNuisanceVariation}{\instr{\SIrange{7}{9}{\percent}}} 
\newcommand{\SnolabDepth}{\SI{2}{\km}}
\newcommand{\SnolabDepthKWE}{\SI{6}{\km\ \mbox{water-equivalent}}}

\newcommand{\larmasserror}{\SI{3279 \pm 96}{\kg}} 
\newcommand{\larmasserrortable}{3279 $\pm$ 96}    
\newcommand{\larmass}{\SI{3279}{\kg}}             
           
\newcommand{\larlevelcm}{\SI{55}{\cm}}           
\newcommand{\GArThickness}{\SI{30}{\cm}}          
\newcommand{\LGLength}{\SI{45}{\cm}}

\newcommand{\AVThickness}{\SI{5}{\cm}}            
\newcommand{\AVInnerDiameter}{\SI{1.7}{\m}}       
\newcommand{\TPBThickness}{\SI{3}{\micron}}       
\newcommand{\ResurfacerThickness}{\SI{0.5}{\mm}}  
\newcommand{\PMTCoverage}{\SI{76}{\percent}}      
\newcommand{\PMTTemperatureRange}{\SIrange{240}{290}{\kelvin}}
\newcommand{\QPEWindow}{\SI{10}{\us}}

\newcommand{\LArFilmThickness}{\SI{50}{\micro m}}

\newcommand{\FastDigitizerRate}{\SI{250}{\mega\sample\per\second}} 
 
\newcommand{\SlowDigitizerRate}{\SI{62.5}{\mega\sample\per\second}}

\newcommand{\SPEChargePrecisionStat}{\instr{\SI{\pm0.3}{\percent (stat)}}} 
\newcommand{\SPEChargePrecisionSyst}{\SI{\pm3}{\percent (syst)}} 
\newcommand{\SPERelativeRMS}{\SI{\sim43}{\percent}} 
\newcommand{\ZLEPresampleTime}{\SI{80}{\ns}}

\newcommand{\PrescaleFactor}{\SI{99}{\percent}}

\newcommand{\PrescaleRangeEne}{\SIrange{50}{565}{\keVee}}

\newcommand{\qNarrow}{\mbox{$Q_n$}}
\newcommand{\qWide}{\mbox{$Q_w$}}
\newcommand{\qRatio}{\mbox{$Q_n/Q_w$}}
\newcommand{\qNarrowTime}{\SI{177}{\ns}}

\newcommand{\qWideTime}{\SI{3.1}{\micro s}}
\newcommand{\DAQWindow}{\SI{16}{\micro s}}

\newcommand{\PreTriggerWindow}{\SI{2.4}{\micro s}}

\newcommand{\FpromptSixtyIntegrationWindow}{[\SI{-28}{}, \SI{60}{}]\,\SI{}{\ns}}

\newcommand{\LaserBallJitterMeasurement}{\instr{\SI{<1}{\ns}}}

\newcommand{\SPERatio}{\instr{\SI{1.043}{}}}
\newcommand{\SPESpread}{\instr{\SI{3.3}{\percent}}}
\newcommand{\RelativePMTEfficiencyVariation}{\instr{\SI{1}{\percent}}}
\newcommand{\RelativePMTEfficiencyVariationMildlyUnstable}{\instr{\SI{10}{\percent}}}
\newcommand{\RelativePMTEfficiencyVariationVeryUnstable}{\instr{\SI{30}{\percent}}}

\newcommand{\AveAPProbStability}{\instr{\SI{\pm6}{\percent}}}

\newcommand{\PhysicsRunTime}{279.78}
\newcommand{\PhysicsRunTimeAutomatedDAQ}{264.93}
\newcommand{\PhysicsRunTimeStableCryoCooler}{247.12}
\newcommand{\PhysicsRunTimeStablePMTs}{246.91}
\newcommand{\PhysicsRunTimeTriggerEfficiency}{246.64}
\newcommand{\PhysicsRunTimeMuonSignals}{246.24}
\newcommand{\PhysicsLivetimeTwoDP}{230.63}

\newcommand{\PaperTwoMassRCut}{\instr{\SI{1248\pm40}{}}}
\newcommand{\PaperTwoMassRCutCFTTR}{\instr{\SI{921\pm28}{}}}
\newcommand{\PaperTwoFiducialMassNum}{\instr{\SI{824\pm25}{}}}
\newcommand{\PaperTwoFiducialMassNumForTable}{824 $\pm$ 25}
\newcommand{\PaperTwoFiducialMass}{\instr{\SI{824\pm25}{\kg}}}

\newcommand{\PSDFitsMinFPrompt}{\instr{\SI{99.95}{\percent}}} 
\newcommand{\PSDFitsChisqNDF}{\instr{14,329/9380}}

\newcommand{\LongLifetimeSpread}{\instr{\SI{\pm1.0}{\percent}}} 
\newcommand{\PMTAfterPulsingProb}{\instr{\SI{7.6\pm1.9}{\percent}}} 
\newcommand{\PSDThresholdPERange}{\instr{\SIrange{95}{101}{\PE}}}
\newcommand{\PSDThresholdEnergyRange}{\instr{\SIrange{15.6}{16.6}{\keVee}}}
\newcommand{\PSDThresholdLeakageRateAtNRAFifty}{\instr{$1.2^{+0.7}_{-0.3}\E{-9}$}} 
\newcommand{\PSDThresholdLeakageRateAtNRANinty}{\instr{$2.8^{+1.3}_{-0.6}\E{-7}$}} 

\newcommand{\WIMPPERange}{\instr{\SIrange{95}{200}{\PE}}}
\newcommand{\PSDFullEnergyRange}{\instr{\SIrange{15.6}{32.9}{\keVee}}}
\newcommand{\PSDFullLeakageRateAtNRAFifty}{\instr{$3.5^{+2.2}_{-1.0}\E{-11}$}} 
\newcommand{\PSDFullLeakageRateAtNRANinty}{\instr{$4.1^{+2.1}_{-1.0}\E{-9}$}}  
\newcommand{\ChargeBasedPosRec}{\PE-based}
\newcommand{\TimeBasedPosRec}{time residual-based}

\newcommand{\TimeFitTwoWindow}{\SI{40}{\ns}}
\newcommand{\TimeFitMaxResidual}{\SI{8}{\ns}}

\newcommand{\PaperTwoChargeFitterROIRes}{\instr{\SIrange{30}{45}{\mm}}}
\newcommand{\PaperTwoChargeBasedTimeBasedFitterWIMPDZ}{\instr{\SI{35}{\mm}}} 
\newcommand{\PaperTwoChargeBasedTimeBasedFitterNeckAlphaDZ}{\instr{\SI{290}{\mm}}} 
\newcommand{\PaperTwoRadialCut}{\SI{630}{\mm}} 
\newcommand{\MuonVetoDiameter}{\SI{7.8}{\metre}} 
\newcommand{\MuonVetoHeight}{\SI{7.8}{\metre}} 
\newcommand{\MuonVetoLiveTimeLoss}{\instr{\SI{0.16}{\percent}}} 
\newcommand{\MuonVetoWindow}{\instr{[\SI{-0.1}{}, \SI{1}{}]\,\SI{}{\second}}}

\newcommand{\PaperOneExpoNum}{14.8}
\newcommand{\PaperOneExpo}{\SI{\PaperOneExpoNum}{\tonneday}}
\newcommand{\AlphaScintillationFpromptUncertainty}{\SI{3.5}{\percent}} 
\newcommand{\ChargeBasedFiducialDiscrepancy}{\SI{13}{\percent}} 

\newcommand{\PaperTwoStartDate}{November 4, 2016}
\newcommand{\PaperTwoEndDate}{October 31, 2017}
\newcommand{\PaperTwoLiveTimeNum}{231}
\newcommand{\CryoPMTTriggLiveTimeLoss}{\SI{6.9}{\percent}} 
\newcommand{\PileupLiveTimeLoss}{\SI{6.5}{\percent}} 
\newcommand{\PaperTwoLiveTime}{\SI{\PaperTwoLiveTimeNum}{\liveday}}
\newcommand{\PaperTwoExpoNum}{758}
\newcommand{\PaperTwoExpo}{\SI{\PaperTwoExpoNum}{\tonneday}}
\newcommand{\PaperTwoExpoUncertainty}{\SI{2.9}{\percent}} 

\newcommand{\PaperTwoOpMVPMTs}{\SI{45}{PMTs}}

\newcommand{\DEAPOnePSDPower}{\SI{2.7e-8}{}}
\newcommand{\DEAPOnePSDEnergyRange}{\SIrange{44}{89}{\keVee}}

\newcommand{\argon}{Ar}

\newcommand{\nitrogen}{N\mbox{$_{2}$}}
\newcommand{\LN}{LN\mbox{$_{2}$}}
\newcommand{\oxygen}{O\mbox{$_{2}$}}
\newcommand{\radon}{Rn}
\newcommand{\PVC}{PVC}

\newcommand{\hydrogen}{\mbox{$^{1}$}H}
\newcommand{\kforty}{\mbox{$^{40}$}K}

\newcommand{\krthree}{\mbox{$^{83m}$Kr}}
\newcommand{\arnine}{\mbox{$^{39}$Ar}}
\newcommand{\arforty}{\mbox{$^{40}$Ar}}
\newcommand{\poten}{\mbox{$^{210}$Po}}
\newcommand{\potwo}{\mbox{$^{212}$Po}}
\newcommand{\bitwo}{\mbox{$^{212}$Bi}}
\newcommand{\pofour}{\mbox{$^{214}$Po}}
\newcommand{\posix}{\mbox{$^{216}$Po}}
\newcommand{\poeight}{\mbox{$^{218}$Po}}

\newcommand{\bifour}{\mbox{$^{214}$Bi}}

\newcommand{\bipo}{\mbox{$^{214}$Bi-$^{214}$Po}}

\newcommand{\pbten}{\mbox{$^{210}$Pb}}

\newcommand{\rntwo}{\mbox{$^{222}$Rn}}
\newcommand{\rnzero}{\mbox{$^{220}$Rn}}
\newcommand{\thal}{\mbox{$^{208}$Tl}}
\newcommand{\tho}{\mbox{$^{232}$Th}}

\newcommand{\ura}{\mbox{$^{238}$U}}

\newcommand{\uratwo}{\mbox{$^{232}$U}}

\newcommand{\urafive}{\mbox{$^{235}$U}}

\newcommand{\natwo}{\mbox{$^{22}$Na}}

\newcommand{\HOneNeutronCaptureGammaEnergy}{\SI{2.2}{\MeV}}
\newcommand{\ArFortyNeutronCaptureGammaEnergy}{\SI{6.1}{\MeV}}
\newcommand{\KFortyGammaEnergy}{\SI{1.46}{\MeV}}
\newcommand{\BiTwoOneFourGammaEnergy}{\SI{1.76}{\MeV}}

\newcommand{\ArThreeNineQValue}{\SI{565\pm5}{\keV}}

\newcommand{\ArThreeNineHalfLife}{\SI{269}{\year}}

\newcommand{\RnTwoHalfLife}{\SI{3.8}{\day}}
\newcommand{\PoTwoHalfLife}{\SI{299}{\ns}}

\newcommand{\ThTwoZeroEightHighEGammaEnergy}{\SI{2.61}{\MeV}}

\newcommand{\AmBe}{\ce{AmBe}}

\newcommand{\AArArThreeNineActivity}{\SI{0.95\pm0.05}{\becquerel\per\kg}}

\newcommand{\PoTenAlphaEnergy}{\SI{5.3}{\MeV}}
\newcommand{\RnTwoAlphaEnergy}{\SI{5.5}{\MeV}}
\newcommand{\RnZeroAlphaEnergy}{\SI{6.3}{\MeV}}
\newcommand{\PoEightAlphaEnergy}{\SI{6.0}{\MeV}}
\newcommand{\PoSixAlphaEnergy}{\SI{6.8}{\MeV}}
\newcommand{\PoFourAlphaEnergy}{\SI{7.7}{\MeV}}
\newcommand{\PoTwoAlphaEnergy}{\SI{8.8}{\MeV}}
\newcommand{\BiTwoAlphaEnergy}{\SI{6.1}{\MeV}}
\newcommand{\ArThreeNineTotalActivity}{\SI{3.1}{\kilo\becquerel}}
\newcommand{\PoTenAVSurfaceActivity}{\instr{\SI{0.26\pm0.02}{\milli\becquerel\per\square\metre}}}
\newcommand{\PoTenAVSurfaceRate}{\instr{\SI{1.31\pm0.11}{\milli\Hz}}} 
\newcommand{\PoTenAVBulkActivity}{\SI{2.82\pm0.05}{\milli\becquerel}}
\newcommand{\PoTenAVBulkRate}{\SI{0.51\pm0.02}{\milli\Hz}}

\newcommand{\PbTenHalfLife}{\SI{22.3}{\year}}
\newcommand{\PoTenHalfLife}{\SI{138}{\day}}
\newcommand{\SurfaceEventsBeforeCuts}{\SI{<3600}{}} 
\newcommand{\SurfaceTagsBeforeCuts}{\SI{1461}{}} 

\newcommand{\RnTwoLArBulkActivity}{\SI{0.153\pm0.005}{\micro\becquerel\per\kg}}
\newcommand{\PoEightLArBulkActivity}{\SI{0.159\pm0.005}{\micro\becquerel\per\kg}}
\newcommand{\PoFourLArBulkActivity}{\SI{0.153\pm0.005}{\micro\becquerel\per\kg}}

\newcommand{\PoFourSurfaceRateActivityArea}{\SI{<5.0}{\micro\becquerel\per\square\metre}}
\newcommand{\RnZeroLArBulkActivity}{\SI{4.3\pm1.0}{\nano\becquerel\per\kg}}
\newcommand{\PoSixLArBulkActivity}{\SI{4.5\pm0.4}{\nano\becquerel\per\kg}}

\newcommand{\BiTwoLArBulkActivity}{\SI{<5.6}{\nano\becquerel\per\kg}}
\newcommand{\PoTwoLArBulkActivity}{\SI{3.4\pm1.1}{\nano\becquerel\per\kg}}

\newcommand{\SNOLAB}{\mbox{SNOLAB}}
\newcommand{\DEAP}{\mbox{DEAP-3600}}
\newcommand{\AV}{\mbox{AV}}
\newcommand{\MV}{\mbox{MV}}
\newcommand{\NV}{\mbox{NV}}
\newcommand{\LG}{\mbox{LG}}
\newcommand{\LGs}{\mbox{LGs}}
\newcommand{\FG}{\mbox{FG}}
\newcommand{\FGs}{\mbox{FGs}}
\newcommand{\IFG}{\mbox{IFG}}
\newcommand{\IFGIS}{\mbox{IFG-IS}}
\newcommand{\IFGOS}{\mbox{IFG-OS}}
\newcommand{\OFG}{\mbox{OFG}}
\newcommand{\OFGIS}{\mbox{OFG-IS}}
\newcommand{\SCB}{\mbox{SCB}}
\newcommand{\SCBs}{\mbox{SCBs}}
\newcommand{\DTM}{\mbox{DTM}}
\newcommand{\ZLE}{\mbox{ZLE}}

\newcommand{\SNO}{\mbox{SNO}}

\newcommand{\SCENE}{\mbox{SCENE}}

\newcommand{\PMT}{\mbox{PMT}}
\newcommand{\ChannelEfficiency}{\mbox{CE}}
\newcommand{\ChannelEfficiencies}{\mbox{CEs}}
\newcommand{\PMTs}{\mbox{PMTs}}

\newcommand{\UV}{\mbox{UV}}
\newcommand{\UVA}{\mbox{UVA}}

\newcommand{\WIMP}{\mbox{WIMP}}
\newcommand{\RMS}{\mbox{RMS}}
\newcommand{\WIMPs}{\mbox{WIMPs}}

\newcommand{\DAQ}{\mbox{DAQ}}

\newcommand{\TPB}{\mbox{TPB}}

\newcommand{\LAr}{\ce{LAr}}
\newcommand{\GAr}{\ce{GAr}}

\newcommand{\ROI}{\mbox{ROI}}

\newcommand{\CL}{C.~L.}
\newcommand{\insitu}{\emph{in-situ}}

\newcommand{\WIMPMassThirtyGev}{\SI{30}{\GeV\per\square\c}}

\newcommand{\ArWaveLength}{\SI{128}{\nano\meter}}
\newcommand{\ArWaveLengthGVelocity}{\SI{11}{\cm\per\ns}} 
\newcommand{\TPBWaveLength}{\SI{420}{\nano\meter}} 
\newcommand{\TPBWaveLengthGVelocity}{\SI{24}{\cm\per\ns}}
\newcommand{\CR}{\mbox{CR}}
\newcommand{\CRs}{\mbox{CRs}}
\newcommand{\PDF}{\mbox{spectrum}}
\newcommand{\PDFs}{\mbox{spectra}}
\newcommand{\NR}{\mbox{NR}}
\newcommand{\NRs}{\mbox{NRs}}
\newcommand{\ER}{\mbox{ER}}
\newcommand{\ERs}{\mbox{ERs}}

\newcommand{\bgs}{\mbox{$\beta/\gamma$'s}}
\newcommand{\alpp}{\mbox{$\alpha$ particle}}
\newcommand{\alpps}{\mbox{$\alpha$ particles}}
\newcommand{\alpd}{\mbox{$\alpha$-decay}}
\newcommand{\alpds}{\mbox{$\alpha$-decays}}
\newcommand{\betp}{\mbox{$\beta$ particle}}
\newcommand{\betps}{\mbox{$\beta$ particles}}
\newcommand{\betd}{\mbox{$\beta$-decay}}
\newcommand{\betds}{\mbox{$\beta$-decays}}

\newcommand{\gr}{\mbox{$\gamma$-ray}}
\newcommand{\grs}{\mbox{$\gamma$-rays}}
\newcommand{\alphan}{\mbox{($\alpha,n$)}}

\newcommand{\PSD}{\mbox{PSD}}
\newcommand{\FPrompt}{\mbox{F$_{\text{prompt}}$}}

\newcommand{\FNinety}{\mbox{F$_{\text{90}}$}}
\newcommand{\PE}{\mbox{PE}}
\newcommand{\PEs}{\mbox{PEs}}

\newcommand{\AP}{\mbox{AP}}
\newcommand{\APs}{\mbox{APs}}
\newcommand{\SPE}{\mbox{SPE}}

\newcommand{\chisquareperndf}{\mbox{$\chi^2/$NDF}}

\usepackage{array}
\newcolumntype{L}[1]{>{\raggedright\let\newline\\\arraybackslash\hspace{0pt}}p{#1}}
\newcolumntype{C}[1]{>{\centering\arraybackslash}p{#1}}

\begin{document}
\setcounter{secnumdepth}{5}
\title{Search for dark matter with a \PaperTwoLiveTimeNum-day exposure of liquid argon using \DEAP\ at \SNOLAB}
\newcommand{\UofA}{Department of Physics, University of Alberta, Edmonton, Alberta, T6G 2R3, Canada}
\newcommand{\CNL}{Canadian Nuclear Laboratories Ltd, Chalk River, Ontario, K0J 1J0, Canada}
\newcommand{\CU}{Department of Physics, Carleton University, Ottawa, Ontario, K1S 5B6, Canada}
\newcommand{\LNGS}{INFN Laboratori Nazionali del Gran Sasso, Assergi (AQ) 67100, Italy}
\newcommand{\LU}{Department of Physics and Astronomy, Laurentian University, Sudbury, Ontario, P3E 2C6, Canada}
\newcommand{\UNAM}{Instituto de F\'isica, Universidad Nacional Aut\'onoma de M\'exico, A.\,P.~20-364, M\'exico D.\,F.~01000, Mexico}
\newcommand{\TUM}{Department of Physics, Technische Universit\"at M\"unchen, 80333 Munich, Germany}
\newcommand{\INFN}{INFN Napoli, Napoli 80126, Italy}
\newcommand{\Napoli}{Physics Department, Universit\`a degli Studi ``Federico II'' di Napoli, Napoli 80126, Italy}
\newcommand{\PU}{Physics Department, Princeton University, Princeton, NJ 08544, USA}
\newcommand{\PRISMA}{PRISMA Cluster of Excellence and Institut f\"ur Kernphysik, Johannes Gutenberg-Universit\"at Mainz, 55128 Mainz, Germany}
\newcommand{\QU}{Department of Physics, Engineering Physics, and Astronomy, Queen's University, Kingston, Ontario, K7L 3N6, Canada}
\newcommand{\RHUL}{Royal Holloway University London, Egham Hill, Egham, Surrey TW20 0EX, United Kingdom}
\newcommand{\RAL}{Rutherford Appleton Laboratory, Harwell Oxford, Didcot OX11 0QX, United Kingdom}
\newcommand{\SL}{SNOLAB, Lively, Ontario, P3Y 1M3, Canada}
\newcommand{\Sussex}{University of Sussex, Sussex House, Brighton, East Sussex BN1 9RH, United Kingdom}
\newcommand{\TRIUMF}{TRIUMF, Vancouver, British Columbia, V6T 2A3, Canada}

\affiliation{\UofA}
\affiliation{\CNL}
\affiliation{\CU}
\affiliation{\LNGS}
\affiliation{\LU}
\affiliation{\UNAM}
\affiliation{\TUM}
\affiliation{\INFN}
\affiliation{\Napoli}
\affiliation{\PU}
\affiliation{\PRISMA}
\affiliation{\QU}
\affiliation{\RHUL}
\affiliation{\RAL}
\affiliation{\SL}
\affiliation{\Sussex}
\affiliation{\TRIUMF}

\author{R.~Ajaj}\affiliation{\CU}
\author{P.-A.~Amaudruz}\affiliation{\TRIUMF}
\author{G.\,R.~Araujo}\affiliation{\TUM}
\author{M.~Baldwin}\affiliation{\RAL}
\author{M.~Batygov}\affiliation{\LU}
\author{B.~Beltran}\affiliation{\UofA}
\author{C.\,E.~Bina}\affiliation{\UofA}
\author{J.~Bonatt}\affiliation{\QU}
\author{M.\,G.~Boulay}\affiliation{\CU}\affiliation{\QU}
\author{B.~Broerman}\affiliation{\QU}
% ...

\collaboration{DEAP Collaboration}\thanks{Corresponding email: deap-papers@snolab.ca}\noaffiliation

\date{\today}
\begin{abstract}
\DEAP\ is a single-phase liquid argon (\LAr) direct-detection dark matter experiment, operating \SnolabDepth\ underground at \SNOLAB\ (Sudbury, Canada). 
The detector consists of \larmass\ of \LAr\ contained in a spherical acrylic vessel. This paper reports on the analysis of a \PaperTwoExpo\ exposure taken over a period of \PaperTwoLiveTime\ during the first year of operation.
No candidate signal events are observed in the \WIMP-search region of interest, which results in the leading limit on the \WIMP-nucleon spin-independent cross section on a \LAr\ target of \PaperTwoWIMPLimitOneHundredGeV\ (\PaperTwoWIMPLimitOneTeV) for a \SI{100}{GeV\per\square c} (\SI{1}{TeV\per\square c}) WIMP mass at 90\%~C.~L.  
In addition to a detailed background model, this analysis demonstrates the best pulse-shape discrimination in \LAr\ at threshold, employs a Bayesian photoelectron-counting technique to improve the energy resolution and discrimination efficiency, and utilizes two position reconstruction algorithms based on the charge and photon detection time distributions observed in each photomultiplier tube.
\end{abstract}

\maketitle

\section{Introduction}\label{sec:intro}

An abundance of astrophysical observations indicates that dark matter, a non-luminous form of matter not described by the Standard Model of particle physics, comprises approximately \darkmatteredensitypercent\ of the total energy density of the universe~\cite{planck_collaboration_planck_2018}.
By contrast, baryonic matter is estimated to account for \baryonicmatteredensitypercent\ of the energy density.
Despite the significant abundance, dark matter has not yet been directly detected in terrestrial experiments. 
Many theoretical models predict particles with appropriate phenomenological properties, such as those described in~\cite{Bertone:2005bi, Feng:2010dz}. 
One such candidate is the weakly interacting massive particle (\WIMP).
In such models, the elastic scattering of \WIMPs\ with nuclei produces low energy ($\lesssim$\SI{100}{keV}) nuclear recoils (\NRs)~\cite{goodman_detectability_1985}. 
Direct detection experiments seek to observe this signature; current results limit the spin-independent \WIMP-nucleon scattering cross section to be less than \CurrentBestWIMPLimit\ at 100 GeV/c$^{2}$ at 90\%~\CL~\cite{aprile_dark_2018}.

Detecting these rare, low energy signals is facilitated by a large target mass with exceptionally low backgrounds, below 1 event per tonne per year.
Previous experimental results demonstrated the effectiveness of liquid argon (\LAr) for achieving these conditions~\cite{amaudruz_first_2017, the_darkside_collaboration_darkside-50_2018}. 
Ease of purification, high scintillation efficiency and transparency to its own scintillation light makes it well-suited for a multi-tonne \WIMP\ detector.
The \DEAP\ experiment uses the unique scintillation time profile of \LAr\ to achieve pulse shape discrimination (\PSD) \cite{Boulay:2006hu}. 
It has previously been shown that \PSD\ can be used to suppress electronic recoil (\ER) backgrounds by a factor better than \DEAPOnePSDPower, in an energy range of \DEAPOnePSDEnergyRange~\cite{amaudruz_measurement_2016}.

The results presented here are from the \DEAP\ experiment, using non-blinded data collected from \PaperTwoStartDate\ to \PaperTwoEndDate.
\DEAP\ has previously performed the first \WIMP\ search with a single-phase \LAr\ detector (measuring scintillation only), during a \PaperOneExpo\ total exposure~\cite{amaudruz_first_2017}. 
In this paper, the results are updated to a \PaperTwoExpo\ total exposure collected during \PaperTwoLiveTime. 
The result is the most sensitive dark matter search performed using a \LAr\ target for \WIMP\ masses above \WIMPMassThirtyGev. 
This analysis shows the strongest background discrimination using \PSD\ in any dark matter search, achieving an average leakage probability of \PSDFullLeakageRateAtNRANinty\ with \SI{90}{\percent} \NR\ acceptance in the dark matter search region of \PSDFullEnergyRange.

\section{Detector and Data Acquisition}\label{sec:detector}
The \DEAP\ detector is located approximately \SnolabDepth\ (\SnolabDepthKWE) underground at the \SNOLAB\ facility near Sudbury, Ontario, Canada. 
In the current run, the detector has been operating with a \LAr\ target since \PaperTwoStartDate. 
The analysis of data from a previous run is discussed in \cite{amaudruz_first_2017}. 
For the data collection period discussed here, the total mass of the \LAr\ target is \larmasserror. This value of the total \LAr\ mass is calculated using the same method as described in \cite{amaudruz_first_2017}.

\subsection{Detector description}\label{subsec:detector_description}
 \begin{figure}
 \centering
 \includegraphics[width=0.99\linewidth]{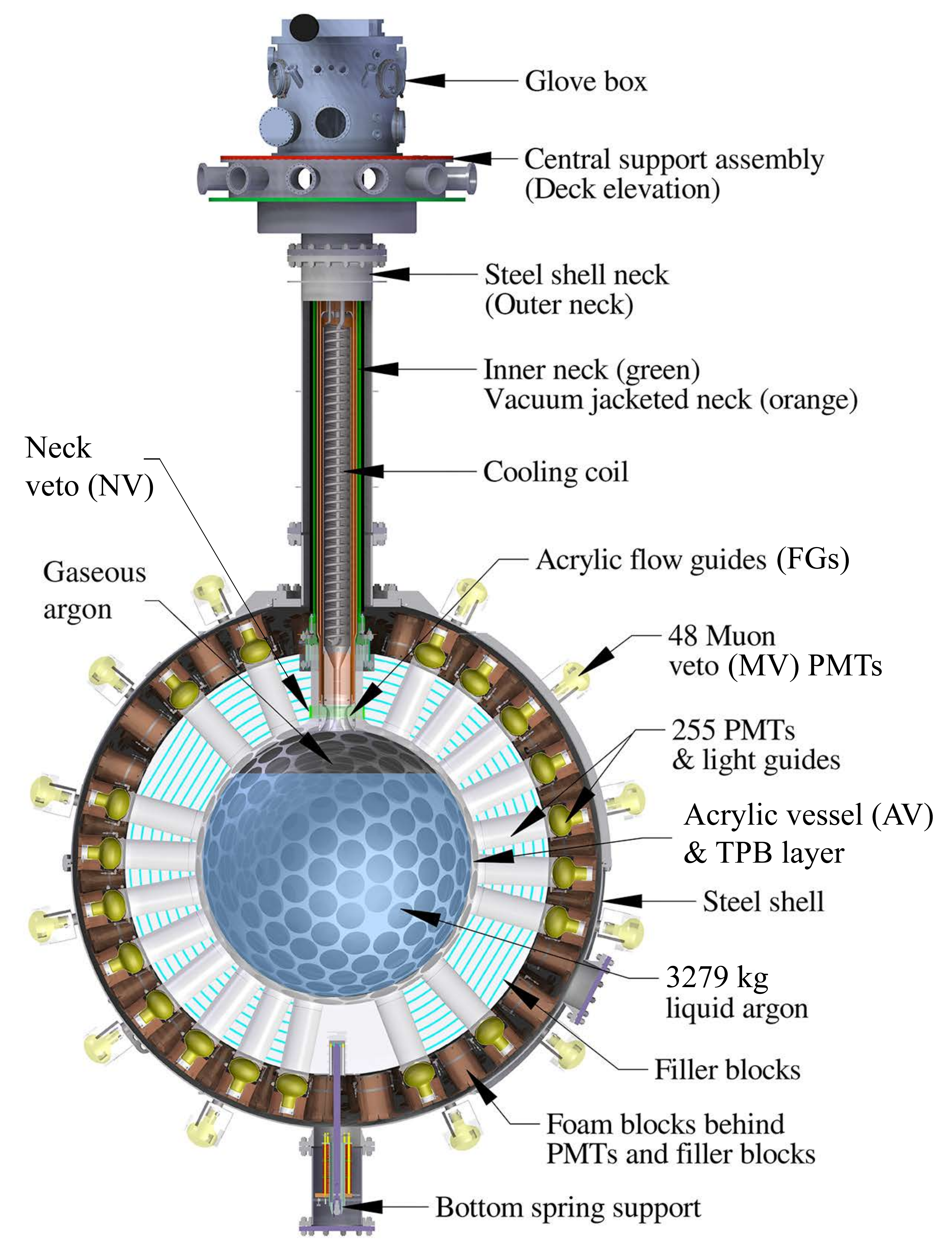}
 \caption{Cross section of the \DEAP\ detector components located inside the water tank (not shown). Inside the steel shell are inward-looking \PMTs, light guides, filler blocks, and the acrylic vessel (\AV), which holds the liquid argon target and the gaseous argon layer. Located on the outer surface of the steel shell are muon veto \PMTs. Above this, a steel neck contains the neck of the \AV, acrylic flowguides and the cooling coil. The neck is coupled to a central support assembly on which the glovebox is located. Shown also is the neck veto fiber system (green).}

 \label{fig:deapSchematic}
\end{figure} 

A cross-sectional diagram of the \DEAP\ detector is shown in Figure~\ref{fig:deapSchematic}. 
The complete design of the detector is detailed in~\cite{amaudruz_design_2017}. 
The detector consists of ultra-pure \LAr\ contained in a \AVThickness\,thick ultraviolet absorbing (\UVA) acrylic vessel (\AV) with an inner diameter of \AVInnerDiameter.
This \UVA\ acrylic was chosen to reduce the amount of Cherenkov light originating from the acrylic.
The top \GArThickness\ of the \AV\ is filled with gaseous argon (\GAr). 
The \GAr/\LAr\ interface is \larlevelcm\ above the equator of the \AV. 
The \GAr\ and \LAr\ regions are viewed by an array of 255 inward-facing \SI{8}{\inch} diameter Hamamatsu R5912 HQE low radioactivity photomultiplier tubes (\PMTs).
The characterization of these \PMTs\ is discussed in~\cite{the_deap_collaboration_-situ_2017}.
These \PMTs\ are optically coupled to \LGLength\ long \UVA\ acrylic light guides (\LGs), which transport visible photons from the \AV\ to the \PMTs.
The volume between the \LGs\ is filled with alternating layers of high density polyethylene and Styrofoam ``filler blocks'', which provide passive shielding of neutrons from detector components such as the \PMTs. 
The filler blocks also provide thermal insulation so that the \PMTs\ operate between \PMTTemperatureRange.

The inner surface of the \AV\ is coated with a \TPBThickness\ layer of 1,1,4,4-tetraphenyl-1,3-butadiene (\TPB) that converts \ArWaveLength\ scintillation light produced by the \LAr\ to visible wavelengths over a spectrum that peaks at \TPBWaveLength~\cite{francini_tetraphenyl-butadiene_2013}.
The \TPB\ was evaporated onto the inner surface of the \AV\ using a spherical source that was lowered in through the \AV\ neck; this process and characterization of the \TPB\ coating is discussed in~\cite{broerman_application_2017}.
At the wavelengths emitted by the \TPB, the light can travel through the \AV\ and \LGs\ and be detected by the \PMTs, near their peak quantum efficiency. 
These \LG-coupled \PMTs\ provide \PMTCoverage\ coverage of the \AV\ surface area. 
There are 11 distinct ``pentagonal'' regions on the \AV\ surface with reduced \LG\ coverage that are each smaller in diameter than an \LG. 
Excluding these pentagonal regions, the \LGs\ are approximately uniformly spaced around the outer \AV\ surface. 
The outer surfaces of the \AV\ between \LGs\ and the \LGs\ themselves are respectively covered with diffuse Tyvek reflectors and Mylar to enhance light collection. 

The spherical symmetry of the detector volume is broken by an opening at the top of the \AV, which leads to a \UVA\ acrylic neck and flange. This flange is connected to a longer stainless steel vacuum-jacketed neck ending in the glovebox. 
The neck contains a stainless steel liquid \nitrogen-filled (\LN) cooling coil, which condenses \GAr\ during filling and operation. 
The condensed \LAr\ enters the \AV, directed by a set of \UVA\ acrylic flowguides (\FGs) located at the opening of the neck. 
These \FGs\ direct the flow of argon to and from the cooling coil during detector operation.

Two bundles of uncladded Kuraray Y11 wavelength shifting optical fibers are wrapped around the base of the outer surface of the \AV\ neck. 
Both ends of each bundle couple to a Hamamatsu R7600-300\,\PMT, for a total of 4 neck veto (\NV) \PMTs. 
They are located above the filler blocks that surround the \AV\ neck at a distance from the \AV\ center equal to the other \AV\ \PMTs. 
The \NV\ is used to tag any visible light produced close to the \AV\ neck, a relatively photon-insensitive region of the detector.

Prior to coating the \AV\ with \TPB\ and filling the detector, a mechanical resurfacer was lowered into the detector under a low-radon atmosphere in order to remove the inner \ResurfacerThickness\ layer of acrylic along with \rntwo\ progeny that either adsorbed to or diffused into the acrylic surface while it was exposed to air during construction \cite{pgiampa_resurface}.

The entire assembly as described is contained in a stainless steel sphere that is purged with a constant flow of \radon-scrubbed \nitrogen\ gas. 
This sphere is submerged in a \MuonVetoHeight\ high by \MuonVetoDiameter\ diameter wide water tank with 48 outward-looking \SI{8}{\inch} diameter Hamamatsu R1408 \PMTs\ mounted on its outer surface. 
Together, these \PMTs\ and the water tank constitute a Cherenkov muon veto (\MV) used for tagging cosmogenically-induced backgrounds, while the shielding water provides suppression of neutron and gamma backgrounds from the cavern.

A series of calibration tubes are placed from the top of the \MV\ at locations around the stainless steel sphere.
These tubes allow radioactive sources to be lowered into the \MV, at various locations around the outside of the detector, so that it may be calibrated with neutron and \gr\ sources.
Calibration sources may be deployed with a set of detectors viewed by an additional pair of calibration \PMTs, allowing tags to be generated for events in coincidence with a radioactive decay of the source.

\subsection{Data acquisition}\label{subsec:detector_daq}

\begin{figure}[htb]
 \centering
 \includegraphics[width=\linewidth]{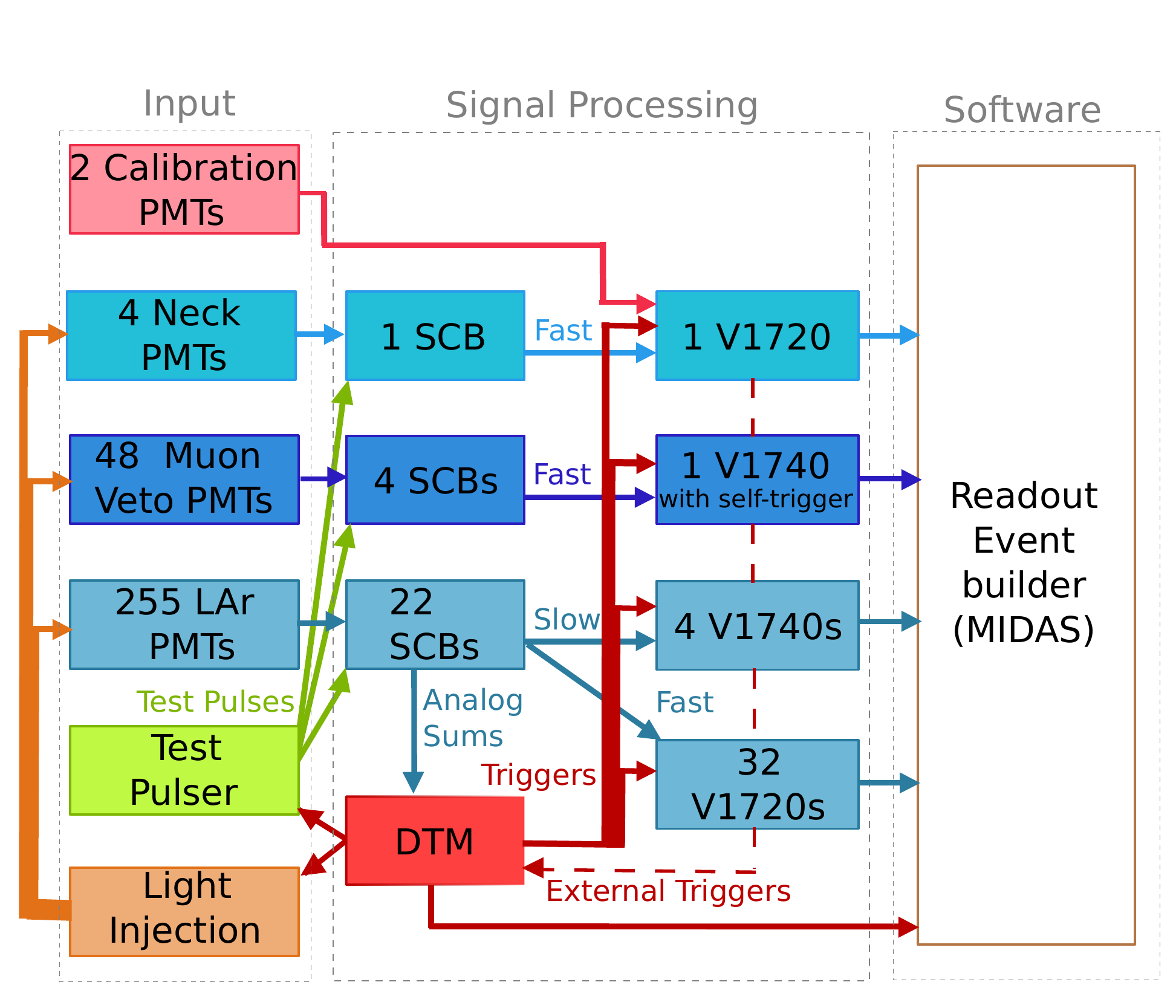}
 \caption{A block diagram of the \DEAP\ data acquisition system, adapted from~\cite{amaudruz_design_2017}. Shown are the \PMTs, the digitizer and trigger module (\DTM), the signal conditioning boards (\SCBs), the event builder, the light injection system, the test pulser systems, the fast high-gain channel digitizers (V1720s), and the slow low-gain channel digitizers (V1740s).
 } 
 \label{fig:daq}
\end{figure}

A block diagram of the data acquisition (\DAQ) system is shown in Figure~\ref{fig:daq}. 
The \DAQ\ is designed to digitize all signals from the inner detector \PMTs\ in order to achieve a timing resolution of \SI{<1}{\ns}. 
Each \PMT\ is connected to one of 12 channels on a custom-built signal conditioning board (\SCB). 
The \SCBs\ decouple the signal from the high voltage and shape the signals to optimize digitization.
A total of 27 \SCBs\ are required for all of the \AV, \MV, and \NV\ \PMTs.

The \SCBs\ output to both high-gain (V1720) and low-gain (V1740) waveform digitizer channels, which sample at \FastDigitizerRate\ and \SlowDigitizerRate\ respectively. 
Signals from the high-gain channels are used by most of the analysis, while those from low-gain channels extend the dynamic range of the detector for high energy events, such as those generated by the \alpds\ of \rntwo\ and \rnzero\ progeny in the \LAr\ target.

Each \SCB\ sums all its inputs and provides that signal to a digitizer and trigger module (\DTM), which determines when trigger conditions have been met.
The \DTM\ defines two rolling charge integrals: \qNarrow, a narrow integral over a \qNarrowTime\ window, and \qWide, a wide integral over a \qWideTime\ window. The \DTM\ then computes \qRatio\ to estimate the prompt fraction of charge. Both \qNarrow\ and \qRatio\ are used by the \DTM\ trigger decision algorithm. 
The \DTM\ pre-scales \PrescaleFactor\ of \ER-like signals (\qRatio$<0.45$) in the energy range \qNarrow$\approx$\PrescaleRangeEne\ which is predominantly populated by \arnine\ decays.
Only the \DTM\ summary information is recorded for these events, including variables such as the trigger time, \qNarrow, and \qWide. For all other kinds of events, the trigger signal is sent to the digitizers.
Special trigger signals can be set for calibration purposes.

When a trigger signal is received by the digitizers, \PMT\ waveforms are recorded on each channel for a total length of \DAQWindow, with \PreTriggerWindow\ before the trigger.

The 48 \MV\ \PMTs\ are independently read out by an additional V1740 digitizer operating in ``self-trigger'' mode. 

Zero-length-encoding (\ZLE) is employed, along with other algorithms, to reduce the volume of data recorded to disk. 
This algorithm implements zero-suppression in the firmware of each channel by ignoring regions of the waveform that are more than \ZLEPresampleTime\ removed from a per-channel voltage threshold, set to $10\%$ of the mean amplitude of a single photoelectron (\SPE).
Individual \PMT\ signals---such as photoelectrons (\PEs)---are identified from these blocks of data.

The data acquisition system is discussed in more detail in~\cite{amaudruz_design_2017}.

\subsection{PMT calibration}\label{subsec:detector_pmtcal}

The following \AV\ \PMT\ characteristics were calibrated before the detector was filled with LAr: (1) the channel-to-channel \PMT\ timing variation, which is constrained to \LaserBallJitterMeasurement, (2) the relative channel efficiencies (\ChannelEfficiencies), which besides the \PMTs' quantum efficiencies include the effect of attenuation in the \LGs\ and \LG-\PMT\ couplings, and (3) the \PMTs' afterpulsing (\AP) rates and time distributions \cite{amaudruz_design_2017, the_deap_collaboration_-situ_2017}. 
Since the \LAr\ fill, the stability of the relative efficiencies and of the afterpulsing rates are monitored continuously, and the \PMT\ single photoelectron (\SPE) charge response is calibrated daily to within \SPEChargePrecisionStat\ \SPEChargePrecisionSyst~\cite{the_deap_collaboration_-situ_2017}. 
The ongoing monitoring and calibration use both an LED light injection system \cite{amaudruz_design_2017, the_deap_collaboration_-situ_2017} and the \LAr\ scintillation light. 
The efficiencies and \SPE\ charge response of the \MV\ \PMTs\ are also calibrated regularly using injected LED light.

The stability of the \PMTs\ is discussed in more detail in Section~\ref{sec:stability}.
Further details on the \PMT\ calibration and stability monitoring techniques are discussed in~\cite{the_deap_collaboration_-situ_2017}.

\section{Data processing and reconstruction}\label{sec:recon}
Data are recorded using \MIDAS~\cite{lindner_deap-3600_2015}. 
Data analysis and Monte Carlo simulations are performed using the \RAT\ framework~\cite{rat_github}, based on \CERNRoot~\cite{brun_root_1997} and \Geant~\cite{Agostinelli:2003fg}. 

Binary files produced by \MIDAS\ are processed with \RAT\ to produce a list of \ZLE\ waveforms for each channel, with identified \PE\ detection times in the corresponding \PMT.
These values are calibrated for channel timing offsets, time-of-flight, and \PMT\ gains, and they are used to compute analysis variables, such as the energy estimator and \PSD\ parameter described below.

\subsection{Time-of-flight corrections}
Due to its size and time resolution, \DEAP\ is sensitive to the time-of-flight of photons from the scintillation vertex to the \PMTs.
To correct for this, an algorithm is employed to estimate the true event time and position.
This algorithm considers a test position $\vec{x}_0$ and event time $t_0$. 
For each \PE\ detected, the ``time residual'' $t_{\text{res}}$ is calculated as the difference between the \PE\ detection time and $t_0$ in excess of the time-of-flight implied by the straight-line distance from $\vec{x}_0$ to the relevant \PMT.
Values for $\vec{x}_0$ and $t_0$ are chosen to minimize $\sum t_{\text{res}}^2$ for pulses with $t_{\text{res}} < \TimeFitMaxResidual$. 
The best fit value of $t_0$ is then subtracted from each \PE\ detection time.

\subsection{Photoelectron counting}

A first order estimate of the number of \PEs\ detected by a \PMT\ can be found by integrating the observed charge and dividing by the mean \SPE\ charge for that \PMT.
This method was used in the first \DEAP\ result \cite{amaudruz_first_2017}.
This technique is subject to two factors which degrade the energy resolution: the width of the \SPE\ charge distribution and the presence of \AP\ charges. 
Since \PSD\ relies on measuring the number of prompt and late scintillation \PEs, mitigating these uncertainties can improve \PSD\ effectiveness. 
The root-mean-square (\RMS) of the \SPE\ charge distribution is measured to be \SPERelativeRMS\ of the mean for the \AV\ \PMTs, using laser calibration data~\cite{the_deap_collaboration_-situ_2017}.
Similarly, the mean probability of a \PE\ in an \AV\ \PMT\ generating an \AP\ is measured to be \PMTAfterPulsingProb.

The \PE\ measurement is improved by using a Bayesian \PE-counting algorithm, which determines the most likely number of \PEs\ in a \PMT\ pulse, factoring out charge produced by \APs~\cite{akashi-ronquest_improving_2015,butcher_method_2017}.
This algorithm uses a prior distribution based on the number of \PEs\ and \APs\ preceding a pulse, given its charge, the \LAr\ scintillation time profile, and the \APs' time and charge distribution for the relevant \PMT. 
The prior and the \SPE\ charge distribution are used to compute the posterior distribution of the number of \PEs.
Instead of using the  most likely number of \PEs\ in a pulse, as described by~\cite{butcher_method_2017}, the mean of the posterior distribution is used, as it was found to more accurately reproduce the tail of the pulse shape~\cite{burghardt_2018}. 

\begin{figure}[htb]
 \centering
 \includegraphics[width=1.01\linewidth]{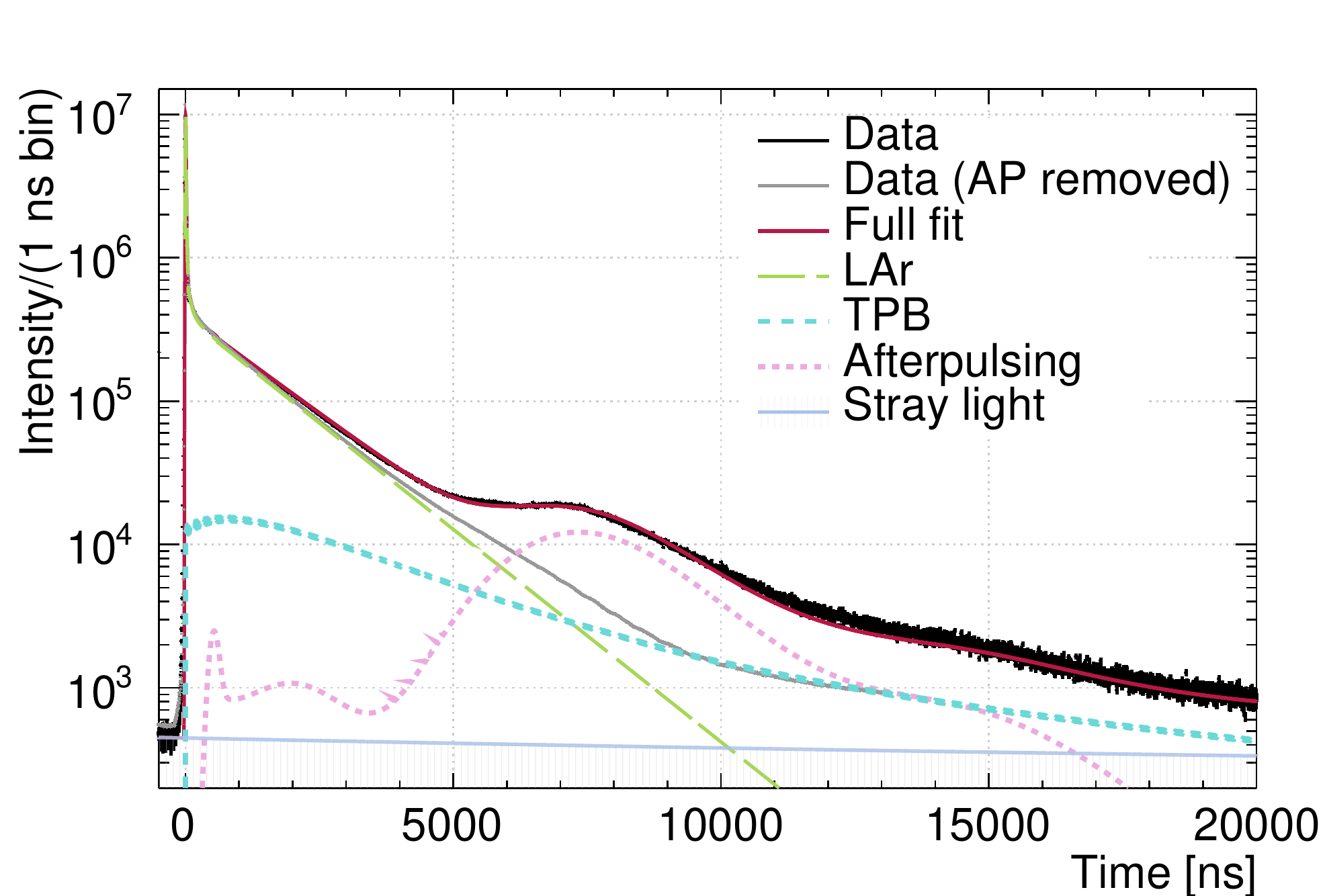}
 \caption{Average \arnine\ pulseshape before correction of instrumental effects (black) shown together with a model fit (red). The fit accounts for the following effects, which are shown individually: LAr singlet, triplet, and intermediate~\cite{heindl_scintillation_2010} light emission (green dashed), TPB prompt and delayed light emission~\cite{stanford_surface_2018} (blue dash-dotted), afterpulsing following all the previous components (pink dotted), and stray light (grey filled), which accounts for dark noise and the delayed TPB emission from previous events.
The pulse shape made from pulses that use the pulse-by-pulse AP removal algorithm (see text) is also shown (grey solid). 
}
 \label{fig:pecounting}
\end{figure}

This algorithm is applied to each \SPE-calibrated \PMT\ signal and is summed over all such signals in the first \QPEWindow\ of an event to determine the expected number of \PEs. 
The specific implementation of this algorithm and a description of its effects on \PSD\ are discussed separately in more detail in~\cite{butcher_2015,burghardt_2018}. Figure~\ref{fig:pecounting} illustrates how this algorithm separates \PEs\ and \AP\ charges, on average.
After \AP-removal, the pulse shape can be seen to closely follow the \LAr\ scintillation and \TPB\ fluorescence time profiles.

\subsection{PSD Parameter}
The parameter \FPrompt\ is defined as the fraction of \PE\ detected in a prompt window around the event time. The maximum separation between \ER\ and \NR\ events was obtained with a prompt window spanning \FpromptSixtyIntegrationWindow\ around the event time. \FPrompt\ is therefore calculated by,

\begin{equation}
\FPrompt = \frac{\sum_{t=\SI{-28}{\ns}}^{\SI{60}{\ns}} \PE(t)}{\sum_{t=\SI{-28}{\ns}}^{\SI{10}{\us}} \PE(t)}.
\label{eq:fprompteqn}
\end{equation}

\section{Detector response calibration}\label{sec:calibration}

The light yield, energy resolution, and \FPrompt\ distributions are calibrated using external radioactive sources lowered into one of the calibration pipes running along the outside of the stainless steel sphere, or using internal radioactivity naturally present in the detector, such as \arnine. Monoenergetic gamma lines are used as a cross-check.

\subsection{Light yield and energy resolution}

The detector energy response is calibrated using the \ER\ events generated by the \betds\ of the trace \arnine\ isotope in the \LAr\ target. 
The calibration is compared to measurements with a \natwo\ source and naturally occurring \gr\ lines from detector materials.
\arnine\ is naturally present in atmospherically-derived \LAr\ and \betds\ with a half-life of \ArThreeNineHalfLife\ and an endpoint of \ArThreeNineQValue~\cite{wang_ame2016_2017}. 
It has been measured to have a specific activity of \AArArThreeNineActivity~\cite{calvo_backgrounds_2018}.

A parameterization of the \arnine\ spectrum to \ER\ data describes a response function that relates the energy deposited in the detector, $E$ to the number of detected \PEs. It assumes a Gaussian response with mean $\mu$ and variance $\sigma$ defined as follows,
\begin{gather}
\begin{aligned}
 \mu&=\langle N_{\text{DN}}\rangle + Y_{\text{PE}}\cdot E, \\
 \sigma^2&=\sigma^2_{\text{PE}}\cdot\mu+\sigma^2_{\text{rel, LY}}\cdot\mu^2,
\end{aligned}
\label{eq:erespModel}
\end{gather} 
where $\langle N_\text{DN}\rangle$ is the average number of \PEs\ produced by dark noise and uncorrelated photons in the \PE\ integration window, $Y_\text{PE}$ is the light yield of the detector, $\sigma^2_{\text{PE}}$ is a resolution scaling factor that accounts for effects such as the Fano factor and \PE\ counting noise, and $\sigma^2_{\text{rel, LY}}$ accounts for the variance of the light yield relative to its mean value.

$Y_\text{PE}$, $\sigma^2_\text{PE}$, and $\sigma^2_\text{rel, LY}$ are treated as fit parameters.
$\langle N_\text{DN}\rangle$ is constrained by looking at \PMT\ signals preceding scintillation events. 
When performing spectral fits, $\langle N_\text{DN}\rangle$ is allowed to float, while a penalty term maintains that it stay within uncertainty of its nominal value.
The value of $\langle N_\text{DN}\rangle$ is found to be \AveDarkNoisePhysics\ in standard physics runs. 
For data taken with a \natwo\ source, it is measured to be \AveDarkNoiseSodium.
$\langle N_\text{DN}\rangle$ is higher when a calibration source is present due to the higher scintillation rate during these runs producing uncorrelated background photons from slow \TPB\ fluorescence on millisecond time-scales~\cite{stanford_surface_2018}. 

The \arnine\ \betd\ spectrum used in this analysis was calculated in~\cite{kostensalo_spectral_2017}, in which the shape factor is computed using nuclear shell model and the Microscopic Quasiparticle-Phonon Model codes.
This spectrum was fit to the observed \PE\ distribution, with additional contributions from \arnine\ pileup events and \gr\ backgrounds, generated by Monte Carlo simulations. 
The \gr\ spectrum is normalized to the observed rates of events coming from decays of \kforty, \bifour, and \thal\ seen at higher energies.

Uncertainties in the spectral shape of the \arnine\ energy spectrum were probed by fitting spectra evaluated from~\cite{davidson_first_1951,morita_theory_1963,hayen_high_2018} to the data.
These calculations approximate the shape factor following the prescription in~\cite{davidson_first_1951} while making additional finite nuclear size and mass corrections and radiative corrections.
The best fit was obtained using the spectrum from Kostensalo \etal~\cite{kostensalo_spectral_2017}, which converged with \chisquareperndf$=$\ArThreeNineFitChisqNDF\ in the \ArThreeNineFitRange\ range.
Further studies to better understand the \arnine\ spectral shape are currently planned.
These efforts include studying the effects of additional nuclear effects such as weak magnetism, as alluded to in~\cite{hayen_high_2018}, and applying additional radiative corrections to the spectrum computed in~\cite{kostensalo_spectral_2017}.

\begin{figure}[htb]
 \includegraphics[width=1.01\linewidth]{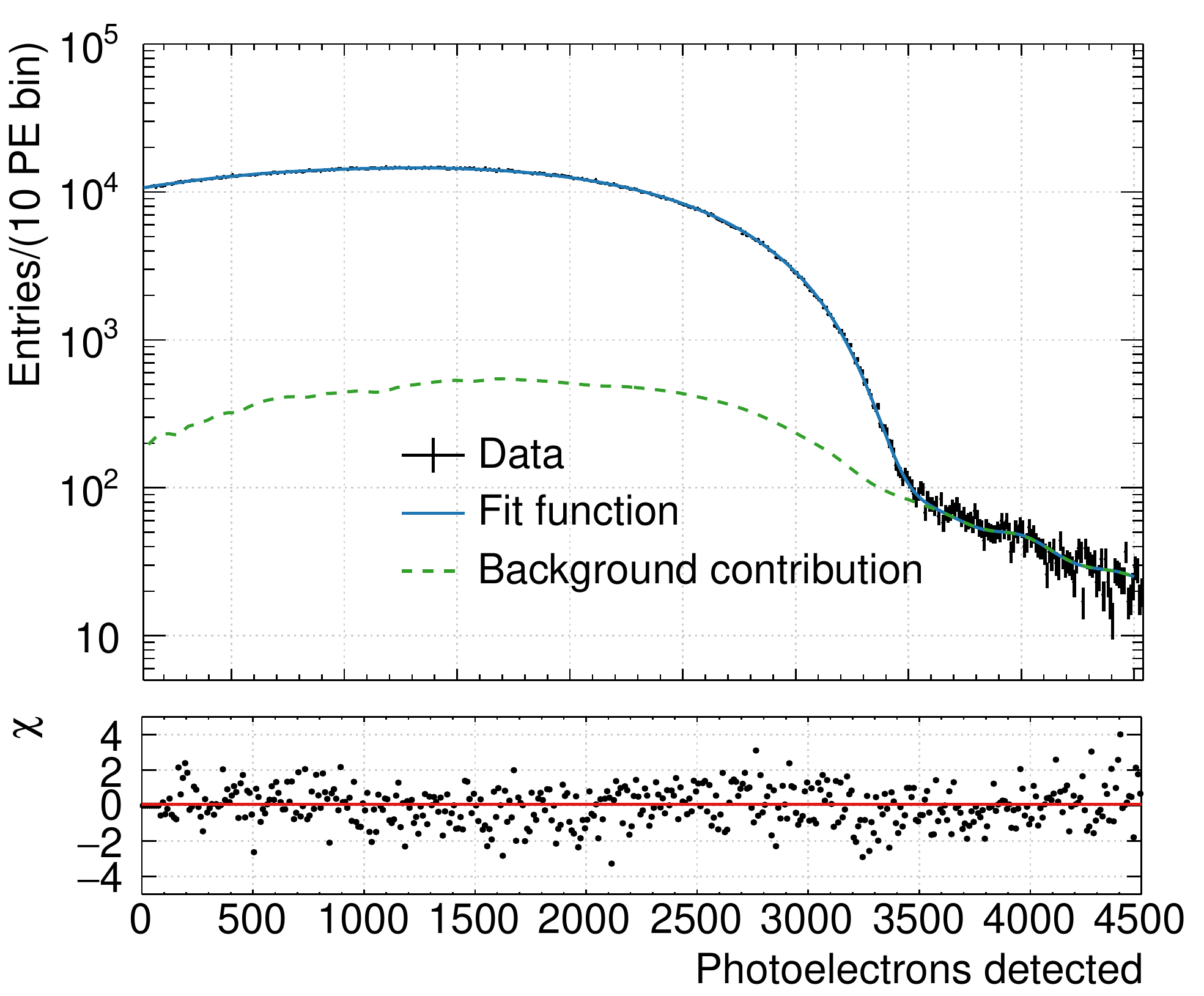}
 \caption{\arnine\ model (blue line) fit to data (black). Included in the fit is the expected background contribution from \grs\ and \arnine\ pile-up events (green). 
 }
 \label{fig:ar39}
\end{figure}

To account for potential mis-modeling uncertainty, additional fits were performed, allowing for first-order corrections to the \arnine\ spectrum $S_{\text{Ar}}(E)$ with a slope treated as a nuisance parameter $a_0$.
An additional penalty term of $(a_0/0.01)^2$ was added to $\chi^2$, to constrain its value close to 0. 
The modified $\beta$-spectrum is described by 
\begin{equation}
S^\prime_{\text{Ar}}(E) = \left(1-a_0\left(1-2E/500\right)\right)S_{\text{Ar}}(E), 
\end{equation}
where $E$ is the energy of the \betp\ in units of keV.
While such excursions may be due to deviations in the \arnine\ spectrum from the tested models, further studies are needed before a physical interpretation can be assigned to the value of $a_0$.

\begin{table}[htb]
 \centering
 \setlength\extrarowheight{2pt}
 \caption{
  Best fit response function parameters from a fit to \arnine\ events collected throughout the data collection period. The fit converged with \chisquareperndf\ of \ArSpectrumFitChiSqNDF.
  The value shown for $\langle N_\text{DN}\rangle$ is derived from direct measurements, as described in the text.
  }
 \begin{tabular}{lcc}\hline\hline
  \multirow{2}{*}{PE mean}    & $\langle N_\text{DN}\rangle$ & $Y_\text{PE}$                    \\
                              & \DarkNoiseWithErr            & \LightYieldWithErr               \\\hline
  \multirow{2}{*}{Resolution} & $\sigma^2_{\text{PE}}$       & $\sigma^2_{\text{rel, LY}}$       \\
                              & \ResFactorWithErr            & \RelLYVarWithErr                 \\
  \hline\hline
 \end{tabular}
 \label{tab:fitresults}
\end{table}

With this nuisance parameter, the fit was found to converge with \chisquareperndf$\approx$\ArSpectrumFitChiSqNDF, with a \ArNuisanceVariation\ deviation from the spectrum derived in~\cite{kostensalo_spectral_2017}.
The origin of this deviation is not yet understood, and is still being investigated. It is found to have little effect on the best fit values of the response function parameters or on the final \WIMP\ search result. The results of this fit are shown in Figure~\ref{fig:ar39}.
The differences between the best fit values for each parameter with and without the nuisance parameter are propagated into the parameters' uncertainties.
The best fit response function parameters are shown in Table~\ref{tab:fitresults}.

\begin{figure}[htb]
 \includegraphics[width=1.01\linewidth]{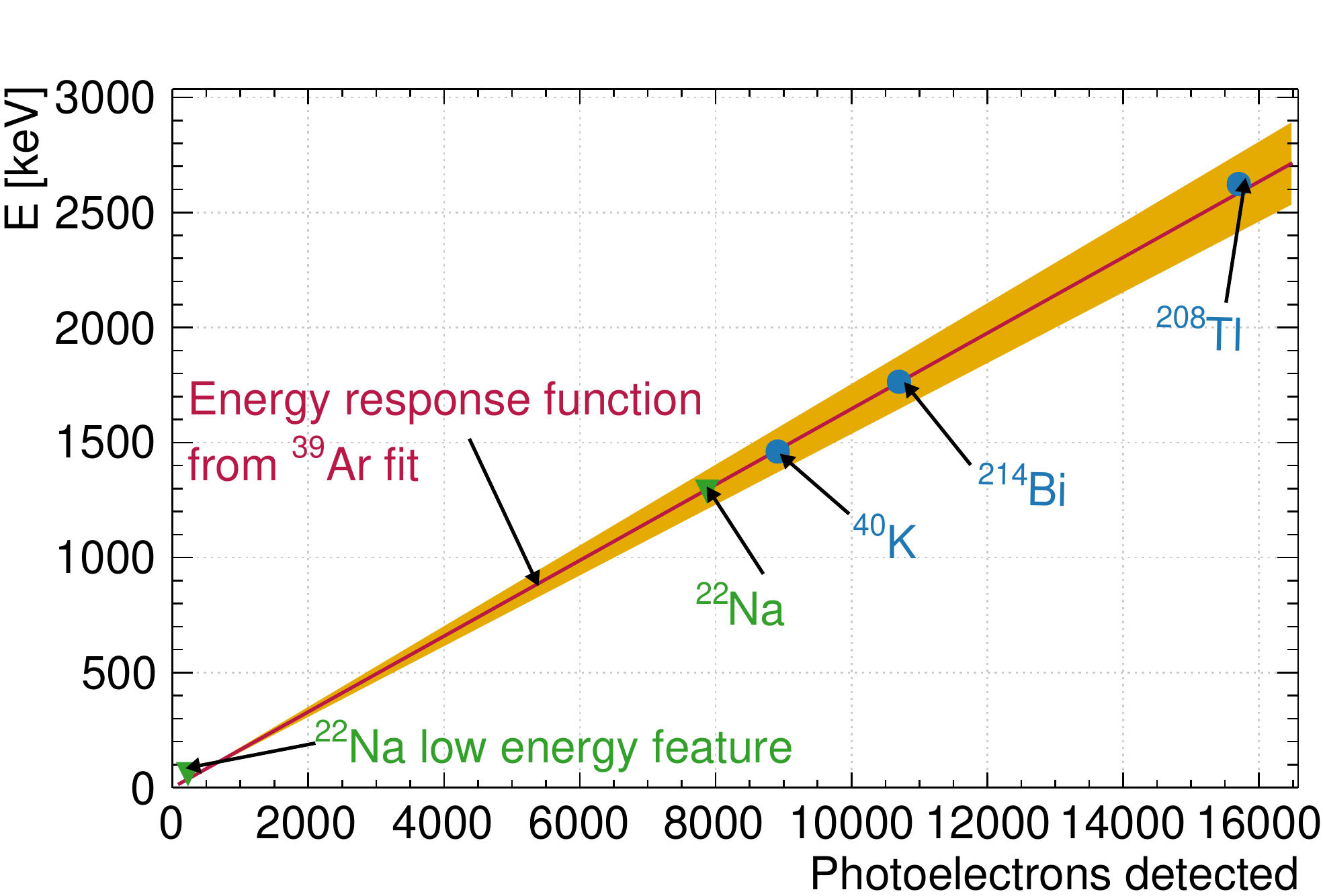}
 \caption{The energy response function (red), showing the number of detected \PE\ for an event depositing energy $E$ in the \LAr. The uncertainties of the response function are also shown (yellow band). The response function agrees with the number of \PE\ detected from known mono-energetic sources of \grs\ from the detector materials. 
 }
 \label{fig:ecal}
\end{figure}

A \natwo\ source was lowered into the calibration tubes outside the stainless steel shell to compare the consistency of the response function calibrated with \arnine\ to the spectrum produced by events from tagged \natwo\ decays, which contains a prominent 1.27\,MeV \gr\ and a low energy spectrum feature resulting from \grs\ attenuating in acrylic~\cite{amaudruz_first_2017}. 
A cross-check using the \gr\ lines from \kforty\ (\KFortyGammaEnergy), \bifour\ (\BiTwoOneFourGammaEnergy), and \thal\ (\ThTwoZeroEightHighEGammaEnergy) is also performed. 
These isotopes are naturally present in detector materials and are visible in standard physics runs.
Figure~\ref{fig:ecal} shows the estimated number of detected \PEs\ using the light yield from \arnine\ extrapolated out to these energies.
As shown in this figure, the energy response function remains very linear over a wide range of energies, with non-linearities starting to arise above \KFortyGammaEnergy\ due to digitizer saturation.

Data were also collected with an \AmBe\ neutron source deployed in order to validate the \NR\ quenching and \PSD\ models. 
\NR\ quenching factors were derived from \SCENE\ measurements~\cite{Cao:2015ks}, using the Lindhard-Birks model fit to the measured \NR\ light yields relative to \krthree\ \ERs. 
The estimated uncertainties for these quenching factors were dominated by the uncertainty in the Birks factor. 

\begin{figure}[htb]
 \includegraphics[width=1.01\linewidth]{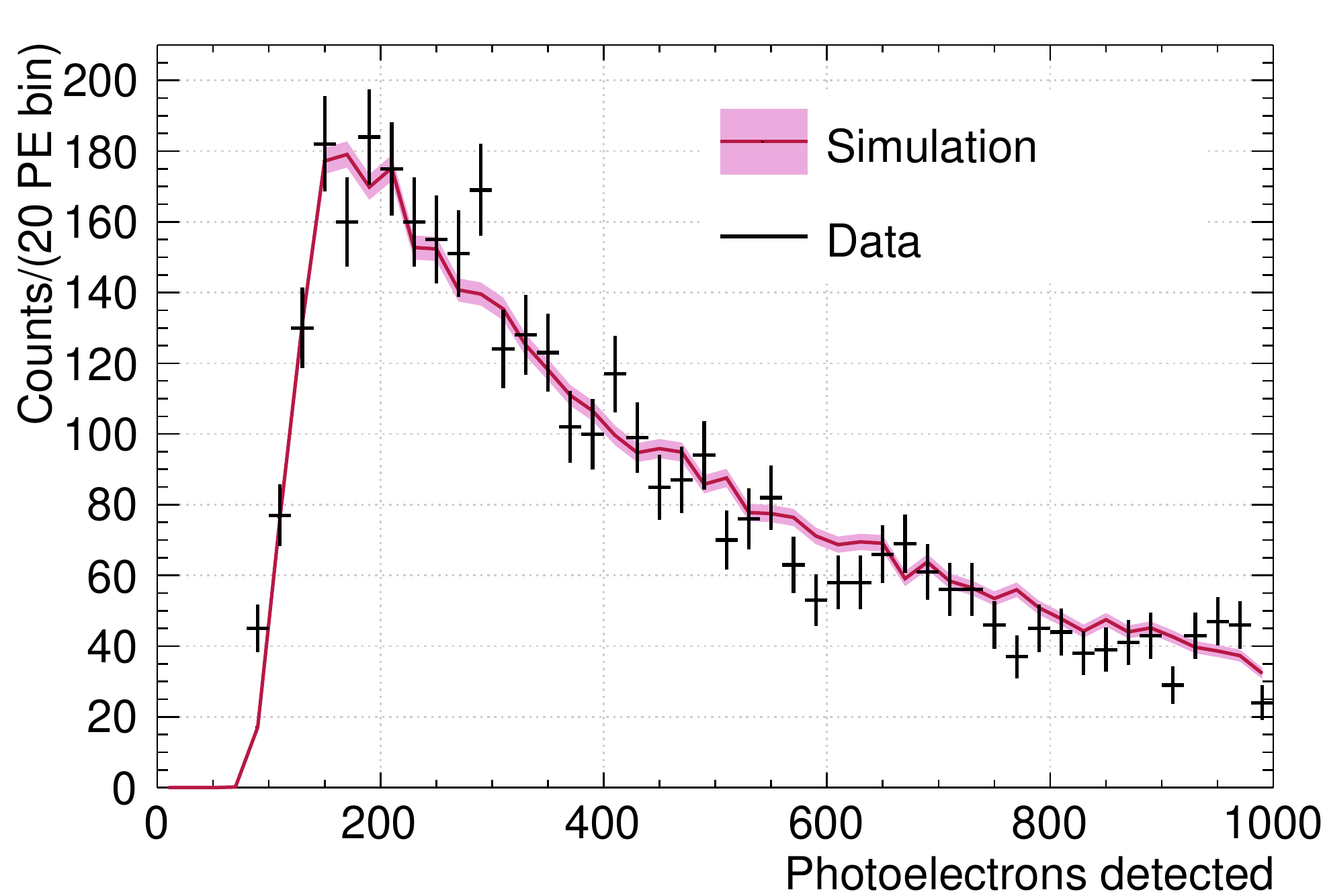}
 \caption{Comparison of the detected \PE\ distribution for high-\FPrompt\ \NR-like events from an \AmBe\ neutron source simulation (pink) and data (black). The peak at low \PE\ is due to the \FPrompt\ cut that removed \ERs; due to the high rate of \ERs\ in coincidence with neutron-induced \NRs\ in the calibration data, particularly strong cuts are needed to obtain a clean \NR\ spectrum.
Uncertainties shown are all statistical, using nominal values for the quenching and detector optics models.
} 
 \label{fig:ambespec}
\end{figure}

This model is implemented in the simulation and validated by comparing the observed \PE\ spectrum of neutron-induced \NRs\ in the \AmBe\ neutron source data to the simulated one.
The agreement between the model and data can be seen in Figure~\ref{fig:ambespec}.

\subsection{\FPrompt\ distributions} \label{subsec:calibration_psd}

Following a particle interaction, excimers form in the \LAr, and the singlet/triplet population ratio is a function of the nature and the energy of the interaction. 
Due to the different decay times of the two types of excimers, different particles produce different \FPrompt\ distributions that vary with their energy. 
In this analysis, \PSD\ is used to differentiate between \NRs, \ERs, and \alpp\ interactions.

\subsubsection{Electronic recoils}

An empirical function has been developed that characterizes the \FPrompt\ distribution for \ERs; this function was chosen as it was found to describe the data well over a wide range of energy. For an \ER\ event in which $q$ \PE\ are detected, the probability of observing an \FPrompt\ value of $f$ is described by,
\begin{gather}
\begin{aligned}
 F^{\text{ER}}(f,q) &= \Gamma(f;\bar{f},b)\ast\text{Gauss}(f;\sigma), \\
 \bar{f}(q)         &= a_0 + \frac{a_1}{q-a_2} + \frac{a_3}{(q-a_4)^2},               \\
 b(q)               &= a_5 + \frac{a_6}{q}     + \frac{a_7}{q^2}   ,                    \\
 \sigma(q)          &= a_8 + \frac{a_9}{q}     + \frac{a_{10}}{q^2},
\end{aligned}
\label{eq:psdmodel}
\end{gather}
where $\Gamma(f;\bar{f},b)$ is the Gamma distribution with mean $\bar{f}$ and shape parameter $b$, and $\text{Gauss}(f,\sigma)$ is a Gaussian distribution with standard deviation $\sigma$ and a mean of 0. 
The parameters $a_i$ are fit parameters that describe how $\bar{f}$, $b$, and $\sigma$ vary with $q$.

\begin{figure}[htb]
 \centering
  \includegraphics[width=1.01\linewidth]{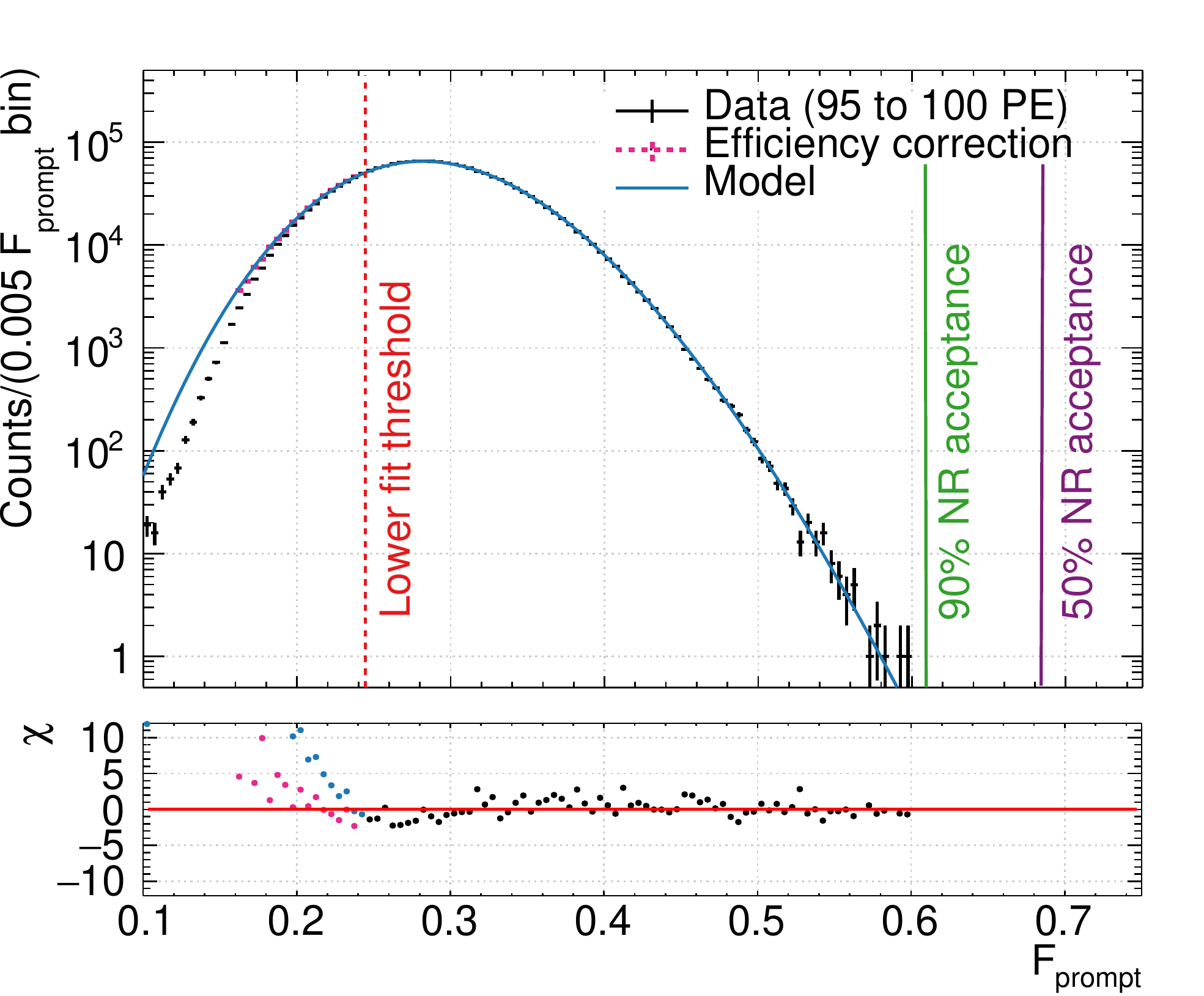}
 \caption{\FPrompt\ distribution for \ERs\ from standard physics data in the lowest \SI{1}{\keVee} energy bin in the \WIMP-search region of interest. The \PSD\ model is fit to the data to the right of the vertical dashed line, where the trigger efficiency is approximately unity. Agreement between the best fit model and the data can be seen; below the \FPrompt\ fit region, trigger efficiency corrections to the data show that the model agrees when extrapolated to lower values. Solid vertical lines show the \FPrompt\ value in the corresponding \PE\ bins above which \SI{90}{\percent} or \SI{50}{\percent} of \NRs\ are expected to be found.  In the bottom plot, black points show residuals over the range where the model is fit; blue points compare the extrapolated model directly to the observed data prior to correcting for the decreasing trigger efficiency, while pink points compare the extrapolated model to the data after making these corrections.
 }
 \label{fig:rpromptFit}
\end{figure}

The parameters $a_i$ are fit to the distribution of \FPrompt\ vs. \PE.
Within each \PE\ bin, the resulting values of $\bar{f}$, $b$, and $\sigma$ describe the shape of the \FPrompt\ distribution, neglecting trigger efficiency effects.
In each \PE\ bin, the fit considers values of \FPrompt\ for which the trigger efficiency is estimated to be greater than \PSDFitsMinFPrompt.
The resulting fit is well-constrained and converges with \chisquareperndf$=$\PSDFitsChisqNDF. 
In effect, $\bar{f}$, $b$, and $\sigma$ are the physically relevant parameters while $a_i$ parameterize their energy dependence, forcing them to vary smoothly across \PE\ bins and allowing \FPrompt\ distributions to be interpolated.
An example of this fit in a single \PE\ bin is shown in Figure~\ref{fig:rpromptFit}. 
The validity of this fit has been tested by performing it over a limited range of \FPrompt\ and comparing extrapolated values to the data outside the fit range.
These tests show that extrapolated expectations agree with the data, indicating the robustness of this method.

Since the \DTM\ triggers on the number of prompt \PEs, low \FPrompt\ events at low \PE\ are less likely to produce a trigger signal. 
A software correction has been developed to account for the reduced trigger efficiency for these events, following the procedure described in~\cite{pollmann_estimating_2019}. 
Data with this correction applied are shown in Figure~\ref{fig:rpromptFit}. 
While $F^{\text{ER}}(f,q)$ is only fit over the range where the trigger efficiency is near unity, the extrapolated model agrees better with the efficiency-corrected \FPrompt\ distribution.

\subsubsection{Nuclear recoils}

Mean \FPrompt\ values for \NRs\ are determined from measurements reported by the \SCENE\ collaboration~\cite{Cao:2015ks}.
\SCENE\ reports median values of \FNinety, defined as the fraction of charge observed in the first \SI{90}{\ns} of an event, for different \NR\ energies.
Equivalent singlet/triplet ratios are determined for each median \FNinety\ value, which are used as input to a Monte Carlo simulation of \DEAP.
This simulation propagates the detector timing response, including photon times-of-flight and \PMT\ effects such as \AP\ into the resulting \FPrompt\ distribution.
Uncertainties in the extracted singlet/triplet ratio are determined from uncertainties reported by \SCENE\ as well as uncertainties in the singlet and triplet lifetimes.
Uncertainties from the \AP\ rates and triplet lifetime in \DEAP\ are also propagated into the uncertainty on the mean \FPrompt\ values.

For \NRs, it is assumed that the spread of the \FPrompt\ distribution around the mean is governed by the same effects that drive the spread in the \ER\ distribution, with an inverted skew.
The \FPrompt\ distribution for \NRs\ with $q$\,\PE\ is then given by,
\begin{gather}
\begin{aligned}
 F^{\text{NR}}(f,q) &= \Gamma(1-f;1-\bar{f},b)\ast\text{Gauss}(f;\sigma),    \\
 b(q)               &= a_5 + \frac{a_6}{q}     + \frac{a_7}{q^2},                     \\
 \sigma(q)          &= a_8 + \frac{a_9}{q}     + \frac{a_{10}}{q^2},
\end{aligned}
\label{eq:psdnrmodel}
\end{gather}
where $\bar{f}(q)$ is the mean \FPrompt\ value for \NRs\ at $q$, predicted by the simulation, and $b(q)$ and $\sigma(q)$ are governed by the fit parameters $a_i$ in Equation~\ref{eq:psdmodel}.

\begin{figure}[htb]
 \centering
 \includegraphics[width=1.01\linewidth]{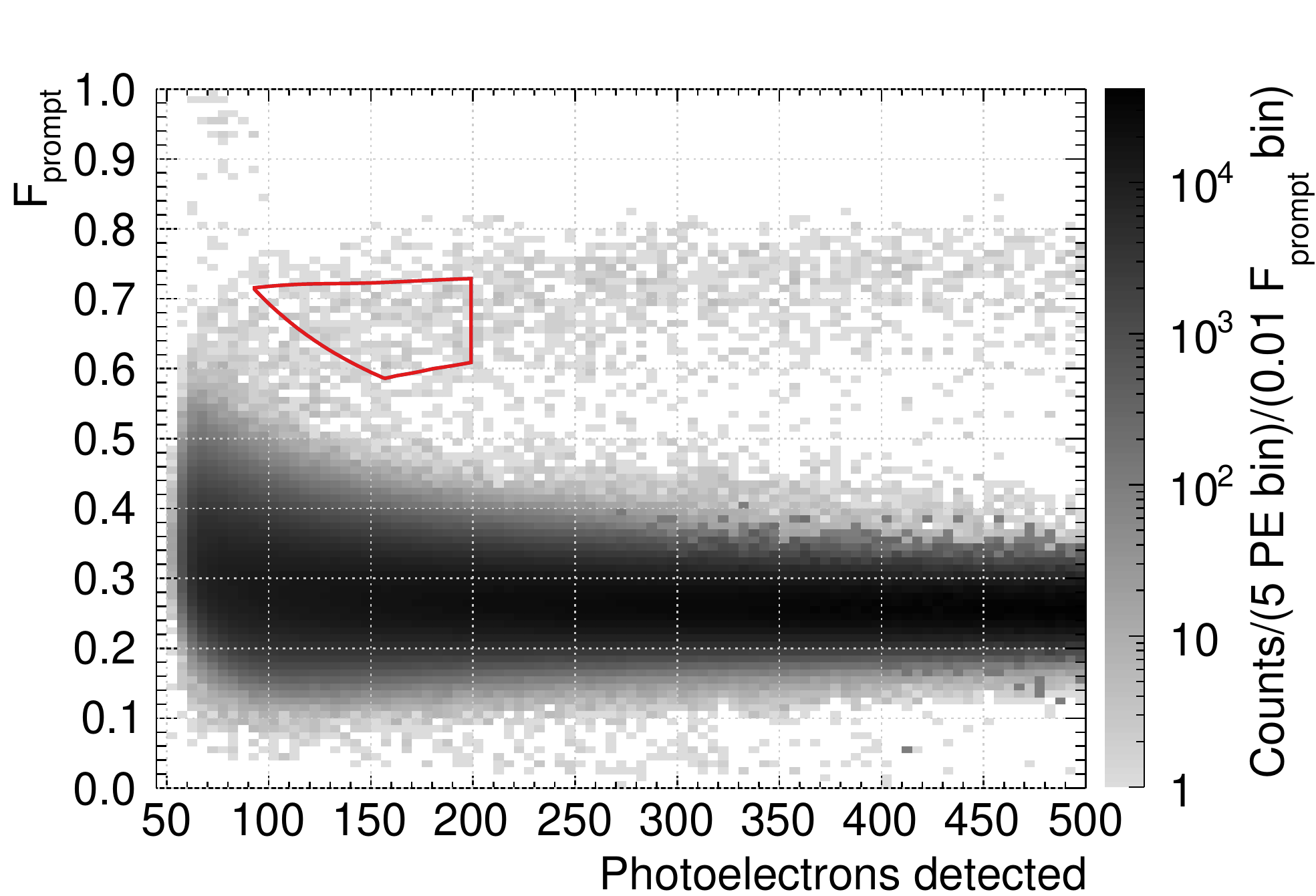}
 \caption{\FPrompt\,vs.\,\PE\ distribution for data taken with an \AmBe\ neutron source near the equator of the stainless steel shell. The \WIMP-search \ROI\ is shown in red. Separation can be seen between the \NR\ band (\FPrompt\SI{\sim0.7}{}) and \ER\ band (\FPrompt\SI{\sim0.3}{}). Events in the prescaled region (\ER\ band with \PE$\gtrsim$300) are weighted with a correction factor of \SI{100}{}, corresponding to the prescale factor. Note that multiple neutron scatters and pileup with Cherenkov and \ER\ signals bias the \NR\ spectrum away from the expected distribution for pure single-scatter \NRs\ and populate the region between both bands.
 }
 \label{fig:ambePSD}
\end{figure}

\begin{figure}[htb]
 \centering
 \includegraphics[width=1.01\linewidth]{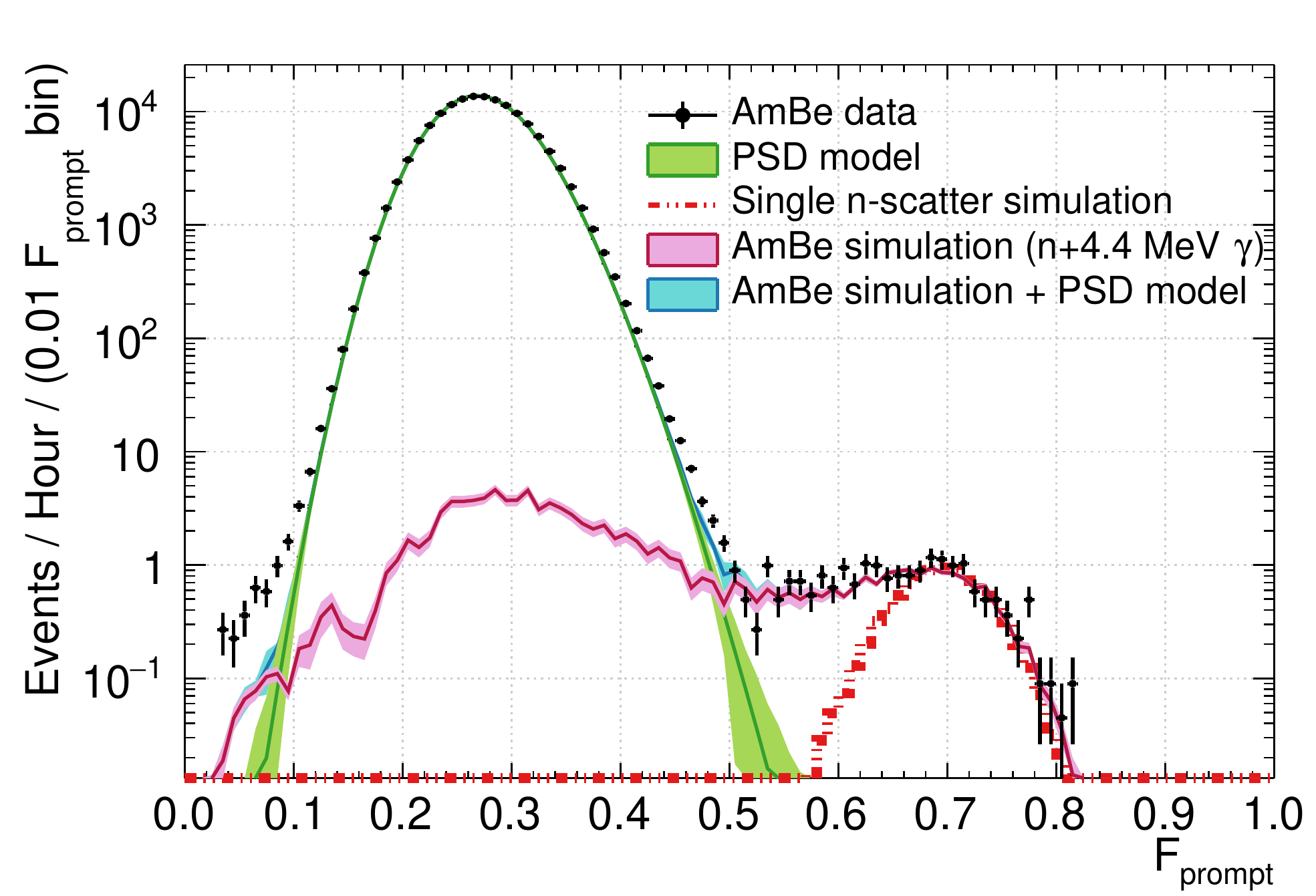}
 \caption{\FPrompt\ distribution in the \SIrange{120}{200} \PE\ range of events from AmBe data (black) and simulations of single-scatter neutrons (red dashed). Also shown are simulated events from an AmBe source (pink), the \ER\ \PSD\ model (green) and their sum (blue). 
 }
 \label{fig:ambeRprompt}
\end{figure}

The \NR\ \FPrompt\ distribution is validated using \AmBe\ calibration data. 
An \AmBe\ neutron source was lowered into a calibration tube outside of the stainless steel shell of the detector.
The resulting signal seen in the detector is shown in Figure~\ref{fig:ambePSD}, with the \WIMP-search \ROI\ shown for comparison.
Many of the neutron-induced \NRs\ are accompanied by \ER\ or Cherenkov pileup from \grs\ correlated with neutron production in the \AmBe\ source, while others are biased by multiple scatter events.
As a result, we do not expect the \AmBe\ data to directly reproduce the \FPrompt\ distribution predicted for single scatter \NRs.
Instead, we simulate the \AmBe\ source and compare the simulated and observed \FPrompt\ distributions.
Figure~\ref{fig:ambeRprompt} shows this comparison; agreement between data and simulation to within uncertainties indicates the validity of the model.

\subsubsection{$\alpha$ decays }

\DEAP\ detects full energy \alpd\ events produced by \rntwo, \rnzero, and their progeny from within the \LAr. 
These events reconstruct above \SI{\sim23000}{}~\PE\ and are subject to digitizer and \PMT\ saturation effects that reduce the number of detected \PE\ and the value of \FPrompt\ when using the normal high-gain scheme intended for low \PE\ events.
This effect broadens \PE\ and \FPrompt\ distributions, biasing their values downward by preferentially causing the number of prompt \PE\ to be underestimated. 

The three most frequent \alpds\ in the \LAr\ are \rntwo, \poeight\ and \pofour\ (\alpp\ energies of \RnTwoAlphaEnergy, \PoEightAlphaEnergy, and \PoFourAlphaEnergy\ respectively). 
Signals observed using the low-gain channels are used to apply digitizer and \PMT\ saturation corrections to signals observed in the high-gain channels, as described in~\cite{jmclaughlin_2018}. 
These corrections allow for more accurate \FPrompt\ and \PE\ values to be calculated and a parametrization between the mean \FPrompt\ as a function of \alpp\ energy.
This parametrization is implemented into the simulation for \alpp\ scintillation in \LAr, and extrapolated across the energy range \SIrange{5.0}{10.0}{\MeV} such that \poten\ (\PoTenAlphaEnergy) and higher energy \rnzero\ daughters like \potwo\ (\PoTwoAlphaEnergy) can be modelled. At \PoTenAlphaEnergy, the model uncertainty corresponds to a \AlphaScintillationFpromptUncertainty\ uncertainty in the mean \FPrompt\ value.

Understanding the relationship between the mean \FPrompt\ and energy for \alpps\ allows for modelling of high energy \alpds\ in the \AV\ neck region. These events are shadowed and reconstruct with low \PE.
As will be discussed in Section~\ref{sec:neck-alphas}, such events are caused by the absorption of ultraviolet (\UV) scintillation by acrylic components in the \AV\ neck. 
These events are not affected by digitizer clipping or \PMT\ saturation effects and hence the \FPrompt\ of these events preserves information about the \alpp\ energy that produced them.

\section{Position reconstruction}\label{sec:position}

\DEAP\ utilizes two complementary position reconstruction algorithms: one using the spatial distribution of \PMT\ hits (\ChargeBasedPosRec\ algorithm) and one that also includes timing information (\TimeBasedPosRec\ algorithm).

The \ChargeBasedPosRec\ algorithm computes the likelihood $\mathcal{L}(\vec{x})$ that the scintillation event happened at some test position $\vec{x}$ as,
\begin{gather}
 \begin{aligned}
 \ln\mathcal{L}(\vec{x}) &= \sum_{i=1}^{N_{\text{PMTs}}}\ln\text{Poisson}\left(q_i;\lambda_i\right), \\
 \lambda_i               &= \lambda_i\left(|\vec{x}|,\frac{\vec{x}\cdot\vec{r}_i}{|\vec{x}||\vec{r}_i|},q_{\text{total}}\right),
 \end{aligned}
\end{gather}
where $\text{Poisson}(q_i;\lambda_i)$ is the Poisson probability of observing $q_i$\,\PE\ in \PMT\ $i$ at position $\vec{r}_i$ over the full \QPEWindow\ event window. 
The expected number of \PE\ in \PMT\ $i$ is given by $\lambda_i$, which is a function of the radius of the test position $|\vec{x}|$, the angle between the test position and PMT$_i$, and total PE integrated over all PMTs $q_\mathrm{total}$.

Values for $\lambda_i$ are calculated based on a Monte Carlo simulation of the detector, including the full optical model.
These simulations assume a completely filled detector, with scintillation events generated inside the \LAr\ along three distinct axes: one collinear with the axis of the \AV\ neck and two perpendicular axes within the equatorial plane of the \AV. 
A set of splines is then used to generate tables of $\lambda_i$ values.
This algorithm does not account for timing information within the \QPEWindow\ event window.
The position returned by this algorithm is the one that maximizes $\ln\mathcal{L}(\vec{x})$.

In contrast, the \TimeBasedPosRec\ algorithm uses both charge and time information of early pulses in an event to calculate the position. 
As with the time-of-flight corrections used to correct \PE\ detection times, time residuals are defined as the time at which a \PE\ was detected in excess of what the time-of-flight would suggest.
However, this algorithm uses a more precise, albeit slower method for determining the time residuals.
Prior to data processing, a grid of test positions $\vec{x}_j$ is defined inside the \LAr\ relative to the \PMT\ location, and the time residual distribution $\mathcal{L}^{t\text{ res}}(\Delta t; \vec{x}_j, \text{PMT}_i)$ is calculated.
These calculations utilize a simplified optical model of the detector, including the group velocities of \UV\ photons emitted by \LAr\ (\ArWaveLengthGVelocity\ at \ArWaveLength) and visible photons emitted by \TPB\ (\TPBWaveLengthGVelocity\ at \TPBWaveLength), as well as the \LAr\ scintillation and \TPB\ fluorescence time constants, the average travel time of visible photons in the \LG\ and \AV\ acrylic, and the average \PMT\ response time.
The group velocities used for these calculations were determined based on measured \LAr\ refractive indices at various wavelengths, as reported by~\cite{sinnock_refractive_1969}, following the procedure described in~\cite{grace_index_2015}.

In the simplified optical model, reflections and scattering of visible photons in \TPB\ are neglected.
Rayleigh scattering in the \LAr\ is neglected as well; in the \ChargeBasedPosRec\ alogithm, scattering lenghts of \SI{1.65}{\m} and \SI{1082}{\m} are assumed for wavelengths of \SI{128}{\nm} and \SI{420}{\nm}, respectively, following the procedure outlined in~\cite{grace_index_2015}. 
The time response of the \LGs\ and \PMTs\ from the initial calibration of the detector as discussed in Section~\ref{subsec:detector_pmtcal} is assumed. 

The likelihood $\mathcal{L}(t_0,\vec{x}_0)$ of a given event time $t_0$ and test position $\vec{x}_0$ is computed as
\begin{equation}
 \ln\mathcal{L}(t_0,\vec{x}_0) = \sum_{i=1}^{N_{\text{PE}}}\ln\mathcal{L}^{t\text{ res.}}(t_i-t_0;\vec{x}_0,\text{PMT}_i),
\end{equation}
where $t_i$ is the time at which the $i^\text{th}$ \PE\ was detected in channel $\text{PMT}_i$; the $N_{\text{PE}}$ in the first \TimeFitTwoWindow\ are considered for this calculation.
This algorithm returns the values of $\vec{x}_0$ and $t_0$ that maximize $\ln\mathcal{L}(t_0,\vec{x}_0)$.

\subsection{Validation}\label{sec:position-validation}

The \WIMP-search analysis presented here relies primarily on the \ChargeBasedPosRec\ algorithm for fiducialization, though it also requires that both algorithms converge and agree with each other. 
Doing so allows for the rejection of events whose position are mis-reconstructed, as may be the case for events originating outside of the \LAr, where the assumptions underlying both algorithms are not realized. 

\begin{figure}[htb]
 \centering
 \includegraphics[width=1.01\linewidth]{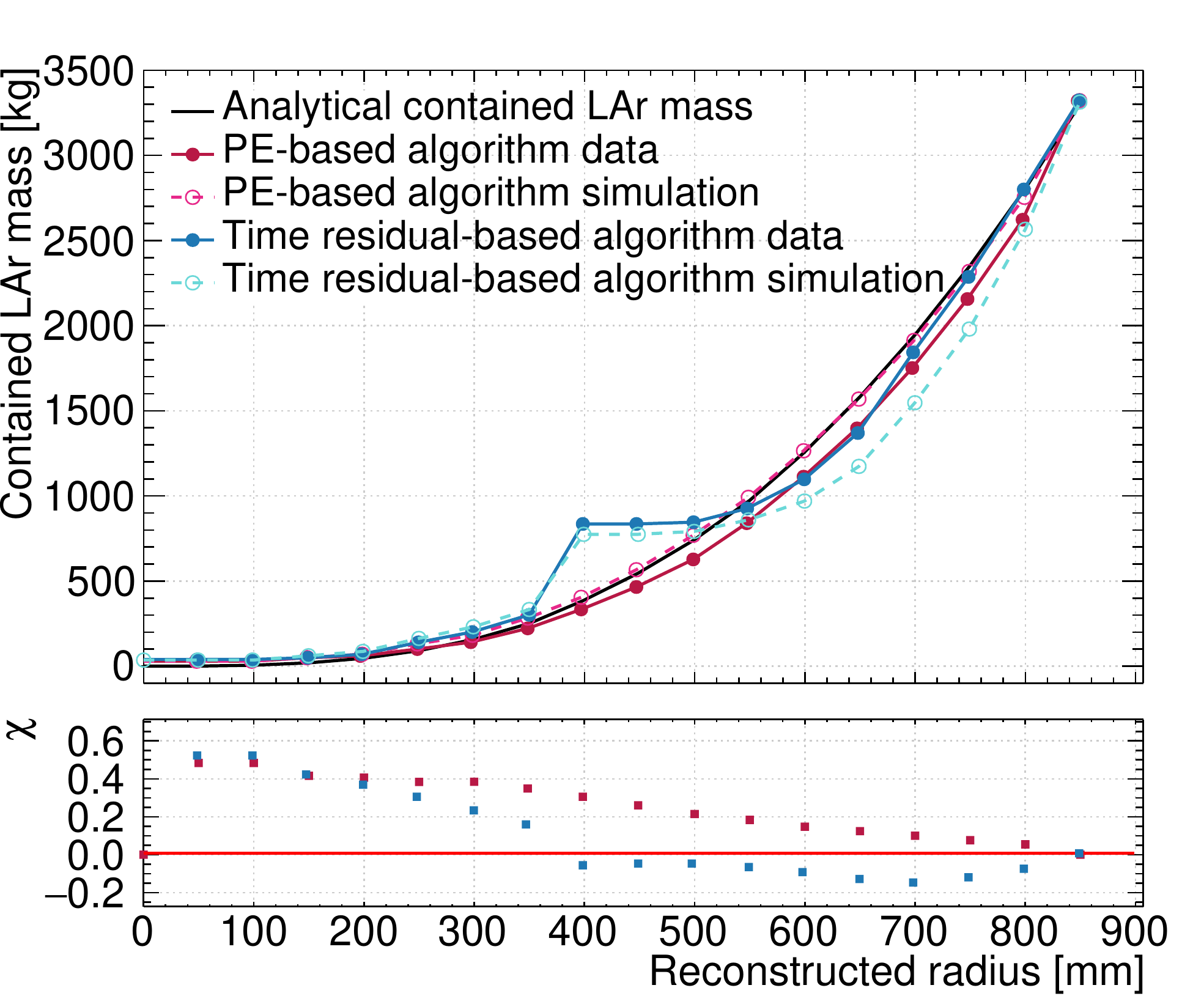}
 \caption{Estimates from the \ChargeBasedPosRec\ (red) and \TimeBasedPosRec\ (blue) algorithms of the contained mass of \LAr\ within a radius of the reconstructed position. The estimate is based on the total fraction of \arnine\ \ER\ signals in the 95--200 \PE\ range reconstructing within a given radius. Values from both data (solid) or simulations (dashed) are shown. It is assumed that the true positions of the \arnine\ nuclei are uniformly distributed throughout the \LAr\ target. Shown also is the estimated contained mass calculated by considering only the geometric volume (black). 
 The bottom inset shows the difference between data and simulation for each algorithm.
 }
 \label{fig:positionFitterUniformity}
\end{figure}

Both algorithms are validated using \arnine\ \betds, uniformly distributed in the \LAr.
Non-uniformities in their reconstructed positions therefore provide a measure of the algorithms' bias.
Figure~\ref{fig:positionFitterUniformity} demonstrates the uniformity of the \ChargeBasedPosRec\ algorithm. 
The \TimeBasedPosRec\ algorithm, which provides an additional test for mis-reconstruction, exhibits a sharp change between reconstructed radii values \SIrange{350}{400}{mm}. This non-uniformity is an artifact of the time-residual calculations used by this algorithm, and is subject to refinement in a future analysis. 
The fiducial radius used in this analysis is \PaperTwoRadialCut\ based on the returned value of the \ChargeBasedPosRec\ algorithm; at this value data and simulation agree to within \ChargeBasedFiducialDiscrepancy.

\begin{figure}[htb]
 \centering
 \includegraphics[width=1.01\linewidth]{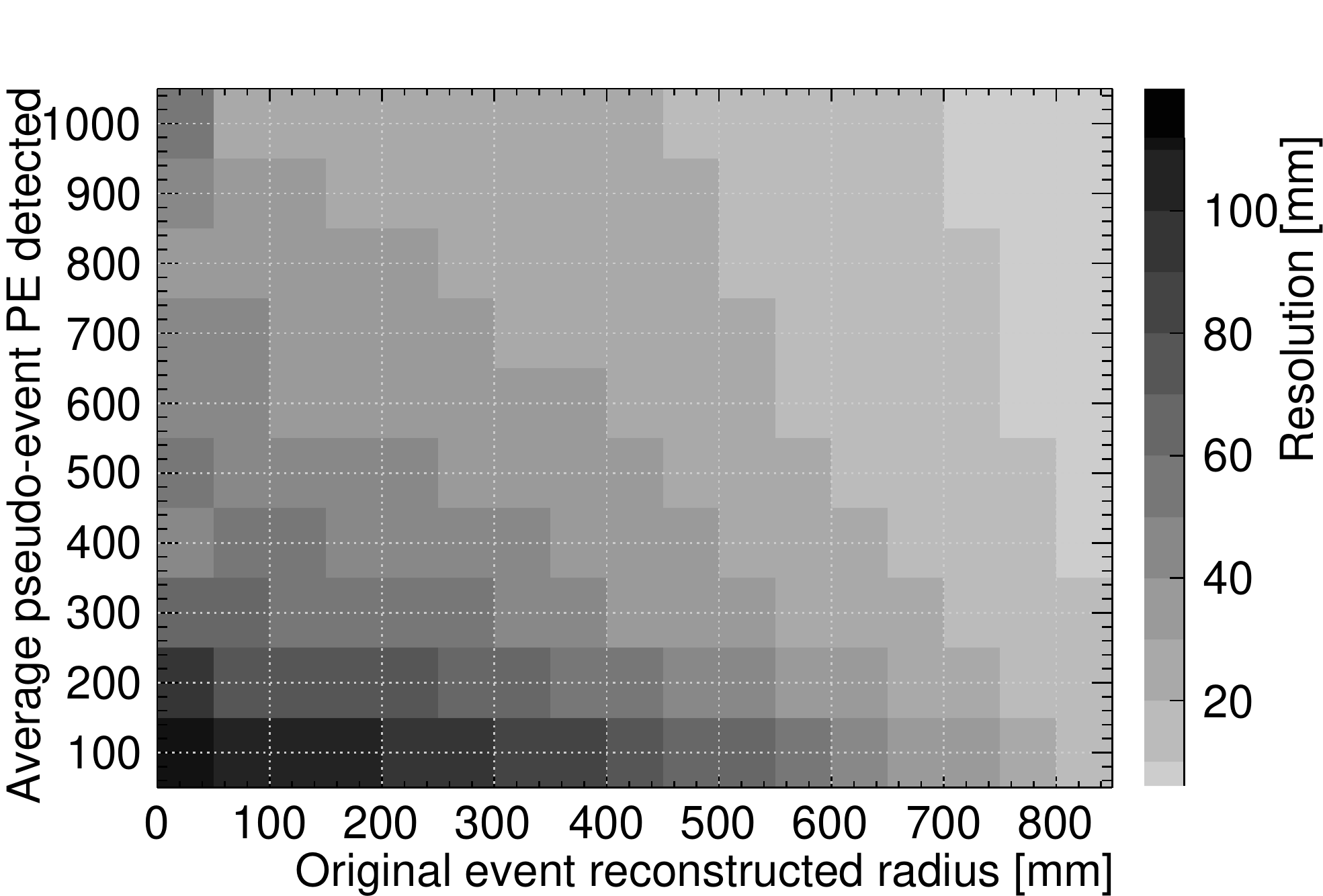}
 \caption{Position resolution evaluated using the data-driven pseudo-event method, as a function of the average number of \PE\ in both pseudo-events and the reconstructed radius drawn from the same original event, as returned by the \ChargeBasedPosRec\ algorithm. The $z$-axis scale denotes the resolution, defined as the characteristic width of the distribution of distances between reconstructed pseudo-events drawn from the same original event.}
 
 \label{fig:positionSplitterResolution}
\end{figure}

A data-driven method is used to estimate the position resolution. 
First, \arnine\ \betd\ events are split into two ``pseudo-events'': each \PE\ from an event is independently assigned to each of the two pseudo-events with a 50\% probability. 
Doing so results in both pseudo-events having approximately half the number of \PEs\ as the original event, from the same true position. 
The position resolution is determined from the distribution of reconstructed distances between pseudo-events, in bins of average pseudo-event \PE\ and original event reconstructed radius; the characteristic width of each such distribution is shown in Figure~\ref{fig:positionSplitterResolution}. 
Within the \WIMP-search \PE\ region, near the \PaperTwoRadialCut\ radial cut used in this analysis, pseudo-events typically reconstruct within \PaperTwoChargeFitterROIRes\ of each other.

\begin{figure}[htb]
 \centering
 \includegraphics[width=1.01\linewidth]{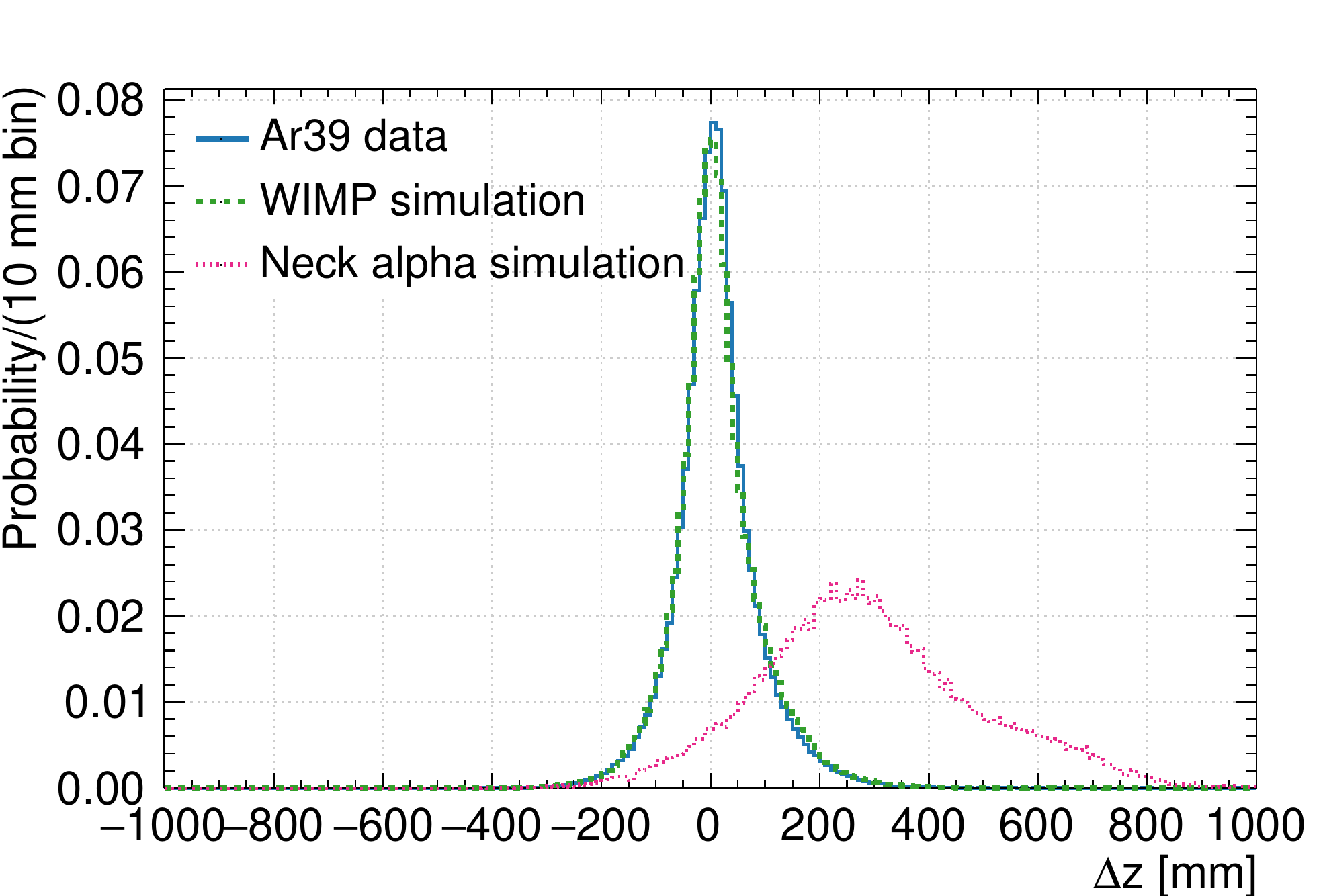}
 \caption{
 The $z$-coordinate of the reconstructed position (along the axis parallel to the \AV\ neck) as estimated by the \TimeBasedPosRec\ algorithm minus the $z$-coordinate estimated by the \ChargeBasedPosRec\ algorithm. 
 \arnine\ data (solid blue) and a simulation of \WIMP\ \NRs\ (dashed green) show similar distributions, centered around the origin. A significant offset is seen for simulated \alpds\ in the neck (dotted magenta). 
 }
 \label{fig:positionFitterCompZ}
\end{figure}

Figure~\ref{fig:positionFitterCompZ} shows the difference between the $z$-coordinates reconstructed by both algorithms, where the $z$-axis runs parallel to the \AV\ neck. 
Both algorithms typically agree for \arnine\ events in data and for simulations of \arforty\ recoils, returning $z$-coordinates that agree to within \PaperTwoChargeBasedTimeBasedFitterWIMPDZ\ for 50\% of such events. 
For simulated events generated by \alpds\ through a \LAr\ film on the surface of the \AV\ neck (to be discussed in Section~\ref{sec:neck-alphas}), this distribution is very different. 
The neck directs light to the bottom of the detector, causing the \ChargeBasedPosRec\ algorithm to reconstruct it with a low $z$-coordinate, while the \TimeBasedPosRec\ algorithm systematically reconstructs these events closer to the top of the detector. 
As a result, the \TimeBasedPosRec\ algorithm reconstructs these events an average of \PaperTwoChargeBasedTimeBasedFitterNeckAlphaDZ\ higher than the \ChargeBasedPosRec\ algorithm.
A similar shift is observed for neck \alpds\ when comparing the distance between reconstructed positions.

\section{Detector stability and run selection}\label{sec:stability}

The state of the \DAQ\ and process systems is continuously monitored and the quality of the data is assessed during collection and after processing. This allows for different levels of data quality checks. 
Data from the detector and from the \DAQ\ and process system sensors are continuously monitored by automated processes and by the person on shift. Any anomalous behaviour is flagged. This data includes, but is not limited to: \PMT\ rates, \PMT\ bias voltages, \PMT\ baselines, \AV\ pressure and \DAQ\ rack temperature.

For the dataset discussed here, the \LAr\ is not recirculated. 
Hence, the primary function of the process system is to maintain the \LAr\ target inside the \AV\ at a constant temperature and pressure. This is achieved through continuous circulation of \LN\ in the cooling coil. For further details on the process system, see~\cite{amaudruz_design_2017}.

Impurities (e.g. \oxygen, \nitrogen) can decrease the scintillation yield \cite{Jones:2013ef, QinghaoChen:2013vta, Acciarri:2009dj} of \LAr\ by absorbing the scintillation light and, for electronegative impurities, by capturing the excitation energy from the \argon\ excimers. 
Electronegative impurities thus preferentially suppress the triplet scintillation component and affect the \PSD\ distributions \cite{Amsler:2008jq, Collaboration:2016ws}. 

The purity of the \LAr\ target, and thus the stability of analysis inputs, is monitored by examining calibrated \PMT\ waveforms from \arnine\ \ER\ events and other detector backgrounds. 
This yields the \LAr\ long lifetime component (which includes detector effects such as the slow component of the \TPB\ response, as described in~\cite{stanford_surface_2018}, and is not a direct measure of the decay constant of the triplet state of the \argon\ dimer) and the light yield at a granularity of better than \SI{1}{\hour}. 
In the same processing step, self-diagnostic pulses injected into the data stream by the \DAQ\ system are evaluated to verify proper behaviour of each \PMT\ channel. 
Any anomalous behaviour is again flagged.

\begin{figure}[htb]
 \centering
 \includegraphics[width=1.01\linewidth]{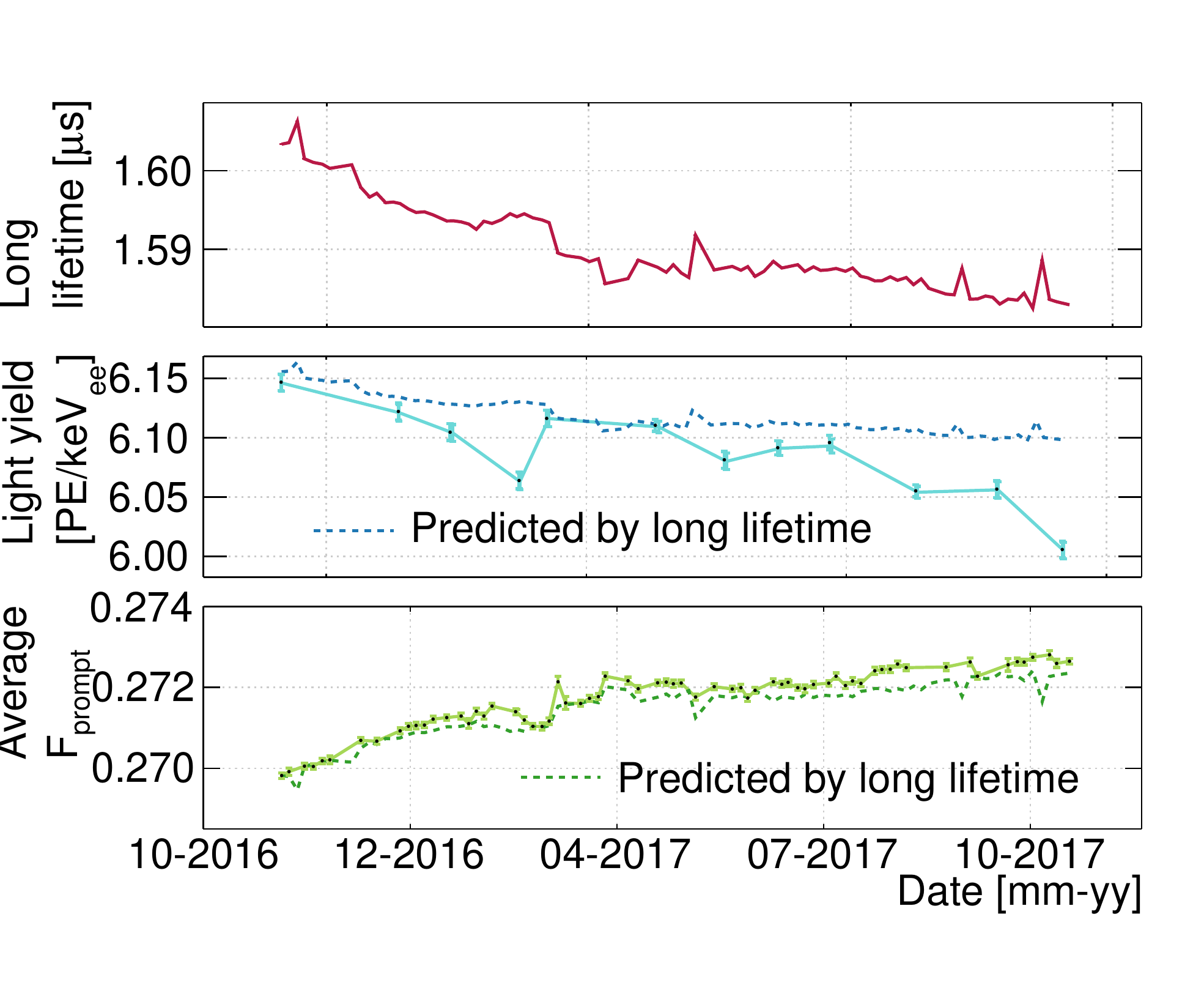}
 \caption{Shown over the time period spanned by this dataset is the long lifetime component of \LAr\ scintillation (top), the detector light yield (middle) and mean \FPrompt\ of \ER\ signals generated by \ThTwoZeroEightHighEGammaEnergy \grs\ in the \LAr\ target throughout the run. For the latter two plots, the predicted value based on the long lifetime value at that time is also shown (dashed).
 }
 \label{fig:larStability}
\end{figure}

As shown in Figure~\ref{fig:larStability}, throughout the time period discussed, the long lifetime and light yield were stable to within \LongLifetimeSpread\ and \instr{\LightYieldSpread}, respectively. 
The high value of the long lifetime shown here is indicative of a high level of chemical purity with regard to electronegative contaminants, in accordance with the design goals as described in~\cite{amaudruz_design_2017}.
The mean of the \FPrompt\ distribution at high energies is directly affected by changes to the long lifetime component of \LAr\ scintillation. 
The variation of \FPrompt\ from high energy \ER\ events is shown also, and is found to be stable to within \FpromptSpread. 
Given this level of stability, no corrections are applied to account for temporal variations in the light yield or long lifetime. 
The dashed lines in Figure~\ref{fig:larStability} show what the light yield and mean \FPrompt\ values would be if the decrease in the long lifetime was the only factor reducing their values.

Changes in the \PMT\ response over the data taking period are accounted for in this analysis. 
For 250 of the 255 \AV\ \PMTs, the \ChannelEfficiencies\ are constant to within \RelativePMTEfficiencyVariation. 
Two \PMTs\ have changes of less than \RelativePMTEfficiencyVariationMildlyUnstable. 
Three have changes in excess of \RelativePMTEfficiencyVariationVeryUnstable. 
One of the three is stable for the first two-thirds of the data collection period, after which it is removed from the analysis and omitted from calculations of analysis variables. 
The two remaining \PMTs\ with large changes in \ChannelEfficiency\ are located about the pentagonal region at the bottom of the \AV. 
Position reconstruction is particularly sensitive to changes in \ChannelEfficiency, as it relies on signals measured in individual \PMTs.

The gain of each \PMT\ is measured in the form of the mean \SPE\ charge. The mean \SPE\ charge, averaged over the 254 \PMTs\ used throughout the entire data collection is \SPERatio\ times larger at the end of the data collection than at the beginning. The RMS of this mean \SPE\ charge ratio is \SPESpread. These changes are also propagated through the analysis. The probability of afterpulsing is found to be stable to within \AveAPProbStability\ of the quoted value and is fixed in the analysis throughout the data collection.

The 4 neck veto \PMTs\ remained operational throughout the time period of this dataset. 
In the \MV, \PaperTwoOpMVPMTs\ remained stable and \SI{3}{} failed.

\subsection{Run selection and live time determination}\label{sec:run-selection}

Selection criteria are applied to each run to remove periods where instabilities could affect the dark matter search. 
These criteria include the stability of the \AV\ cooling system, stability of the \PMT\ charge distributions, and the trigger efficiency.

The first requirement is that the difference between the maximum and minimum values of the \AV\ pressure recorded for the run corresponds to less than a \SI{10}{\mm} variation in the \LAr\ fill level. 
Such variations are expected if maintenance is performed on the process system or when replenishing the L\nitrogen\ in circulation. 
The second requirement is based on the charge readout of each \PMT\ channel, taken in 5 minute samples. 
Runs are omitted if at least one \PMT\ exhibits intermittent behavior, defined as reading less than \SI{50}{\percent} of its nominal average charge at any stage throughout the run. 
While such excursions are rare and only occur in certain \PMTs, they indicate temporary malfunctioning behavior in the corresponding \PMTs.
Finally, to maintain good calibration of the \PSD\ model and its prediction throughout the dataset, the last requirement is enforced based on whether the trigger efficiency can be determined for the run. 
Due to drifts in the \PMT\ gain and the details of how the \DTM\ receives \PMT\ signals, the trigger efficiency can vary slightly from run to run. 
The trigger efficiency is determined for each run using the method described in~\cite{pollmann_estimating_2019}.
This procedure requires a large enough data sample in regions with low trigger efficiency.  
Runs that are shorter than approximately \SI{1}{\hour} do not have enough statistics and are omitted.
The point corresponding to \SI{50}{\percent} trigger efficiency varies by \SI{10}{\percent} between runs; these variations primarily affect \ER\ events and are negligible for \NRs\ in the \ROI.
These run selection criteria and their impact on the total live time are summarized as ``Stable cryocooler'', ``Stable \PMTs'' and ``Trigger efficiency obtained'' in Table~\ref{tab:lowLevelCuts-fiducial-cuts}, resulting in a live time loss of \CryoPMTTriggLiveTimeLoss\ after automatic \DAQ\ and shifter checks.

The total live time is also affected by events in the \MV\ passing the veto threshold (``Muon veto events'') and by \DAQ\ self-diagnostic triggers, the removal of pile-up with \arnine, and Cherenkov events in the \LGs. 
When an event passes the veto threshold of the \MV, all \AV\ events within a \MuonVetoWindow\ window around the trigger are vetoed; noise and \grs\ causing the \MV\ to pass the vetoing threshold therefore reduce the total live time.
The three latter conditions are low-level cuts factored into the ``Dead time'' entry of Table~\ref{tab:lowLevelCuts-fiducial-cuts}, resulting in a live time loss of \PileupLiveTimeLoss\ after applying run selection criteria. 
Cherenkov events generated in the \LGs\ are one of two Cherenkov populations discussed in Section~\ref{subsubsec:backgrounds_cherenkov}.
They are readily removed without affecting the \WIMP\ acceptance and hence are factored into the dead time.

\section{Background Model \& Cut Selection}
\label{sec:bkgds}
\label{sec:backgrounds}

\WIMP-like events may be produced in the detector by a variety of background sources that include \betp\ and \gr\ interactions in the \LAr\ and acrylic, neutron-induced nuclear recoils in the \LAr, and \alpds\ from surfaces in contact with \LAr. 
In this analysis, the total number of predicted background events after applying all event selection cuts in the \WIMP\ search region of interest (\ROI), $N^{\text{ROI}}_{\text{bkg}}$ is expressed as follows,
\begin{gather}
\begin{aligned}
 N^{\text{ROI}}_{\text{bkg}} &= N^{\text{ROI}}_{\text{\ER}} + N^{\text{ROI}}_{\text{Cher}}\\ 
&\quad+ N^{\text{ROI}}_{\text{$n$, rdg}} + N^{\text{ROI}}_{\text{$n$, csg}}\\
&\quad+ N^{\text{ROI}}_{\text{$\alpha$, AV}} + N^{\text{ROI}}_{\text{$\alpha$, neck}},
\end{aligned}
\label{eq:background-sum}
\end{gather}
where the individual terms are the expected number of background events from \ERs\ ($N^{\text{ROI}}_{\text{\ER}}$), Cherenkov light produced in acrylic ($N^{\text{ROI}}_{\text{Cher}}$), radiogenic neutrons ($N^{\text{ROI}}_{\text{$n$, rdg}}$), cosmogenic neutrons ($ N^{\text{ROI}}_{\text{$n$, csg}}$), and \alpds\ from both the \AV\ surface ($N^{\text{ROI}}_{\text{$\alpha$, AV}}$) and the \AV\ neck flowguides ($N^{\text{ROI}}_{\text{$\alpha$, neck}}$). 
The rest of this section focuses on characterizing the background models to determine each $N^{\text{ROI}}_i$.

\subsection{Methodology}
\label{sec:background-characterization}

The components of the background model are constructed using various combinations of calibration data, sidebands in the physics data, and simulations.
For each background component, a control region (\CR) is defined by an event selection in the physics data.
Each \CR\ uses different cuts, which are detailed in Sections~\ref{subsec:backgrounds_betagamma}--\ref{subsec:backgrounds_alphas} in the context of the relevant backgrounds. 
Background models are tuned based on these \CRs\ and calibration data.
In addition to the low-level event selection cuts and fiducial cuts listed in Table~\ref{tab:lowLevelCuts-fiducial-cuts}, an \ROI\ is defined in \FPrompt vs.~\PE\ space and a set of background rejection cuts are designed to remove additional backgrounds in the \WIMP\ \ROI.
Target upper limits were chosen for the expectation value of each component of the background model to achieve a total expectation of $N^{\text{ROI}}_{\text{bkg}}<1$. 
The bounds of the \ROI, background rejection cuts, and fiducial cuts were tuned on the background models to satisfy the targets while maintaining the highest achievable \WIMP\ acceptance.

The \WIMP\ \ROI\ is a region in \FPrompt\ vs.~\PE\ space designed for sensitivity to low energy nuclear recoils; it is defined in Section~\ref{sec:roi-definition} and spans the \WIMPPERange\ range.

The background rejection cuts are introduced in Sections~\ref{subsubsec:backgrounds_cherenkov} and~\ref{sec:neck-alphas}, and they are summarized in Section~\ref{sec:bkgds-summary}. 
These cuts and their effects on the \WIMP\ acceptance, background model, and data are summarized in Table~\ref{tab:cuts}.

Fiducialization is achieved with a set of three cuts.
First, only events that reconstruct below the \LAr\ fill level ($z<550$ mm) and within a \PaperTwoRadialCut\ radius are accepted.
Two additional fiducial cuts are applied based on the fraction of total event charge in two sets of \PMT\ rows: the bottom three rows and the top two rows, closest to the opening of the \AV\ neck. 
These two cuts are discussed in Sections~\ref{sec:alphas-long-lived} and \ref{sec:neck-alphas} respectively.

The fiducial mass is determined using \arnine\ \ER\ events in the \WIMPPERange\ range.
After applying all fiducial cuts, it is measured to be \PaperTwoFiducialMass. 
The uncertainty on this value accounts for the uncertainty on the mass of the \LAr\ target and the relative difference seen when applying these cuts to \arnine\ \betd\ and \arforty\ \NR\ simulations.
Table~\ref{tab:lowLevelCuts-fiducial-cuts} shows the estimated contained \LAr\ mass as each of the three fiducial cuts are sequentially applied.
The final fiducial mass is the value after all three cuts.

\begin{table}[htb]
 \centering
 \caption{The cumulative impact of the run selection criteria on the data live time is shown. Below this, total fiducial \LAr\ mass is shown after applying each fiducial cut cumulatively.}
 \begin{tabular}{cL{4cm}L{2cm}}\hline\hline
                                                               & Selection criteria                         & Live time                    \\
                                                               &                                            & [days]                       \\\hline
  \multirow{6}{*}{\rotatebox[origin=c]{90}{\tiny run}}         & Physics runs                               & \multirow{1}{*}{\instr{\PhysicsRunTime}} \\
  							       & Pass automatic DAQ \& shifter checks       & \multirow{2}{*}{\instr{\PhysicsRunTimeAutomatedDAQ}} \\
                                                               & Stable cryocooler                          & \multirow{1}{*}{\instr{\PhysicsRunTimeStableCryoCooler}} \\
                                                               & Stable \PMTs                               & \multirow{1}{*}{\instr{\PhysicsRunTimeStablePMTs}} \\
                                                               & Trigger efficiency obtained           & \multirow{1}{*}{\instr{\PhysicsRunTimeTriggerEfficiency}} \\ \rule{0pt}{4ex}    
  \multirow{2}{*}{\rotatebox[origin=c]{90}{\tiny event}}       & Muon veto events                           & \multirow{1}{*}{\instr{\PhysicsRunTimeMuonSignals}} \\  
                                                               & Dead time                                   & \instr{\PhysicsLivetimeTwoDP}                    \\\hline
                                                               & \textbf{Total}                             & \instr{\PhysicsLivetimeTwoDP}                    \\ \hline \hline
                                                               & Fiducial cut                                   & Contained \LAr\ mass [kg] \\ \hline
 	                                                       & No fiducial cuts                           & \larmasserrortable \\
 	                                                       & Reconstructed position $z<$~550 mm \& radius~$<$~630 mm & \multirow{3}{*}{\PaperTwoMassRCut} \\
 	                                                       & Charge fraction in top 2 rows of \PMTs\    & \multirow{2}{*}{\PaperTwoMassRCutCFTTR} \\
 	                                                       & Charge fraction in bottom 3 rows of \PMTs\ & \multirow{2}{*}{\PaperTwoFiducialMassNum} \\ \hline
                                                               & \textbf{Total}                             & \PaperTwoFiducialMassNumForTable \\
  \hline\hline
 \end{tabular}
 \label{tab:lowLevelCuts-fiducial-cuts}
\end{table}

The number of events in the \WIMPPERange\ range of each \CR\ ($N^{\text{CR}}_{i}$) and the number of events in the \WIMP\ \ROI\ after low-level cuts ($N^{\text{ROI,LL}}_i$) and after all background rejection and fiducial cuts ($N^{\text{ROI}}_i$) are shown in Table~\ref{tab:bkgdsummary}.
For background models using simulations, the values of $N^{\text{ROI}}_i$ include systematic uncertainties that are derived from multiple simulations of the background source with variations in the optical model and detector response parameters. 
These include variations in the following: (1) the refractive index of \LAr\ and its corresponding relationship to the scattering length and group velocity of light traveling in it, (2) the scattering length of photons in \TPB, (3) the \PMTs' \AP\ probabilities, (4) the light yield of the detector, and (5) the relative \PMT\ efficiencies.
Uncertainties in the bias and resolution of the position reconstruction algorithms and the level of agreement between data and simulation for these quantities, as shown in Figure~\ref{fig:positionFitterUniformity}, are also considered.
For simulated \alpd\ background sources, the systematic uncertainty also includes contributions from variations in the parameters describing \alpp\ scintillation in \LAr, the light yield of \alpps\ in \TPB\ (for \AV\ surface components), and the thickness of a \LAr\ film (for neck \FG\ components). 

The value of each $N^{\text{ROI}}_i$ term in Equation~\ref{eq:background-sum} is determined using these tuned models by applying all \WIMP\ selection cuts to them. 
The remainder of this section discusses how each specific $N^{\text{ROI}}_i$ term is determined.

\begin{table}[htb]
 \centering
 \setlength\extrarowheight{2pt}
 \caption{Predicted number of events from each background source in the \WIMPPERange\ of its respective CR, $N^{\text{CR}}$ and the total number in the \WIMP\ \ROI\ after only low-level cuts, $N^{\text{ROI, LL}}$ and after applying both fiducial and background rejection cuts, $N^{\text{ROI}}$. Upper limits are quoted at 90\%~\CL}
 \begin{tabular}{clc|c|c}\hline\hline
                                                       & Source            & $N^{\text{CR}}$            &  $N^{\text{ROI, LL}}$                    & $N^{\text{ROI}}$ \\\hline 
  \multirow{2}{*}{\rotatebox[origin=c]{90}{\bgs}}      & \ERs              & \PaperTwoERBkgdBeforeCuts                      & \PaperTwoERBkgdAfterLLCuts               & \PaperTwoERBkgdAfterCuts                      \\
                                                       & Cherenkov         & \PaperTwoCherenkovEventsBeforeCuts             & \PaperTwoCherenkovEventsAfterLLCuts      & \PaperTwoCherenkovEventsAfterCuts             \\ 
  \multirow{2}{*}{\rotatebox[origin=c]{90}{$n$'s}}     & Radiogenic        & \PaperTwoTotalRadioNeutronBkgdBeforeCutsWIMPPE & \PaperTwoTotalRadioNeutronBkgdLL         & \PaperTwoTotalRadioNeutronBkgd                \\
                                                       & Cosmogenic        & \PaperTwoTotalCosmoNeutronBkgdBeforeCuts       & \PaperTwoTotalCosmoNeutronBkgdBeforeCuts & \PaperTwoTotalCosmoNeutronBkgd                \\ \rule{0pt}{4ex} 
  \multirow{2}{*}{\rotatebox[origin=c]{90}{$\alpha$'s}}& \AV\ surface      & \SurfaceEventsBeforeCuts                       & \AVSurfaceEventsAfterLLCuts              & \AVSurfaceEventsAfterCuts                     \\
                                                       & \AV\ Neck \FG\    & \PaperTwoTotalNeckAlphaBkgdBeforeCuts          & \PaperTwoTotalNeckAlphaBkgdBeforeCutsLL  & \PaperTwoTotalNeckAlphaBkgd                   \\\hline
                                                       & \textbf{Total}    & N/A                                            & \TotalBackgroundExpectationAfterLLCuts   & \TotalBackgroundExpectationAfterCuts \\
  \hline\hline
 \end{tabular}
 \label{tab:bkgdsummary}
\end{table}

\subsection{\betps\ and \grs}\label{subsec:backgrounds_betagamma}
\betps\ and \grs\ both trigger events in the detector, either by producing scintillation light in the \LAr\ or by creating Cherenkov light in the acrylic.

\subsubsection{Scintillation in \LAr}
High energy electrons, produced by \betds\ of radioisotopes in \LAr\ or by \gr\ interactions in the \LAr, ionize and produce scintillation characterized by low \FPrompt\ \ER\ events.

The dominant source of \ER\ events is from \betds\ of \arnine, as can be seen in the \PE\ distribution shown in Figure~\ref{fig:ar39}.
Due to its long half-life, \arnine\ is present with a near-constant activity of \ArThreeNineTotalActivity\ throughout the dataset.
Low energy \arnine\ \ER\ events are efficiently mitigated with \PSD, using the \FPrompt\ parameter defined in Equation~\ref{eq:fprompteqn}.

\begin{figure}[htb]
 \centering
 \includegraphics[width=1.01\linewidth]{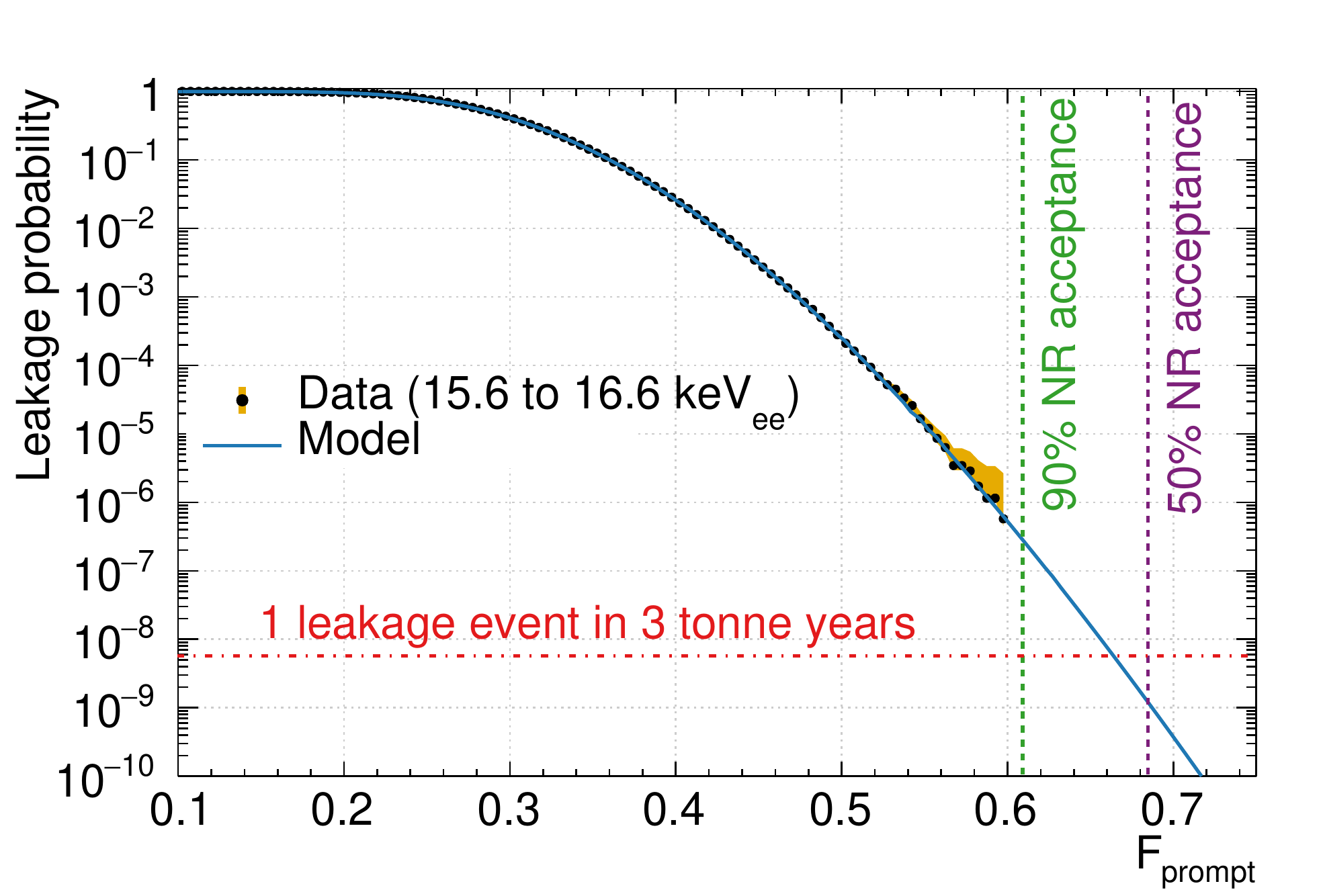}
 \caption{Probability of an \ER\ being detected above a given \FPrompt\ value in the lowest \SI{1}{\keVee} bin in the \WIMP-search region of interest. For comparison, vertical lines show the values above which 90\% or 50\% of nuclear recoils are expected to be found.}
 \label{fig:psdLeakage}
\end{figure}

The \ER\ and \NR\ \PSD\ models in Equations~\ref{eq:psdmodel} and~\ref{eq:psdnrmodel} are used to calculate the number of \ER\ events expected to leak past a given \FPrompt\ value and to determine the \WIMP\ acceptance at that value.

The \CR\ for \ER\ events is defined by the set of events passing low-level event selection cuts in the \WIMPPERange\ range. 
No explicit \FPrompt\ cut is applied to the \CR\ definition, though events whose \FPrompt\ values appear as outliers at a given \PE\ are excluded.
The expected number of events in the \CR\ is $N^{\text{CR}}_{\text{\ER}}=$\PaperTwoERBkgdBeforeCuts. 

Figure~\ref{fig:psdLeakage} shows the fraction of \ER\ events expected above a given \FPrompt\ value, showing the 50\% and 90\% \NR\ acceptance values.
Leakage probabilities are shown for a \SI{1}{\keVee}-wide window near the \WIMP\ search threshold, corresponding to the range \PSDThresholdPERange\ (\PSDThresholdEnergyRange).
In this range, a leakage fraction of \PSDThresholdLeakageRateAtNRANinty\ (\PSDThresholdLeakageRateAtNRAFifty) is predicted for cut values with \SI{90}{\percent} (\SI{50}{\percent}) \NR\ acceptance.
Averaged over the full \WIMP\ search energy range, the leakage probability is projected to be \PSDFullLeakageRateAtNRANinty\ (\PSDFullLeakageRateAtNRAFifty) with \SI{90}{\percent} (\SI{50}{\percent}) \NR\ acceptance. 
The low leakage probabilities at these values demonstrate the power of \PSD\ to efficiently reject \ER\ background events.

The uncertainty in these leakage predictions is driven by uncertainty in the \FPrompt\ values corresponding to the quoted \NR\ acceptance values.
Since the leakage probability decreases rapidly with \FPrompt, small variations in the \FPrompt\ value lead to relatively large variations in the leakage probability.
These uncertainties therefore have little effect on the ultimate \ER\ background prediction.

The \ER\ spectrum is uniformly sampled throughout the data taking period and is measured in the \WIMPPERange\ range approximating the effects of \WIMP\ search cuts and correcting the observed spectrum for the trigger efficiency.
Using this spectrum and the leakage probability estimates, the total number of leakage events above a given \FPrompt\ value is predicted as a function of \PE.

The \FPrompt\ vs.~\PE\ \ROI\ is shown in Figure~\ref{fig:roi_psdbands}; the lower left edge of the \ROI, below \SI{160}{\PE}, is selected for an expectation of \ERLeakage\ \ER\ leakage events in the dataset, with uniform leakage expectation over that edge. 
The acceptance of the \ROI\ for \NRs\ is shown in Figure~\ref{fig:acceptance}.

After applying all fiducial and background rejection cuts, $N^{\text{ROI}}_{\text{\ER}}=$\PaperTwoERBkgdAfterCuts\ \ER\ events are expected to reconstruct inside the \ROI.
The uncertainty on this estimate is dominated by systematic uncertainties in the \ER\ model fits relating to the sample size and range of the fits.

\subsubsection{Cherenkov in acrylic}\label{subsubsec:backgrounds_cherenkov}

Electrons, either from \betds\ or the scattering of \grs, may produce Cherenkov events in the acrylic or \PMT\ glass. 
Since Cherenkov light has a significant UV component, the \UVA\ acrylic of the \AV\ and \LGs\ reduce this background.
As a result, most Cherenkov light generated by \ura\ and \tho\ progeny in the detector materials produce too few \PEs\ to pass the \WIMP\ search \PE\ threshold. 
Due to the short time-scales (\SI{<1}{\ns}) 
over which Cherenkov light is produced, the majority of these events have high \FPrompt\ values, reconstructing above the upper \FPrompt\ bound of the \WIMP\ ROI. 

The detector response to Cherenkov light is characterized using a series of dedicated \uratwo\ source calibration runs taken during the data collection period. 
In these calibration runs, the dominant production mechanism for Cherenkov events is from \ThTwoZeroEightHighEGammaEnergy\ \grs\ emitted by \thal\ at the bottom of the chain.
In each calibration run, the source was deployed through one of several calibration tubes located close to the outside of the stainless steel shell. 
Two primary locations are used: the equator of the stainless steel shell and close to the bottom of the \AV\ neck. 
Two Cherenkov populations are identified based on the ratio $N_{\text{hit}}$/PE where $N_{\text{hit}}$ is the number of \PMTs\ registering hits in the event; this ratio provides a measure of how diffuse the light is. 
The two characteristic types of Cherenkov events are from those produced in the acrylic of \LGs\ and those produced in the \AV\ neck and pentagonal regions between \LGs. 
The distributions of detected \PEs\ across \PMTs\ in these two populations are different and inform the cuts defined to remove them.
 
Light from Cherenkov in the \LGs\ is highly localized; such events are mitigated by removing events with more than 40\% of the total event charge in one \PMT. 
This cut has a negligible effect on the \WIMP\ acceptance. However, it contributes to a small live time loss in the analysis, as described in Section~\ref{sec:run-selection}.

The \AV\ neck and pentagonal regions are less photon-sensitive than other regions of the detector; Cherenkov light produced in these regions can appear more diffuse than in the \LGs.
Events generated in pentagonal regions are rejected by the fiducial radius cut. 
Visible Cherenkov light produced in the neck region can travel through the acrylic of the neck and reach the optical fibers of the \NV. 
Events that generate a signal in at least one \NV\ \PMT\ are cut. 
The probability that light from \LAr\ scintillation triggers the \NV\ after being shifted by the \TPB\ is factored into the total \WIMP\ acceptance as shown in Table~\ref{tab:cuts}.

The \WIMP\ \ROI\ is used as the \CR\ to characterize these events, after only applying cuts to remove pileup events and self-diagnostic triggers from the \DAQ.
Each of the two Cherenkov populations are studied in \uratwo\ calibration runs.
For both populations, the ratio of events generated in the \WIMP\ \ROI\ to those in a higher \FPrompt\ sideband is measured.
This sideband is defined as the region of the \FPrompt\ vs.~\PE\ plane above the \WIMP\ \ROI\ extending to \FPrompt$=1$, across the same \WIMPPERange\ range. 
The rates of both Cherenkov populations in this sideband are measured in the physics data, and this ratio is used to estimate their leakage rates into the \WIMP\ \ROI.

The predicted number of \LG\ Cherenkov events in the \WIMP\ \ROI\ after only applying the \CR\ cuts is \PaperTwoCherenkovEventsLGBeforeCuts\ (90\%~\CL). 
For the \AV\ neck and pentagonal Cherenkov events it is \PaperTwoCherenkovEventsNeckBeforeCuts\ (90\%~\CL). 
Combined, the total is $N^{\text{CR}}_{\text{Cher}}$\PaperTwoCherenkovEventsBeforeCuts\ (90\%~\CL).

To determine the respective fractions of events that survive all \WIMP\ selection cuts, all cuts are applied to both populations in the high \FPrompt\ sideband. 
An upper limit on the fraction of LG Cherenkov events surviving all cuts is determined to be \PaperTwoCherenkovEventsLGLF\ (90\%~\CL). For the \AV\ neck and pentagonal Cherenkov events an upper limit of \PaperTwoCherenkovEventsNeckLF\ (90\%~\CL) is calculated.
The number of events in the \WIMP\ \ROI\ after all event selection cuts for each population is \PaperTwoCherenkovEventsLGAfterCuts\ (90\%~\CL) for LG Cherenkov events, and \PaperTwoCherenkovEventsNeckAfterCuts\ (90\%~\CL) for \AV\ neck and pentagonal Cherenkov events. These combine to produce an expectation of $N^{\text{ROI}}_{\text{Cher}}$\PaperTwoCherenkovEventsAfterCuts\ events (90\%~\CL).

\subsection{Neutrons}\label{subsec:backgrounds_neutrons}
Neutrons can be produced by radiogenic and cosmogenic processes. 
A neutron can scatter on an \argon\ nucleus and produce a \NR\ event, exactly like that expected from a \WIMP.
However, these recoils generally reconstruct with different \PE\ and position distributions than expected from \WIMPs.
These and other differences make it possible to study neutrons outside of the \WIMP\ \ROI\ to inform a prediction of their background rate.

\subsubsection{Radiogenic neutrons}

Radiogenic neutrons can be produced by the \alphan\ reaction induced by \alpds\ in the \ura, \urafive, and \tho\ decay chains, or by the spontaneous fission of \ura.
These isotopes are present in trace quantities in detector components.
The neutron production rate was controlled by careful material selection and an extensive material assay campaign.
The assay results for the materials used in \DEAP\ are given in~\cite{amaudruz_design_2017}.

Based on these assays, the neutron flux and energy spectra from each detector component is determined using \SOURCES~\cite{wilson2002sources} and \neucbot~\cite{westerdale_radiogenic_2017}.
\Geant\ simulations propagate neutrons through the detector to predict the number that will induce \WIMP-like backgrounds from each source.

\begin{table}[htb]
 \centering
 \caption{Predicted number of neutron backgrounds from simulations, using \alphan\ yields calculated by either \SOURCES\ or \neucbot, for the dominant sources. All fission yields are calculated using \SOURCES. Background rates are calculated within a \CR\ used for validating the neutron background model \insitu, and within the \WIMP\ ROI.}
 \setlength\extrarowheight{3pt}
 \begin{tabular}{lcc}\hline\hline
                & \multicolumn{2}{c}{CR prediction}                                                         \\
  Component     & (\SOURCES)                                  & (\neucbot)                                  \\\hline 
  \PMT\ glass   & \PaperTwoPMTGlassRadioNeutronsSOURCESCR     & \PaperTwoPMTGlassRadioNeutronsNeuCBOTCR     \\
  \PMT\ ceramic & \PaperTwoPMTCeramicRadioNeutronsSOURCESCR   & \PaperTwoPMTCeramicRadioNeutronsNeuCBOTCR   \\
  \PMT\ mounts  & \PaperTwoPMTMountRadioNeutronsSOURCESCR     & \PaperTwoPMTMountRadioNeutronsNeuCBOTCR     \\
  Filler blocks & \PaperTwoFillerBlocksRadioNeutronsSOURCESCR & \PaperTwoFillerBlocksRadioNeutronsNeuCBOTCR \\
  Filler foam   & \PaperTwoFillerFoamRadioNeutronsSOURCESCR   & \PaperTwoFillerFoamRadioNeutronsNeuCBOTCR   \\
  Neck \PMTs\   & \PaperTwoNVPMTRadioNeutronsSOURCESCR        & \PaperTwoNVPMTRadioNeutronsNeuCBOTCR        \\\hline
  \textbf{Total}& \PaperTwoTotalRadioNeutronsSOURCESCR        & \PaperTwoTotalRadioNeutronsNeuCBOTCR        \\
  \hline\hline
                & \multicolumn{2}{c}{ROI prediction}                                                          \\
  Component     & (\SOURCES)                                   & (\neucbot)                                   \\\hline 
  \PMT\ glass   & \PaperTwoPMTGlassRadioNeutronsSOURCESROI     & \PaperTwoPMTGlassRadioNeutronsNeuCBOTROI     \\
  \PMT\ ceramic & \PaperTwoPMTCeramicRadioNeutronsSOURCESROI   & \PaperTwoPMTCeramicRadioNeutronsNeuCBOTROI   \\
  \PMT\ mounts  & \PaperTwoPMTMountRadioNeutronsSOURCESROI     & \PaperTwoPMTMountRadioNeutronsNeuCBOTROI     \\
  Filler blocks & \PaperTwoFillerBlocksRadioNeutronsSOURCESROI & \PaperTwoFillerBlocksRadioNeutronsNeuCBOTROI \\
  Filler foam   & \PaperTwoFillerFoamRadioNeutronsSOURCESROI   & \PaperTwoFillerFoamRadioNeutronsNeuCBOTROI   \\
  Neck \PMTs\   & \PaperTwoNVPMTRadioNeutronsSOURCESROI        & \PaperTwoNVPMTRadioNeutronsNeuCBOTROI        \\\hline
  \textbf{Total}& \PaperTwoTotalRadioNeutronsSOURCESROI        & \PaperTwoTotalRadioNeutronsNeuCBOTROI        \\
  \hline\hline
 \end{tabular}
 \label{tab:nbkgdsim}
\end{table}

The polyethylene filler blocks between \LGs\ and borosilicate glass in the \AV\ \PMTs\ are the dominant sources of neutron backgrounds, followed by the polystyrene filler foam between \LGs, the ceramic in the \PMTs, the polyvinyl chloride (\PVC) \PMT\ mounts and the \NV\ \PMTs.
The predicted contributions from these neutron sources are summarized in Table~\ref{tab:nbkgdsim}.
All other detector components are found to have a negligible contribution to the total background rate.
Uncertainties shown in this table are both statistical and systematic, including uncertainties in the assay results and the simulation's optical model and detector response as outlined in Section~\ref{sec:background-characterization}.
The dominant uncertainty comes from the level of \ura\ contamination in the filler blocks.

To validate the neutron background prediction, a \CR\ is defined, extending to \SI{5000}{\pe} with a loose cut requiring \NeutronCRFPromptRange, a radial cut of \NeutronCRRadius, and low-level event selection cuts applied. 
\NR-like events within this \CR\ are identified, and high energy \ER\ events above \SI{1.4}{\MeV} are counted within a \SI{1}{\ms} coincidence window following the \NR.

Due to the abundance of \hydrogen\ in acrylic and the efficient kinematic coupling between \hydrogen\ and neutrons, most neutrons that scatter in the \LAr\ are expected to thermalize within a few of centimeters after leaving the \LAr\ and entering the \AV, while a smaller fraction may thermalize in the \LAr.
Those that thermalize in the acrylic will predominantly capture on \hydrogen\ and produce a \HOneNeutronCaptureGammaEnergy\ \gr, while those that capture on \arforty\ in the \LAr\ will produce multiple \grs\ summing in energy to \ArFortyNeutronCaptureGammaEnergy.
The thermal neutron capture time in acrylic is \SI{250}{\micro\s} and in \LAr\ is \SI{325}{\micro\s}, meaning that over \SI{95}{\percent} of all neutrons that scatter in the \LAr\ will capture within this \SI{1}{\ms} coincidence window following the \NR.

\Geant\ simulations indicate that the probability of detecting a neutron capture event given that the neutron produced a \NR\ in the \CR\ is approximately independent of the origin of the neutron.
By counting the number of \NRs\ followed by a capture \gr\ signal, the number of neutron-induced events within the \CR\ can therefore be determined. 

To account for accidental coincidences, where a \NR\ not caused by a neutron is followed by a random \gr, a ``random coincidence'' sideband window is considered.
This window extends for \SI{19}{\ms}, starting after the end of the coincidence window.
The number of \NRs\ followed by a signal within the random coincidence window is counted and scaled down by a factor of 19 to provide an estimate of the expected number of accidental coincidences during this search.

To determine the tagging efficiency of this method, this coincidence search is applied to \AmBe\ neutron calibration source data, where it is found to tag \NeutronTaggingEfficiency\ of neutron-induced \NRs.
This efficiency is consistent with simulations, which indicate that the primary source of inefficiency is from neutrons that capture on \hydrogen\ in the acrylic and produce a \gr\ that loses more than \SI{800}{\keV} in volumes other than the \LAr.

Applying this search to the data reveals 7 coincidence events in the \CR---none of which fall in the \ROI\ or appear with coincident signals in the \MV---with an expectation of \SI{1.8\pm0.3}{} random coincidences. 
Correcting for the tagging efficiency, this gives a total of \PaperTwoTotalRadioNeutronBkgdBeforeCuts\ neutrons in the \CR, consistent with the prediction in Table~\ref{tab:nbkgdsim}. This corresponds to $N^{\text{CR}}_{\text{$n$, rdg}}=$\PaperTwoTotalRadioNeutronBkgdBeforeCutsWIMPPE\ events in the \WIMPPERange\ range of the \WIMP\ \ROI\ after applying the \CR\ cuts.

Simulations of neutrons coming from the \PMTs\ or from the outer surface of the \AV\ give consistent ratios of events in the \CR\ to those in the \WIMP\ \ROI, after all event selection cuts. 
Scaled by this ratio, the observed number of events in the \CR\ predict $N^{\text{ROI}}_{\text{$n$, rdg}}=$\PaperTwoTotalRadioNeutronBkgd\ neutron-induced backgrounds in the \WIMP\ \ROI\ after all \WIMP\ selection cuts.

\subsubsection{Cosmogenic neutrons}

Cosmogenic neutrons are produced by high energy atmospheric muon interactions with the detector and its environment.
The \SnolabDepthKWE\ overburden of \SNOLAB\ provides a significant reduction to the muon flux experienced by \DEAP. 
The \MV\ allows events induced by muons reaching the detector to be vetoed by the Cherenkov signal they produce in the water of the \MV\ water tank.

Muons are tagged either directly when they pass through the \MV, or indirectly when they produce an electromagnetic shower in the laboratory whose charged products enter the \MV.
These events are identified in \MV\ triggers in which significantly more light is seen in the detector than can be explained by noise or by normally present \grs.
By counting the number of muons passing through the \MV, a flux of \MeasuredMuonFlux\ is measured, consistent with the flux of $(3.31\pm0.10)\E{-10}$\,\SI{}{muons/\square\cm/\s} reported by \SNO~\cite{sno_collaboration_measurement_2009}.

Normalizing to the flux reported in~\cite{sno_collaboration_measurement_2009} and simulating muons with the energy spectra described by~\cite{mei_muon-induced_2006} results in a prediction of $N^{\text{CR}}_{\text{$n$, csg}}$\PaperTwoTotalCosmoNeutronBkgdBeforeCuts\ in the cosmogenic neutron \CR, defined as the \WIMP\ search \ROI\ with only low-level cuts applied. After applying all fiducial and background rejection cuts, this prediction becomes $N^{\text{ROI}}_{\text{$n$, csg}}$\PaperTwoTotalCosmoNeutronBkgd\ prior to any cuts with the \MV\ \PMTs.
These simulations model neutrons below \SI{20}{\MeV} using the ``high-precision'' (HP) \Geant\ physics models, and the default \Geant~9.6 hadronic physics models for higher energy neutrons.

The \MV\ is used to further reduce the rate of these background events. 
A cut has been designed to tag muons passing through the \MV\ based on the number of \PMTs\ hit and the number of \PEs\ detected, in order to identify events in which a significant signal is seen above the baseline background of noise in the \MV.
Events in the \AV\ are vetoed if they fall within a \MuonVetoWindow\ window around the tagged event.
This cut reduces the live time by \MuonVetoLiveTimeLoss, mostly due to accidental veto triggers not caused by muons.

\subsection{\alpps}\label{subsec:backgrounds_alphas}\label{subsubsec:background_alphas_lar}

Signals from \alpds\ from short- and long-lived \rntwo\ progeny as well as short-lived \rnzero\ progeny are observed at several locations inside the detector. 
These include the \LAr\ target, the \LAr/\TPB\ and \TPB/\AV\ surfaces, and the surfaces of the acrylic flowguides in the \AV\ neck. 
A summary of the measured activities or event rates for these \alpds\ is provided in Table~\ref{tab:alphaActivities}. 

\begin{table}[htb]
 \caption{Activity (\SI{}{\becquerel}) or event rate (\SI{}{\Hz}) of different short- and long-lived \alpds\ in the detector. Values are quoted for per \SI{}{\kg} of \LAr\ and per \SI{}{\square\metre} of the \TPB\ or \AV\ surface. Upper limits are quoted as 90\%~\CL\ The value quoted for \bitwo\ accounts for both \alpd\ (36\%) and \betd\ (64\%) modes. Rates on the \FGs\ are quoted for the inner surface (IS) and outer surface (OS).}
 \centering
  \setlength\extrarowheight{4pt}
 \begin{tabular}{cll}\hline\hline
    & Component                         & Activity / Rate                        \\\hline
   \multirow{8}{*}{\rotatebox[origin=c]{90}{\tiny short-lived \alpds}}           
    & \rntwo\ \LAr                      & \instr{\RnTwoLArBulkActivity}          \\
    & \poeight\ \LAr                    & \instr{\PoEightLArBulkActivity}        \\
    & \pofour\ \LAr                     & \instr{\PoFourLArBulkActivity}         \\
    & \pofour\ \TPB\ surface            & \instr{\PoFourSurfaceRateActivityArea} \\
    & \rnzero\ \LAr                     & \instr{\RnZeroLArBulkActivity}         \\
    & \posix\ \LAr                      & \instr{\PoSixLArBulkActivity}          \\
    & \bitwo\ \LAr                      & \instr{\BiTwoLArBulkActivity}          \\
    & \potwo\ \LAr                      & \instr{\PoTwoLArBulkActivity}          \\ \hline
   \multirow{5}{*}{\rotatebox[origin=r]{90}{\tiny long-lived \alpds}}
    & \poten\ \TPB\ \& \AV\ surface     & \instr{\PoTenAVSurfaceActivity}        \\
    & \poten\ \AV\ (bulk)          & \instr{\PoTenAVBulkActivity}           \\
    & \poten\ inner \FG, IS             & \instr{\PaperTwoIFGISRate}             \\
    & \poten\ inner \FG, OS             & \instr{\PaperTwoIFGOSRate}             \\
    & \poten\ outer \FG, IS             & \instr{\PaperTwoOFGISRate}             \\
   \hline\hline
 \end{tabular}
 \label{tab:alphaActivities}
\end{table}

\subsubsection{Short-lived \alpds}

The primary sources of short-lived \alpds\ are from \ura\ progeny: \rntwo, \poeight\ and \pofour\ (and its daughters) with emitted \alpp\ energies of \RnTwoAlphaEnergy, \PoEightAlphaEnergy\ and \PoFourAlphaEnergy\ respectively. 
A sub-dominant event rate is observed from \alpds\ of \tho\ progeny: \rnzero, \posix, \bitwo\ and \potwo\ (and its daughters) with respective emitted \alpp\ energies of \RnZeroAlphaEnergy, \PoSixAlphaEnergy, \BiTwoAlphaEnergy\ and \PoTwoAlphaEnergy.

These \alpds\ occur in the \LAr, appearing as high energy events in the \SIrange{23000}{50000} \PE\ range at high \FPrompt. 
To determine their activities, low-level cuts are applied and a fit to the data is performed by first applying an analytical correction which maps \FPrompt\ and \PE\ to \alpp\ energy in units of keV. 
This mapping corrects for the effect of digitizer and \PMT\ saturation on \PE\ and \FPrompt. 
Simulations of different \alpds\ are used, to which an equivalent correction is applied in order to produce energy spectra.

\begin{figure}[htb!]
 \centering
 \includegraphics[width=1.01\linewidth]{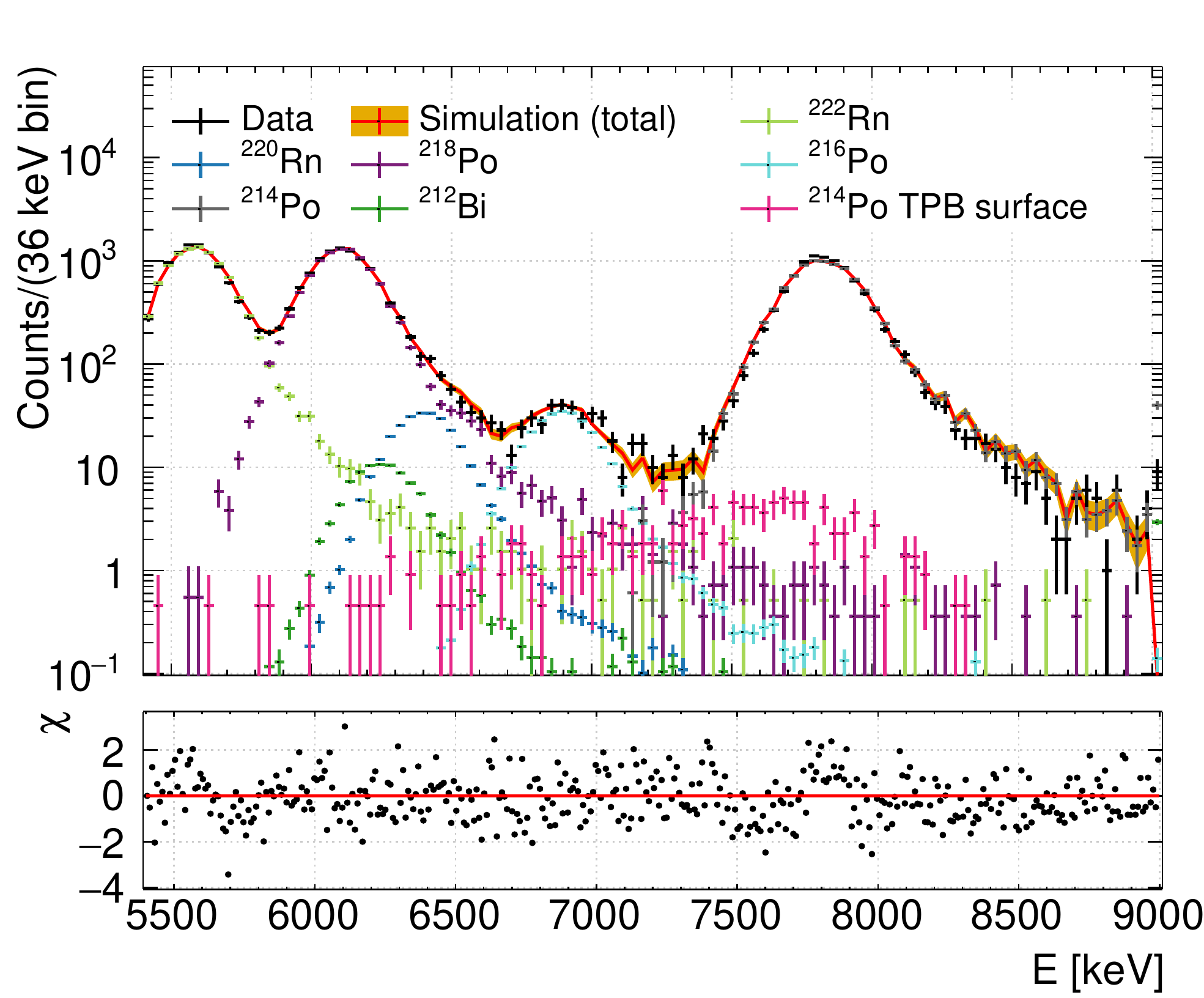}
 \caption{Distribution of mapped \alpp\ energies from \FPrompt\ and \PE\ for short-lived \alpds\ in \LAr. The event selection shown has no cut on position reconstruction and so includes events from $^{214}$Po \alpds\ at the \TPB\ surface (purple).
 }
 \label{fig:bulkAlphaSpec}
\end{figure}

A binned log-likelihood fit of these simulated \PDFs\ to the spectrum observed in data allows for individual activities of each component to be determined. 
These \PDFs\ and the spectrum observed in data are shown in Figure~\ref{fig:bulkAlphaSpec}. 
The fit includes the following \alpd\ components in \LAr\ (in order of increasing \alpd\ energy): \rntwo, \poeight, \bitwo, \rnzero, \posix\ and \pofour. 
In addition, a component of \pofour\ simulated at the \LAr/\TPB\ interface is included to allow for a plate-out fraction of \pofour\ to be resolved. 
The plate-out of nuclei on surfaces such as the \TPB\ is possible over the time-scale in which the parent \rntwo\ nucleus decays ($t_{1/2}=$\RnTwoHalfLife). 

Events caused by the \alpd\ of \potwo\ (\PoTwoAlphaEnergy) are observed in the dataset; however, this component is omitted from the fit. 
Due to its very short half-life ($t_{1/2}=$\PoTwoHalfLife), \potwo\ reliably appears as pile-up with the \betd\ of its parent \bitwo\ within the same time window of a single triggered event. 
This effect smears \PE\ and \FPrompt\ over a broad range of values, migrating the majority of the events out of the fit region. 

The activity of \potwo\ is calculated independently using a simulation of \bitwo-decays to estimate the selection efficiency for observing the \potwo\ \alpd\ in the fit region. 
The estimated activity of \potwo\ using this method is \PoTwoLArBulkActivity\ and is consistent with the activity of \bitwo\ (\BiTwoLArBulkActivity) from the fit. 
Assuming the \bitwo$\rightarrow$\potwo\ \betd\ mode branching fraction of 64\% and secular equilibrium it is also in agreement with the measured rates of \rnzero\ and \posix.

These high energy events in the \LAr\ target do not contribute to backgrounds in the \WIMP\ \ROI.

\subsubsection{Long-lived \alpds: \AV\ surface}\label{sec:alphas-long-lived}

Nuclei that \alpd\ on the inner surfaces of the detector or on visible interfaces between two components (such as the \LAr/\TPB\ and \TPB/\AV\ interfaces) or inside of these components may produce signals from \alpps\ with degraded energies.
As a result, these events may produce fewer \PEs\ and have a larger impact on the \WIMP\ search than those from nuclei that are distributed throughout the \LAr.

\begin{figure}[htb!]
 \centering
 \includegraphics[width=1.01\linewidth]{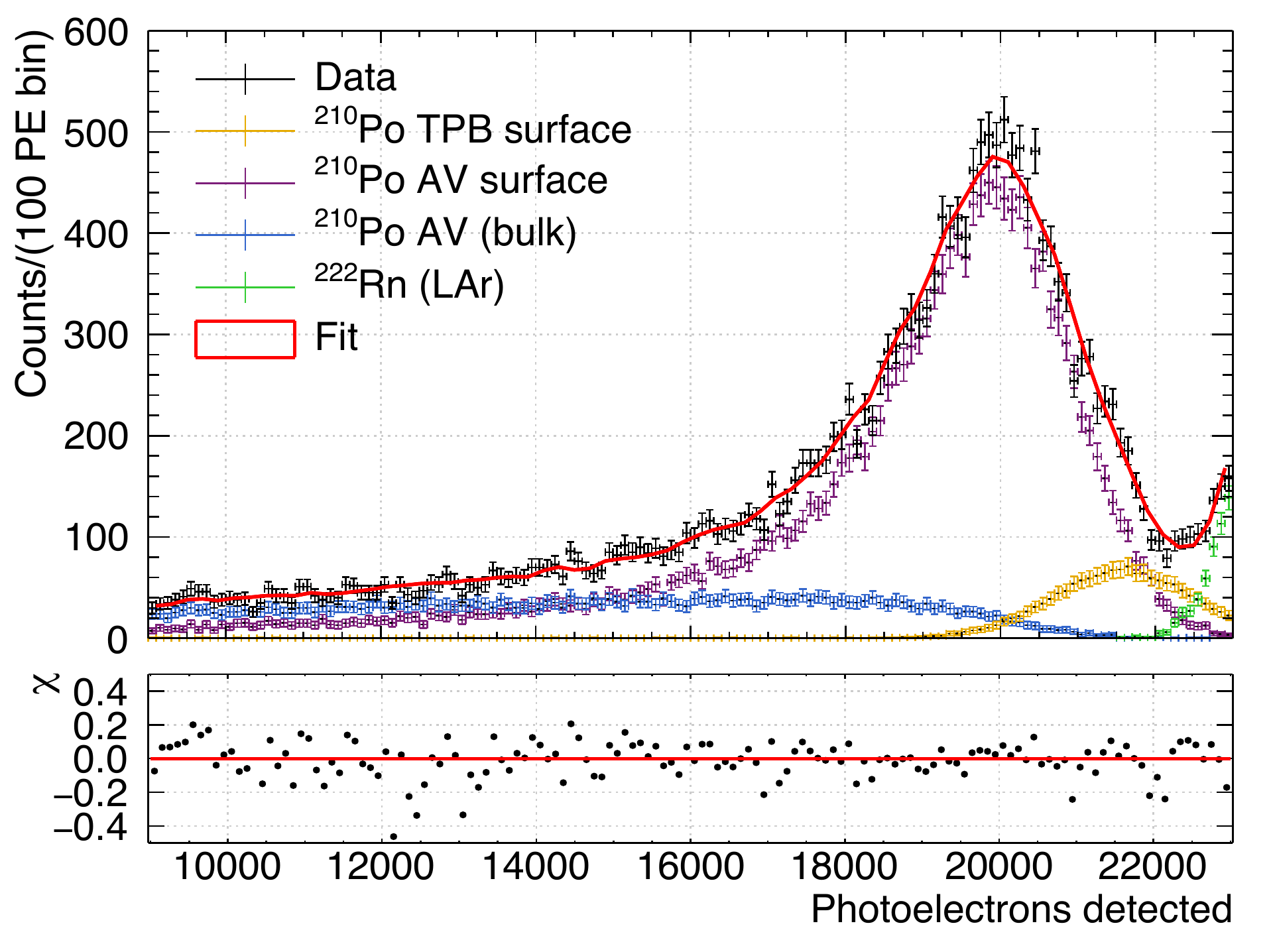}
 \caption{\PE\ distribution from data of \poten\ \alpds\ candidates (black) alongside different simulated \poten\ samples located on the \AV\ surface at the \TPB/\AV\ interface (purple), at the \LAr/\TPB\ interface and within the \TPB\ layer (yellow) and from beneath the inner surface of the \AV\ (blue). Shown also is the neighboring \PE\ distribution from \rntwo\ \alpds\ in \LAr\ (green). Overlaid is the combined fit (red) of each of these \PDFs\ to the observed \PE\ spectrum from data. The event selections shown reconstruct with a $z$-position below the \LAr\ fill level. No cut on the reconstructed radius is applied.}
 \label{fig:surfaceAlpha}
\end{figure}

The primary source of \alpds\ from surfaces is \poten\ ($t_{1/2}=$\PoTenHalfLife) from the inner surface of the \AV. 
\poten\ appears later in the \rntwo\ decay chain than \pbten, which has a half-life of \PbTenHalfLife.
As such, it may appear on detector surfaces out of equilibrium with other isotopes in this decay chain.

Events resulting from these decays peak in the \SIrange{18000}{22000} \PE\ range and extend to lower \PE\ as shown in Figure~\ref{fig:surfaceAlpha}. 
This distribution is obtained by applying low-level cuts and selecting events with \FPrompt$>0.55$ that reconstruct with a $z$-position below the \LAr\ fill level. It is fit with a model of the \poten\ surface activity, which is built with simulated \PDFs, with additional smearing introduced to match the data. 
The model features three components of \poten\ activity: (1) a surface component at the \LAr/\TPB\ interface and throughout the \TPBThickness\ \TPB\ layer, (2) a second surface component at the \TPB/\AV\ interface and (3) a bulk component of \poten\ decays occurring up to \SI{50}{\micron} beneath the inner surface of the \AV. 
Simulations predict no triggers from \poten\ \alpds\ beyond this depth. 
A component of the \rntwo\ \alpds\ \PDF\ in \LAr\ that is close to the \poten\ \PDFs\ in energy is also included in the fit.
This component helps constrain the \poten\ contributions in the \TPB\ layer and at the \LAr/\TPB\ interface.

Not all \alpds\ on detector inner surfaces result in a trigger.
The combined surface activity of \poten\ is measured to be \instr{\PoTenAVSurfaceActivity}. 
From the bulk component it is measured to be \instr{\PoTenAVBulkActivity}. 
This yields event rates of \instr{\PoTenAVSurfaceRate} from all surface decays and \instr{\PoTenAVBulkRate} from the bulk component.

The \CR\ for \poten\ \alpds\ is based on using the fitted values of the activity and the simulations to predict the event rate at low energies in the \WIMPPERange\ range. 
In this \CR, $N^{\text{CR}}_{\text{$\alpha$, AV}}$\SurfaceEventsBeforeCuts\ (90\%~\CL) \poten\ \alpd\ events are predicted before applying any cuts. 
As a cross-check, a search for these events inside the \WIMP\ \ROI\ is performed on the entire dataset, and \SurfaceTagsBeforeCuts\ candidate events are counted. 
None of these events surive the fiducial cuts.

\begin{figure}[htb]
 \centering
 \includegraphics[width=1.01\linewidth]{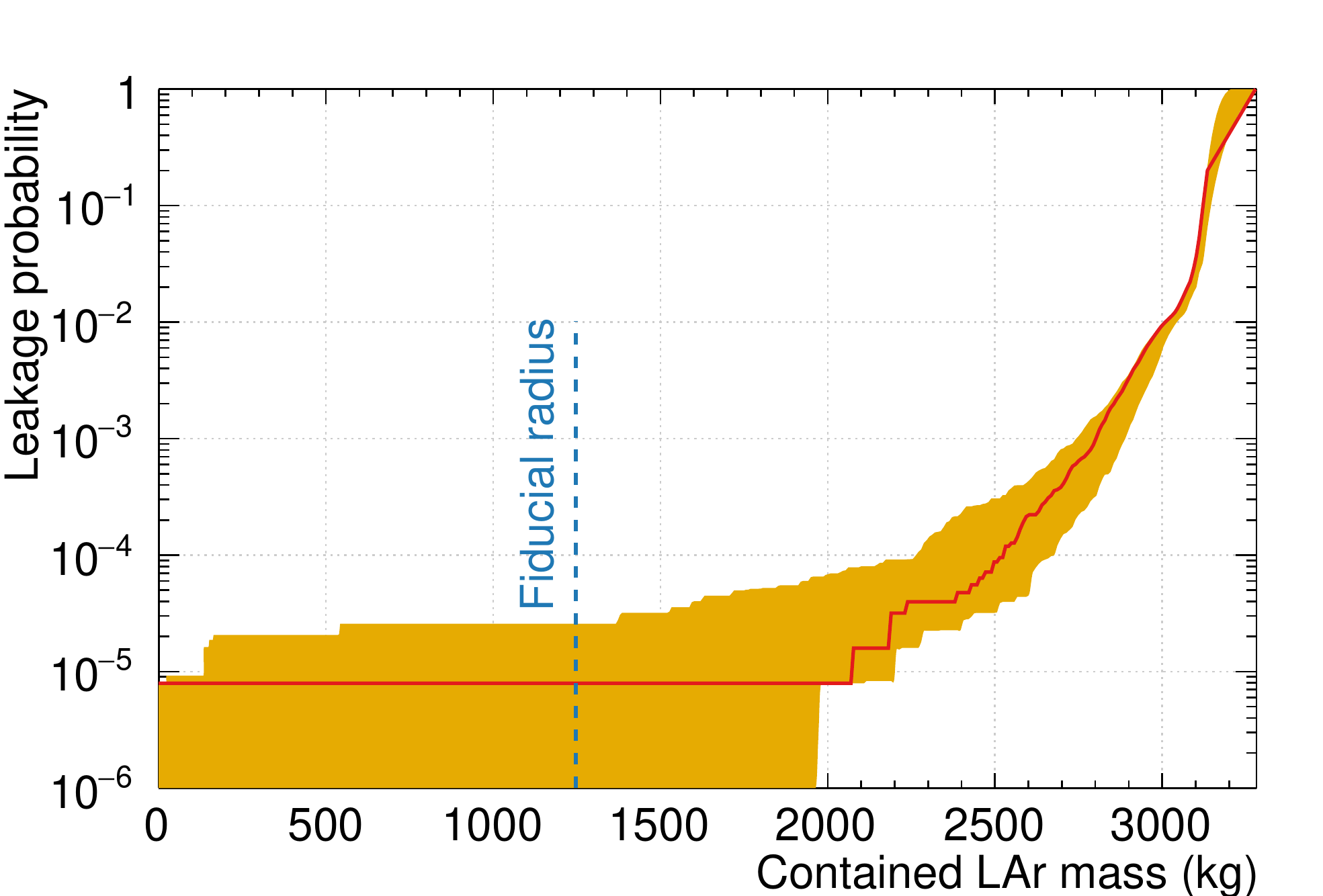}
 \caption{Leakage probability for simulated \alpds\ in the \WIMP\ \PE\ range vs.~the contained \LAr\ mass as determined by events within a given reconstructed radius. The contained \LAr\ mass corresponding to the fiducial cut at 630 mm is shown. The systematic uncertainty on the probability is shown also (yellow band).}
 \label{fig:alphaLeakage}
\end{figure}

A measurement of the leakage probability, defined as the fraction of events that reconstruct withing a given volume is calculated using a simulation of \alpds\ at the \LAr/\TPB\ interface. 
As shown in Figure~\ref{fig:alphaLeakage}, in the \WIMPPERange\ range, the leakage fraction is \PaperTwoLArTPBSurfaceAlphaLeakage\ for the \LAr\ mass contained within the fiducial radius. 
In this analysis, $N^{\text{ROI}}_{\text{$\alpha$, AV}}$\AVSurfaceEventsAfterCuts\ (90\%~\CL) events from \AV\ surface decays are predicted after all event selection cuts in the \WIMP\ \ROI.

In the \SIrange{10000}{20000} \PE\ range of Figure~\ref{fig:surfaceAlpha} two observations are made: (1) 5\% of the events reconstruct with a radius $<$~630~mm and (2) an excess of events is observed reconstructing towards the bottom of the detector. Neither of these observations are predicted by the model in this \PE\ range. 
The excess appears for events reconstructing both inside and outside the fiducial radius. 
A similar excess is observed at lower energies, in a \SIrange{200}{1000} \PE\ sideband of the \WIMP\ \ROI. 
To mitigate potential events of this type from the dark matter search, an additional cut is applied to remove events where 10\% or more of the total event charge is contained in the bottom three rows of \PMTs. 
The cut value is determined from events in the \SIrange{200}{1000} \PE\ sideband, where 99\% of the excess events are removed by the cut. 

This cut fiducializes against events originating from the bottom of the \AV, and hence features in the fiducial cut selection in Table~\ref{tab:lowLevelCuts-fiducial-cuts}. 
It is not solely responsible for removing any events from the \WIMP\ \ROI, and it does not remove all events in the sideband. 
These excess events are being investigated and it is not yet known if the excess at high \PE\ is related to the excess in the low \PE\ sideband. 
One explanation being considered, among others, is a low level of particulate contamination in the \LAr.

\subsubsection{Long-lived \alpds: \AV\ neck}\label{sec:neck-alphas}

\begin{figure}[htb]
 \centering
 \includegraphics[width=0.9\linewidth]{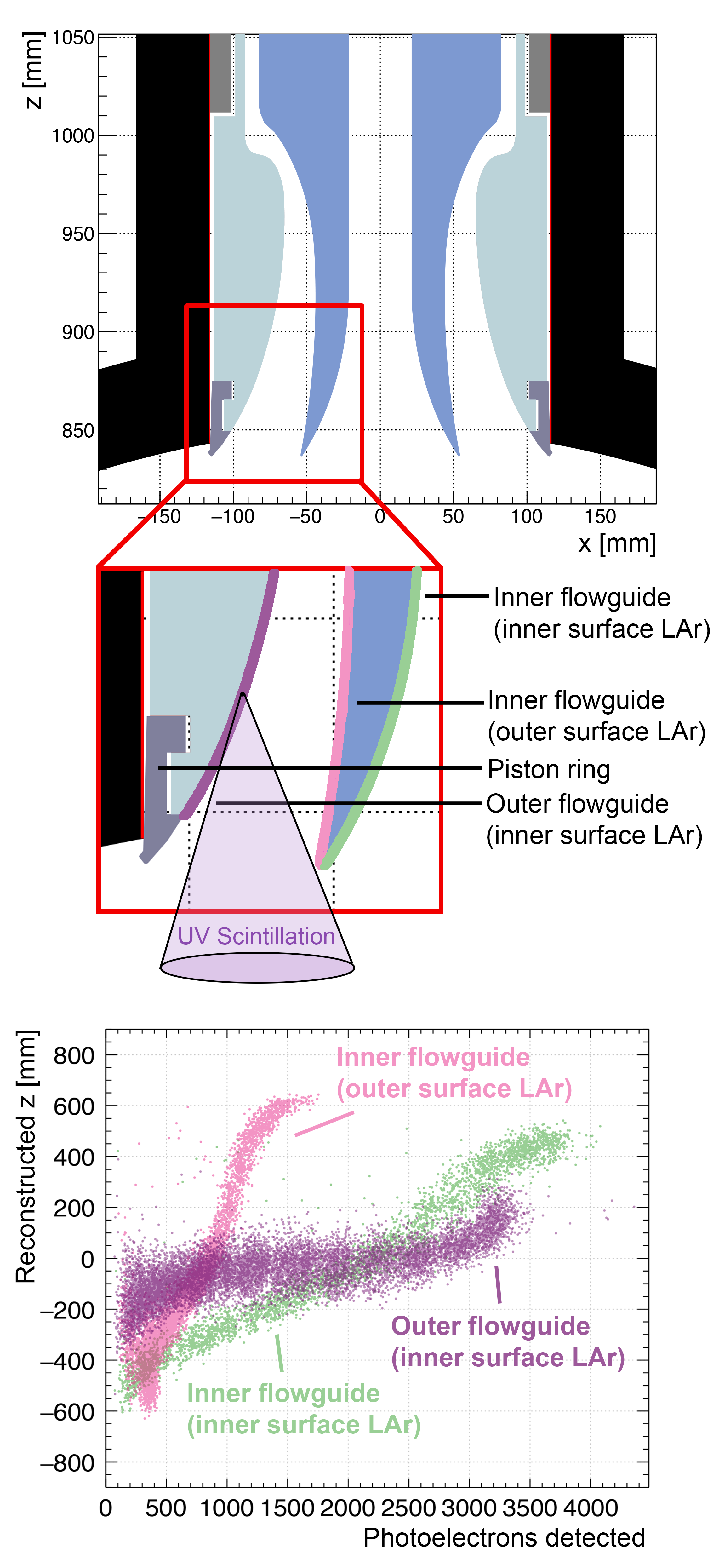}
 \caption{Top \& Middle: Cross-sectional illustration of the \FG\ components in the \AV\ neck. Shown are the three \FG\ surfaces and the piston ring (not coated in \LAr\ for purposes of illustration). Bottom: Simulated relationship in reconstructed $z$~vs.~\PE\ for \alpds\ on the IFG-IS (green), IFG-OS (pink), and OFG-IS (purple).
 }
 \label{fig:neck-alpha-illustration}
\end{figure}

The largest contribution to the background rate after applying fiducial cuts is from \poten\ \alpds\ on the surfaces of the acrylic \FGs\ in the \AV\ neck. 
The neck contains two separate \FGs, referred to as the inner and outer \FGs\ (\IFG\ and \OFG).
There are three distinct surfaces on these components: the \IFG's inner and outer surfaces (\IFGIS\ and \IFGOS) and the \OFG's inner surface (\OFGIS).
These surfaces are illustrated in Figure~\ref{fig:neck-alpha-illustration}. 
The outer surface of the outer \FG\ has no direct line of sight to the \AV; it is coupled to the wall of the \AV\ neck. 
The \FGs\ are located in the \GAr\ buffer region above the fill level of the \LAr\ target. Scintillation light is observed from events in the neck, which are simulated with a thin \LAr\ film coating the \FGs. 
Simulations show that \alpps\ emitted by \poten\ \alpds\ generate up to \SI{5000}{\PE} in the \AV\ \PMTs\ when they scintillate in this film. 
The \FGs\ are not coated in \TPB, and the acrylic absorbs most of the \UV\ scintillation photons incident on their surfaces. 
This results in shadowed event topologies in which only a small fraction of the emitted photons reach the \AV\ \PMTs. 
The number of \PEs\ produced by an \alpd\ on the \FG\ surfaces is therefore determined by the location of the decay.

The mean \FPrompt\ value observed for \alpds\ in the neck is constant up to \SI{5000}{\PE}, consistent with the model prediction for \PoTenAlphaEnergy\ \alpp\ scintilation in \LAr.
This suggests that these events are predominantly from \poten\ \alpds.
As a cross-check, a \bipo\ coincidence search on these events was performed, looking for \bifour\ \grs\ in coincidence with \alpds\ in the neck.
This search provided no evidence that a significant fraction of these events are caused by \alpds\ of other \ura\ progeny.

The shadowing of \UV\ scintillation in the \AV\ neck causes distinct charge distributions across the \PMTs\ for each \FG\ surface.
These distributions produce trends in events' reconstructed $z$-positions that vary with \PE\ in distinct ways, as illustrated in Figure~\ref{fig:neck-alpha-illustration}.

\begin{figure}[htb!]
 \centering
 \includegraphics[width=0.93\linewidth]{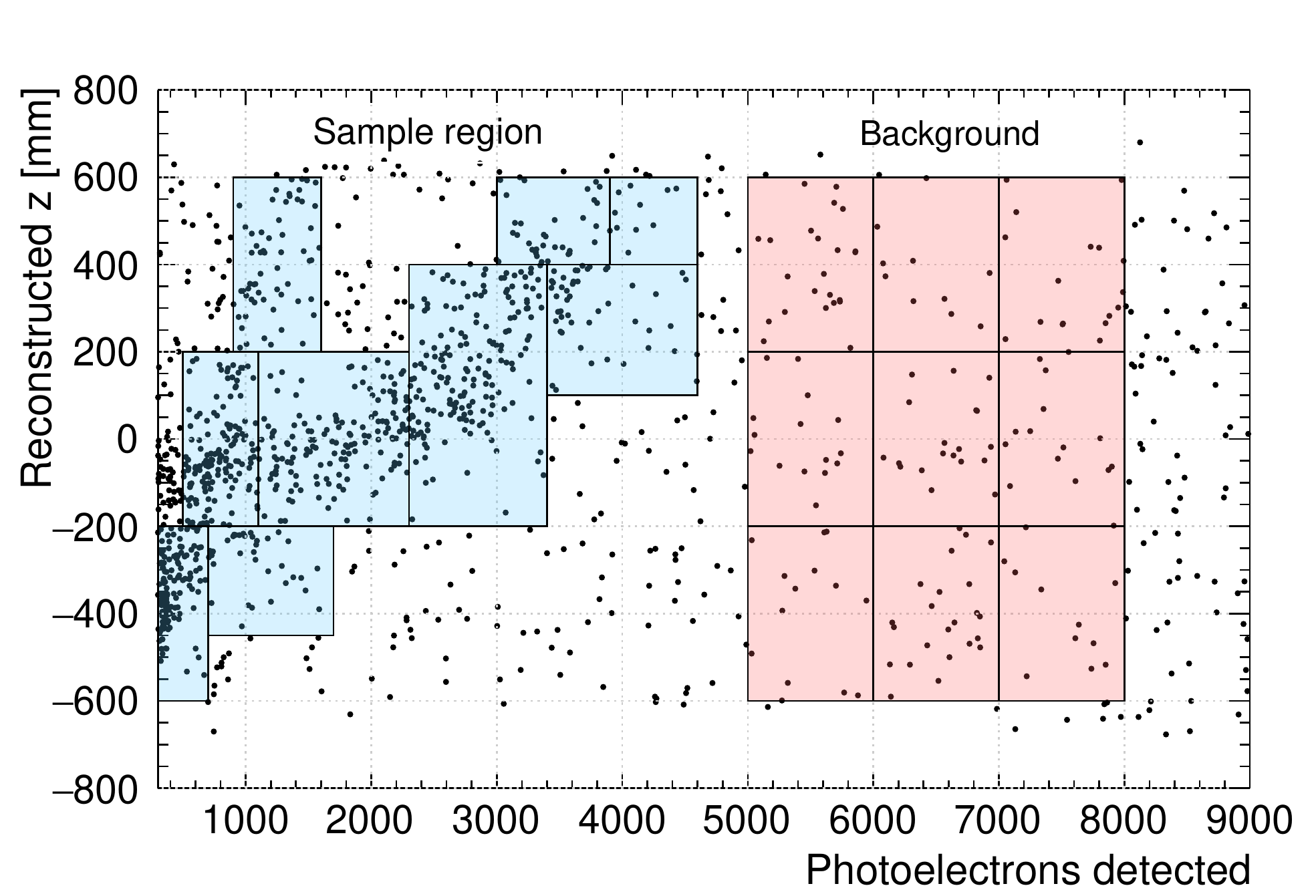}\\
 \includegraphics[width=0.95\linewidth]{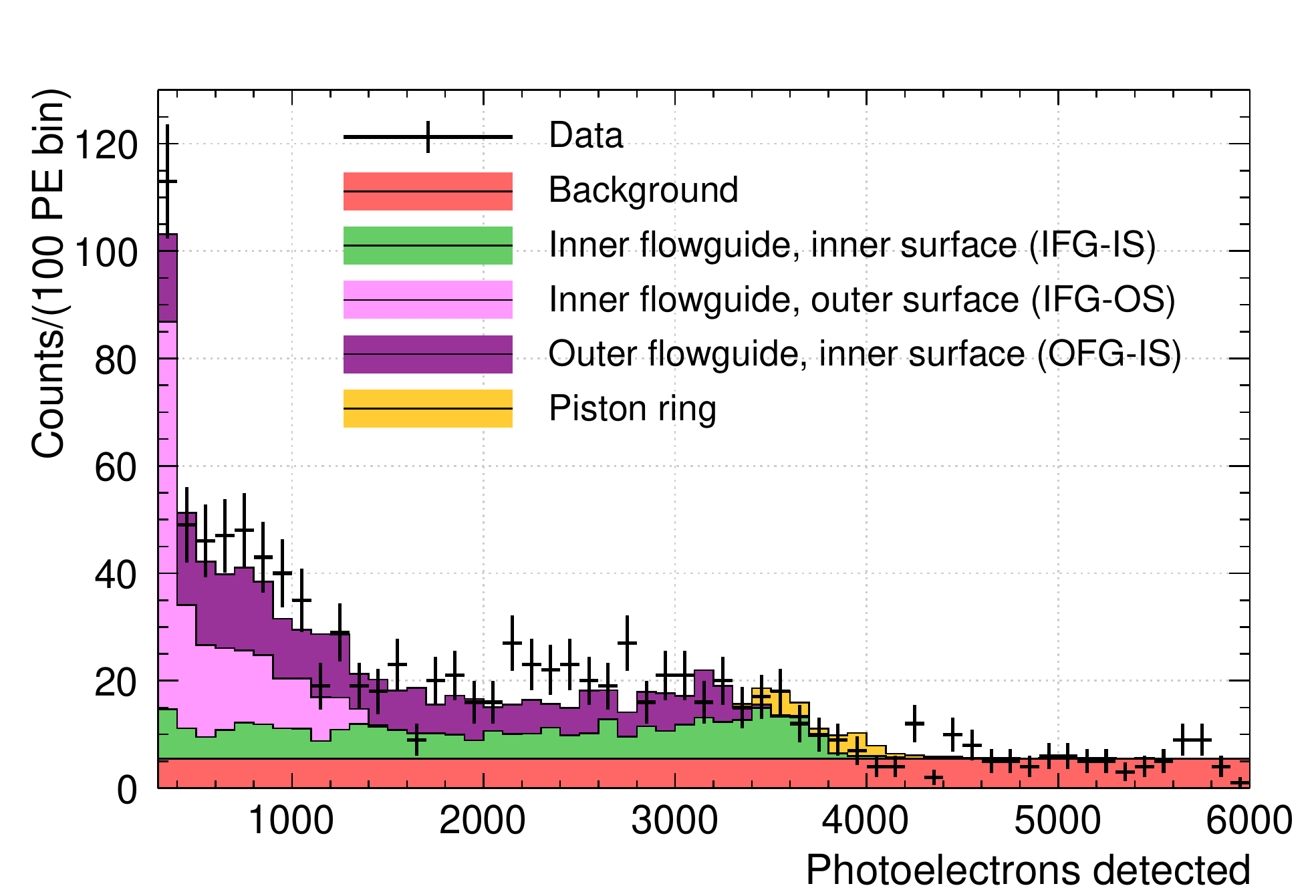}
 \caption{Top: Distribution of reconstructed $z$~vs.~\PE\ for events at high \FPrompt\ in data, using the \ChargeBasedPosRec\ algorithm. Shown are trends from \alpds\ in the neck between \SIrange{300}{5000}{\PE} and a background sideband at higher \PE\ values. The blue boxes denote the sample regions used by the template fit to compare with \PDFs\ from simulations. The red boxes are used to sample a background component. Bottom: 1D-projection of the observed \PE\ spectra (black points) and expected distributions from the background model components, normalized to the best-fit rates. 
 }
 \label{fig:neck-alpha-fit}
\end{figure}

These distributions are used to identify different sample regions within the reconstructed $z$~vs.~\PE\ plane, in which a template fit is performed in the \SIrange{300}{4600}{\PE} range. 
After low-level cuts, events in this energy range with \FPrompt\SI{>0.55}{} are selected and two additional cuts are applied: a loose fiducial cut requiring that the reconstructed $z$-position be \SI{<600}{\mm} and a cut removing events in which more than \SI{3.5}{\percent} of the total event charge is contained within single \PMT. 
The latter cut reduces the number of \AV\ surface events from the event selection.
This fit makes use of simulated distributions for \poten\ \alpds\ on each of the three \FG\ surfaces, shown in Figure~\ref{fig:neck-alpha-fit}. 
A flat background obtained from a \SIrange{5000}{8000}{\PE} sideband is also included in this fit, along with a simulated \PE\ distribution from \poten\ \alpds\ in a \LAr\ film on the \AV\ piston ring. 
The piston ring contributes a small amount to the event rate in the \SIrange{3000}{4000}{\PE} range. 
The \alpds\ on the piston ring were independently studied, and found to have a negligible contribution to the background rate in the \WIMP\ \ROI; scintillation events on this surface are shadowed too little to populate the \WIMP\ \PE\ range.
The fit result predicts the following event rates for each \FG\ component before applying any cuts: \PaperTwoIFGISRate\ (\IFGIS), \PaperTwoIFGOSRate\ (\IFGOS) and \PaperTwoOFGISRate\ (\OFGIS).

The model of the \LAr\ film discussed here assumes that the \FGs\ and piston ring are completely coated with a uniform \LArFilmThickness-thick layer of \LAr. 
This is thick enough for the \alpp\ to stop in the \LAr, resulting in an \FPrompt\ distribution consistent with the one observed in the data. 
Variations on the model with thinner films allow for a contribution of \GAr\ scintillation; these  scenarios are factored into the systematic uncertainty. 
The fraction of the \FG\ surfaces coated in \LAr\ is not yet known, so the event rates cannot be converted into activities. 
This does not impact the prediction of the event rate in the \WIMP\ \ROI\ based on the result of the fit. 

The \CR\ for \alpds\ in the \AV\ neck consists of the \WIMP\ \ROI\ and low-level event selection cuts. 
In the \CR, the model predicts \PaperTwoIFGISAlphaBkgdBeforeCuts, \PaperTwoIFGOSAlphaBkgdBeforeCuts, and \PaperTwoOFGISAlphaBkgdBeforeCuts\ events from the \IFGIS, \IFGOS, and \OFGIS, respectively.
These values give a total expectation of $N^{\text{CR}}_{\text{$\alpha$, neck}}=$\PaperTwoTotalNeckAlphaBkgdBeforeCuts. 
These uncertainties are dominated by systematic uncertainties in the background model; their effects on the background expectation of \alpds\ on the \FGs\ after all background rejection and fiducial cuts are summarized in Table~\ref{tab:neck-systematics}.

\begin{table}[htb]
 \centering
  \setlength\extrarowheight{6pt}
 \caption{Summary of the uncertainty on the overall number of events remaining in the \WIMP\ \ROI\ after applying all background rejection and fiducial cuts. Uncertainties are quoted for each \FG\ surface component.}
 \begin{tabular}{L{3.25 cm}C{1.2cm}C{1.2cm}C{1.2cm}}\hline\hline
 	      \multirow{2}{*}{Systematic} & \multicolumn{3}{c}{Uncertainty [\%]} \\
                                          & IFG-IS                          & IFG-OS                          & OFG-IS                          \\\hline
              Refractive index            & $^{+7}_{-42}$                   & $^{+25}_{-10}$                  & $^{+13}_{-10}$                  \\
  	      \TPB\ scattering length     & \multirow{1}{*}{$^{+86}_{-29}$} & \multirow{1}{*}{$^{+28}_{-21}$} & \multirow{1}{*}{$^{+19}_{-0}$}  \\ 
  	      Afterpulsing prob.          & \multirow{1}{*}{$^{+26}_{-36}$} & \multirow{1}{*}{$^{+0}_{-32}$}  & \multirow{1}{*}{$^{+4}_{-24}$}  \\ 
              Light yield                 & \multirow{1}{*}{$^{+54}_{-0}$}  & \multirow{1}{*}{$^{+0}_{-6}$}   & \multirow{1}{*}{$^{+13}_{-4}$}  \\ 
              Rel. \PMT\ eff.             & $^{+8}_{-0}$                    & $^{+0}_{-13}$                   & $^{+0}_{-29}$                   \\ 
              \alpp\ \FPrompt\            & $^{+83}_{-50}$                  & $^{+58}_{-42}$                  & $^{+80}_{-47}$                  \\
              Reconstructed radius        & $^{+0}_{-75}$                   & $^{+0}_{-31}$                   & $^{+0}_{-26}$                   \\
              LAr film thickness          & $^{+104}_{-0}$                  & $^{+0}_{-49}$                   & $^{+0}_{-66}$                   \\\hline
              \textbf{Combined}	          & $^{+170}_{-110}$                & $^{+69}_{-83}$                  & $^{+85}_{-80}$                  \\\hline\hline
 \end{tabular}
 \label{tab:neck-systematics}
\end{table}

The dominant systematic uncertainties are related to the model of \alpp\ scintillation parameters and the thickness of the \LAr\ film, variations of which can make reconstructed events migrate in or out of the \ROI. 
In addition, changes to the \TPB\ scattering length have a significant impact on the fraction of events that survive fiducial cuts. 
\UV\ photons incident on the \TPB\ layer of the \AV\ can be back-scattered away from the \PMTs, producing an inward bias to the reconstructed radius position. 
It is particularly relevant to sources of collimated \UV\ light such as those generated by \alpds\ in the neck.

The distinctive \FPrompt\ and $z$~vs.~\PE\ distributions produced by \alpds\ in the \AV\ neck distinguish them from \NR\ events originating in the \LAr\ target.  
Cuts have been developed to use this information and were optimized such that the \WIMP\ acceptance is maximized while maintaining a background expectation below the target of \NeckAlphaLeakage\ events from all \FG\ components. 
A summary of the rejection efficiency for each cut removing these events is shown in Table~\ref{tab:neck-alpha-cuts}. 
The reported rejection efficiencies are for events reconstructing inside the \PaperTwoRadialCut\ fiducial radius and in the \WIMPPERange\ range. 
These cuts are described in more detail below.

\begin{table}[htb]
 \centering
  \setlength\extrarowheight{6pt}
 \caption{Predicted rejection efficiency of each cut to remove events generated by \alpds\ from each of the three \FG\ surfaces. 
 The efficiency is calculated for events with a reconstructed radius $<$\PaperTwoRadialCut\ in the range of \WIMPPERange. 
 These efficiencies are determined from simulations. The last row provides an estimate of the combined rejection efficiency after applying all four cuts.}
 \begin{tabular}{L{2.5cm}C{1.2cm}C{1.2cm}C{1.2cm}}\hline\hline
 	      \multirow{2}{*}{Cut name}                & \multicolumn{3}{c}{Neck \alpd\ rejection [\%]} \\
                                                       & IFG-IS              & IFG-OS              & OFG-IS  \\ \hline
              Upper \FPrompt\ cut                      & 73                  & 59                  & 72  \\
  	      Early pulses in \GAr\ \PMTs              & \multirow{2}{*}{80} & \multirow{2}{*}{85} & \multirow{2}{*}{81}  \\ 
  	      Charge fraction in top 2 rows of \PMTs\  & \multirow{3}{*}{57} & \multirow{3}{*}{46} & \multirow{3}{*}{36}  \\ 
              Position reconstruction consistency      & \multirow{3}{*}{90} & \multirow{3}{*}{93} & \multirow{3}{*}{82}  \\ \hline
               \textbf{Combined}	               & 99                  & 99                  & 98  \\ \hline \hline
 \end{tabular}
 \label{tab:neck-alpha-cuts}
\end{table}

Upper \FPrompt\ cut: \alpds\ with initial energies of \PoTenAlphaEnergy\ produce systematically higher \FPrompt\ values than are expected from \NRs\ in the \WIMP\ \ROI, as shown in Figure~\ref{fig:roi_psdbands}.  
The upper \FPrompt\ bound of the \ROI\ therefore removes a significant fraction of \alpds\ in the neck, at the cost of \PaperTwoROIAcceptanceLossTopFP\ acceptance loss to signal events.

Early pulses in \GAr\ \PMTs: The origin of \UV\ scintillation light from these events is above the fill level of the \LAr. 
\PMTs\ whose \LGs\ subtend the \GAr\ region just above the \LAr\ fill level will register \PEs\ from \UV\ photons reflecting from the \GAr/\LAr\ interface before \PMTs\ located further down do so. 
This effect is aided by the fact that the group velocity of \UV\ light in \GAr\ is almost three times that in \LAr. 
The location of \PMTs\ that register the first pulses in the \PE\ integration window can be used to remove \alpds\ in the \AV\ neck.
The rejection power of this cut improves as the number of detected \PEs\ increases. 
At higher \PE\ more \UV\ photons enter the \AV, increasing the probability of \UV\ photons reflecting from the \GAr/\LAr\ interface and reaching a \PMT\ in the \GAr\ region first. 
Events are rejected if any of the first 3 pulses observed in the \PE\ integration window are registered in \PMTs\ that subtend the \GAr\ region. 
Based on simulations, a value of 3 was found to be the minimum number of pulses required to reach the neck \alpd\ background target of $<0.5$ events when combined with all other cuts, without requiring other cuts to induce an even larger acceptance loss.
This cut is the largest source of loss in signal acceptance within the fiducial volume. 
However, it is a simple model-independent method of fiducializing against backgrounds with topologies strongly affected by the \LAr/\GAr\ interface. 

Charge fraction in top 2 rows of \PMTs: Due to the reflection of \UV\ photons at the \GAr/\LAr\ interface, a larger fraction of \PEs\ are seen by \PMTs\ above the fill level for \alpds\ in the neck than for \WIMP-like recoils in the \LAr\ target. 
Events that have \SI{\geq4}{\percent} of the total observed charge seen in the top 2 rows of \PMTs\ (10 \PMTs\ in total) are removed. 
This cut is a charge-based equivalent of the early pulses in \GAr\ \PMT\ cut, removing events close to the \LAr/\GAr\ interface. These two cuts are correlated but are found to both be necessary in order to achieve the desired background level. 

Position reconstruction consistency: Scintillation events originating in the neck produce distinct hit patterns in each channel compared to events originating in the \LAr\ target. 
Since both position reconstruction algorithms are trained on scintillation events in the \LAr\ target, they generally reconstruct such events at similar positions, as demonstrated in Figure~\ref{fig:positionFitterCompZ}. 
However the algorithms give systematically different results for events originating in the neck.
The \ChargeBasedPosRec\ algorithm systematically reconstructs these events lower in the detector than is predicted by the \TimeBasedPosRec\ algorithm. 
To remove events coming from the neck, a consistency cut is used, removing events where the \TimeBasedPosRec\ algorithm reconstructs an event significantly higher in the detector than the \ChargeBasedPosRec\ algorithm did or where both algorithms reconstruct the event far from each other.
Events are rejected if the \TimeBasedPosRec\ algorithm returns a $z$ coordinate higher than the \ChargeBasedPosRec\ algorithm, with a difference more than what would be expected for 90\% of \LAr\ scintillation events.
If an event passes this cut, a further criterion is required: the distance between both estimated positions must be within the expected range for 85\% of \LAr\ scintillation events. 
The cut value for both consistency criteria is a function of prompt \PE.  

After applying all fiducial and background rejection cuts \instr{\PaperTwoIFGISAlphaBkgd}, \instr{\PaperTwoIFGOSAlphaBkgd}, and \instr{\PaperTwoOFGISAlphaBkgd} events from the \IFGIS, \IFGOS\ and \OFGIS\ components are expected in the \WIMP\ \ROI, respectively.
This combines to an overall expectation of $N^{\text{ROI}}_{\text{$\alpha$, neck}}=$\PaperTwoTotalNeckAlphaBkgd\ events in the dataset.

\subsection{Summary of backgrounds}
\label{sec:bkgds-summary}

After applying all fiducial and background rejection cuts to the data, no events remain in the \WIMP\ \ROI. 
Table~\ref{tab:cuts} summarizes the background rejection cuts and their cumulative effect on the \WIMP\ acceptance after applying all fiducial cuts. 
The acceptance is determined using \arnine\ \ER\ signals in the \WIMPPERange\ range, as was done for the calculation of the fiducial mass in Section~\ref{sec:background-characterization}. Similarly, the systematic uncertainty is derived from the level of agreement of these cuts to select simulated \arnine\ \ER\ and \arforty\ \NR\ events.
Table~\ref{tab:cuts} also shows the total predicted number of \ROI\ events, $N^{\text{ROI}}_{\text{bkg}}$ compared to the number observed in the dataset, $N^{\text{ROI}}_{\text{obs}}$.

\begin{table}[htb]
 \centering
  \setlength\extrarowheight{6pt}
 \caption{Cumulative impact of background rejection cuts on the \WIMP\ acceptance, the predicted number of background events, $N^{\text{ROI}}_{\text{bkg}}$ and the total number of observed background events,  $N^{\text{ROI}}_{\text{obs}}$ after applying fiducial cuts to events inside \WIMP\ \ROI. Cuts are grouped by the background they predominantly remove. The value of the acceptance is averaged over the \WIMPPERange\ range.}
 \begin{tabular}{cL{2.5cm}C{1.8cm}C{1.3cm}C{1.3cm}}\hline\hline
                                                                          & Background rejection cut   & \WIMP\ accept. [\%]                             & $N^{\text{ROI}}_{\text{bkg}}$                       & $N^{\text{ROI}}_{\text{obs}}$                         \\\hline
  \multirow{2}{*}{\rotatebox[origin=c]{90}{\tiny Cherenkov}}              & \multirow{2}{*}{Neck veto} & \multirow{2}{*}{\WIMPFiducialAcceptanceNeckVeto}& \multirow{2}{*}{\PredEventsROINeckVeto} & \multirow{2}{*}{\EventsROINeckVeto} \\\\\rule{0pt}{3ex}
  \multirow{4}{*}{\rotatebox[origin=c]{90}{\tiny $\alpha$-decays in neck}}& Early pulses in GAr \PMTs  & \multirow{2}{*}{\WIMPFiducialAcceptancePIFGAR}  & \multirow{2}{*}{\PredEventsROIPIFGAR}   & \multirow{2}{*}{\EventsROIPIFGAR}     \\\rule{0pt}{5ex}
                                                               	          & Position fitter consistency& \multirow{2}{*}{\WIMPFiducialAcceptance}        & \multirow{2}{*}{\TotalBackgroundExpectationAfterCuts} & \multirow{2}{*}{\EventsROI} \\\hline
                                                               	          & \multirow{1}{*}{\textbf{Total}}             & \multirow{1}{*}{\WIMPFiducialAcceptance}                        & \multirow{1}{*}{\TotalBackgroundExpectationAfterCuts}         & \multirow{1}{*}{\EventsROI}                  \\ 
 \hline\hline
 \end{tabular}
 \label{tab:cuts}
\end{table}

\section{\WIMP\ search analysis}~\label{sec:analysis}
Once all \WIMP\ event selection cuts have been applied to the data, the number of events remaining in the \WIMP\ \ROI\ is used to place an upper limit at the 90\% confidence level on the spin-independent \WIMP-nucleon cross section. 

\subsection{ROI definition}
\label{sec:roi-definition}

\begin{figure}[tb]
 \centering
 \includegraphics[width=1.01\linewidth]{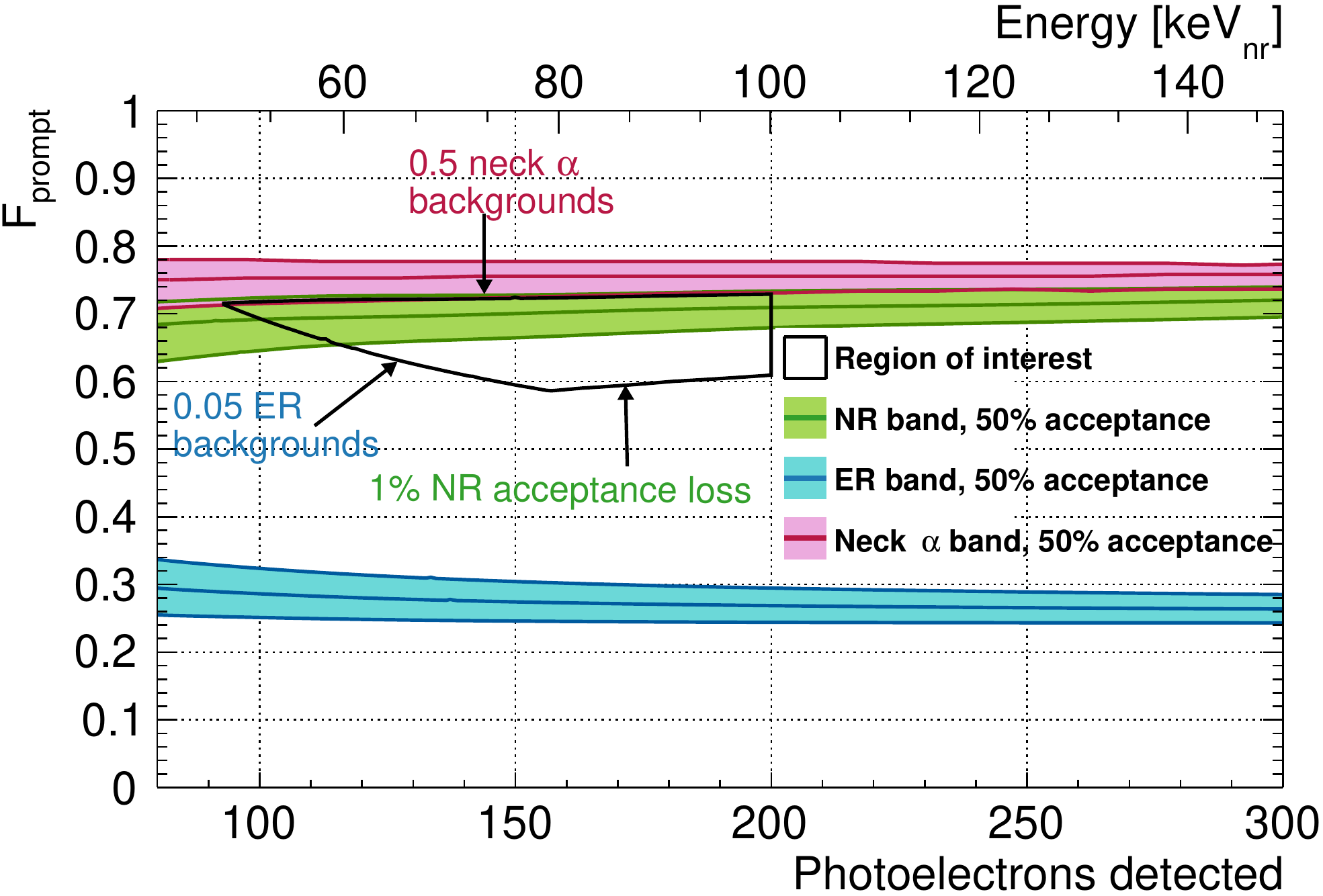}
 \caption{Illustration of the \WIMP\ \ROI\ (black) along with the \ER\ (blue), \NR\ (green) and neck \alpd\ (pink) bands that define the boundaries. Each band is drawn about the median of each class of event, with 25\% of such signals above and 25\% below included in the shaded regions.
 }
 \label{fig:roi_psdbands}
\end{figure}

The \ROI\ used for the \WIMP\ search is driven by the signal and background model. Figure~\ref{fig:roi_psdbands} illustrates how the \ROI\ is defined in the \FPrompt~vs.~\PE\ plane; 50\% acceptance bands are shown for events generated by \ERs, \NRs, and \AV\ neck \poten\ \alpds. The \ER\ and \NR\ bands shown here are from the detector response model described in Section~\ref{subsec:calibration_psd}; the neck \alpd\ band was taken from simulations of \poten\ surface decays on surfaces of the AV neck \FGs.
Since these \alpds\ are at energies of \PoTenAlphaEnergy, their \FPrompt\ values are correspondingly high. 
As such, these events have higher \FPrompt\ values than expected from \NRs\ originating in the \LAr\ target, and they are discriminated against with \PSD.

The lower \FPrompt\ bound of the \ROI\ is defined using two curves. In the \SIrange{95}{160} \PE\ range it is defined to have a constant expected number of leakage events in each \PE\ bin such that the total background rate of \ER\ leakage events after all cuts are applied is \ERLeakage. The curve spanning the \SIrange{160}{200} \PE\ range is defined such that there is a constant 1\% \NR\ acceptance loss.

The upper \FPrompt\ bound of the \ROI\ is defined to have constant \NR\ acceptance loss, with \PaperTwoROIAcceptanceLossTopFP\ of \NRs\ in each \PE\ bin expected to fall above the \ROI. 
As previously described this acceptance loss was chosen because it contributes towards achieving an expectation of \NeckAlphaLeakage\ events from \alpds\ in the \AV\ neck.

The upper \PE\ bound of 200\,\PE\ is chosen to be consistent with the energy of the upper bound used in~\cite{amaudruz_first_2017}, given the different light yields and energy estimators used for both analyses. 
Above this energy, the expected rate of \alpp- and neutron-related background events becomes larger, while a negligible fraction of \WIMP\ events are expected.

\subsection{WIMP acceptance}

\begin{figure}[htb]
 \centering
 \includegraphics[width=1.01\linewidth]{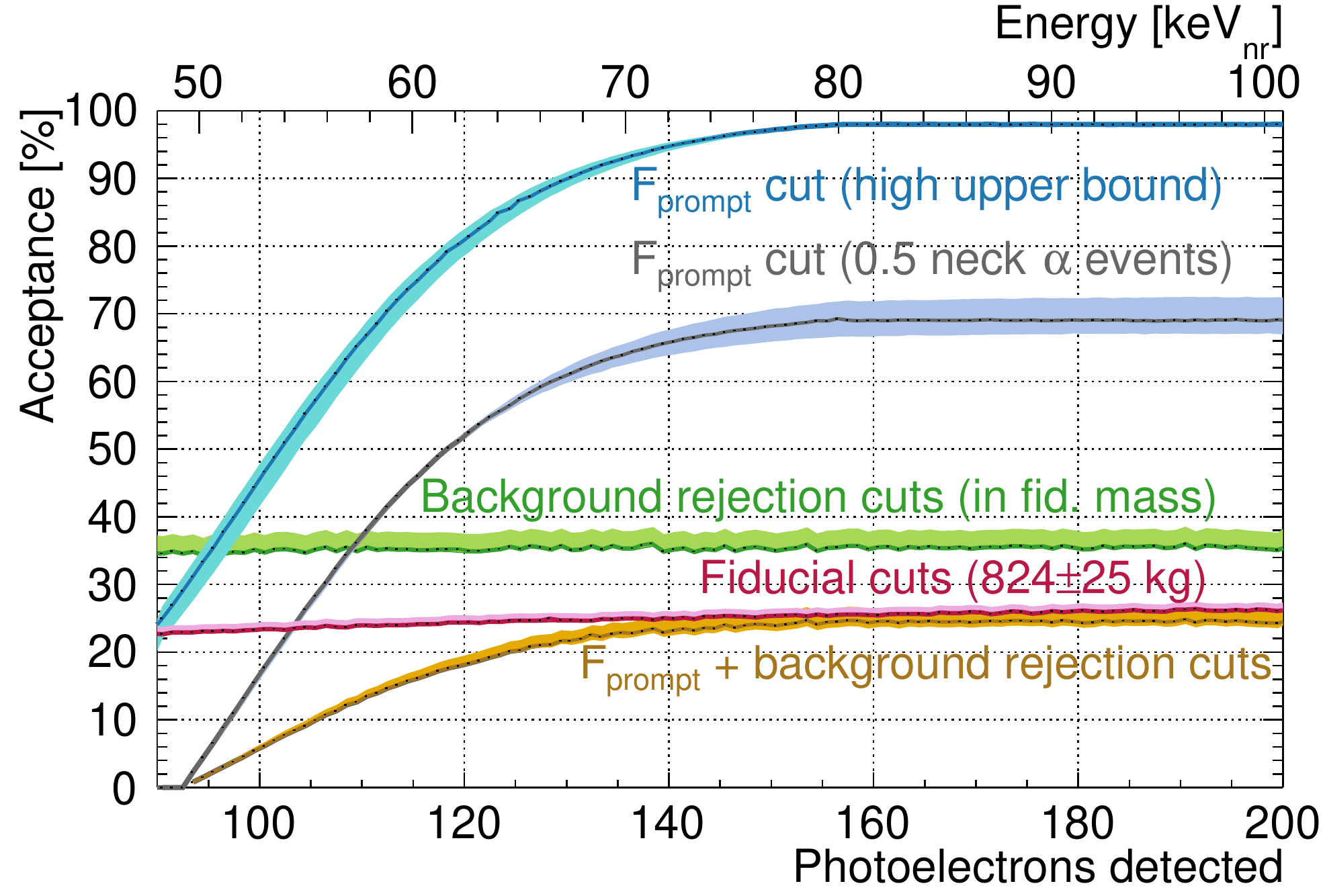}
 \caption{\WIMP\ acceptance as a function of \PE, broken down by cut type. 
 }
 \label{fig:acceptance}
\end{figure}

The \WIMP\ acceptance as a function of \PE\ is shown in Figure~\ref{fig:acceptance}.
``Fiducial cuts'' shows the probability of a \WIMP-like event passing fiducial cuts.
``Background rejection cuts'' refers to the probability of an event passing the cuts listed in Table~\ref{tab:cuts} given that it passed the low-level event selection and fiducial cuts listed in Table~\ref{tab:lowLevelCuts-fiducial-cuts}.
``\FPrompt\ cut'' refers to the probability of an \NR\ appearing in the \ROI.
Figure~\ref{fig:acceptance} also shows the \FPrompt\ cut acceptance for an \ROI\ defined with 1\% \WIMP\ acceptance loss from the upper \FPrompt\ bound, instead of 30\% acceptance loss as in this analysis.
This corresponds to the energy threshold that would be achievable if \alpds\ in the \AV\ neck did not require a tighter cut, but an expectation of \ERLeakage\ \ER\ leakage events were maintained.
This curve demonstrates the power of \PSD\ for discriminating against \ER\ backgrounds while maintaining a low energy threshold.

\subsection{Results}

\begin{figure}[htb]
 \centering
\includegraphics[width=1.01\linewidth]{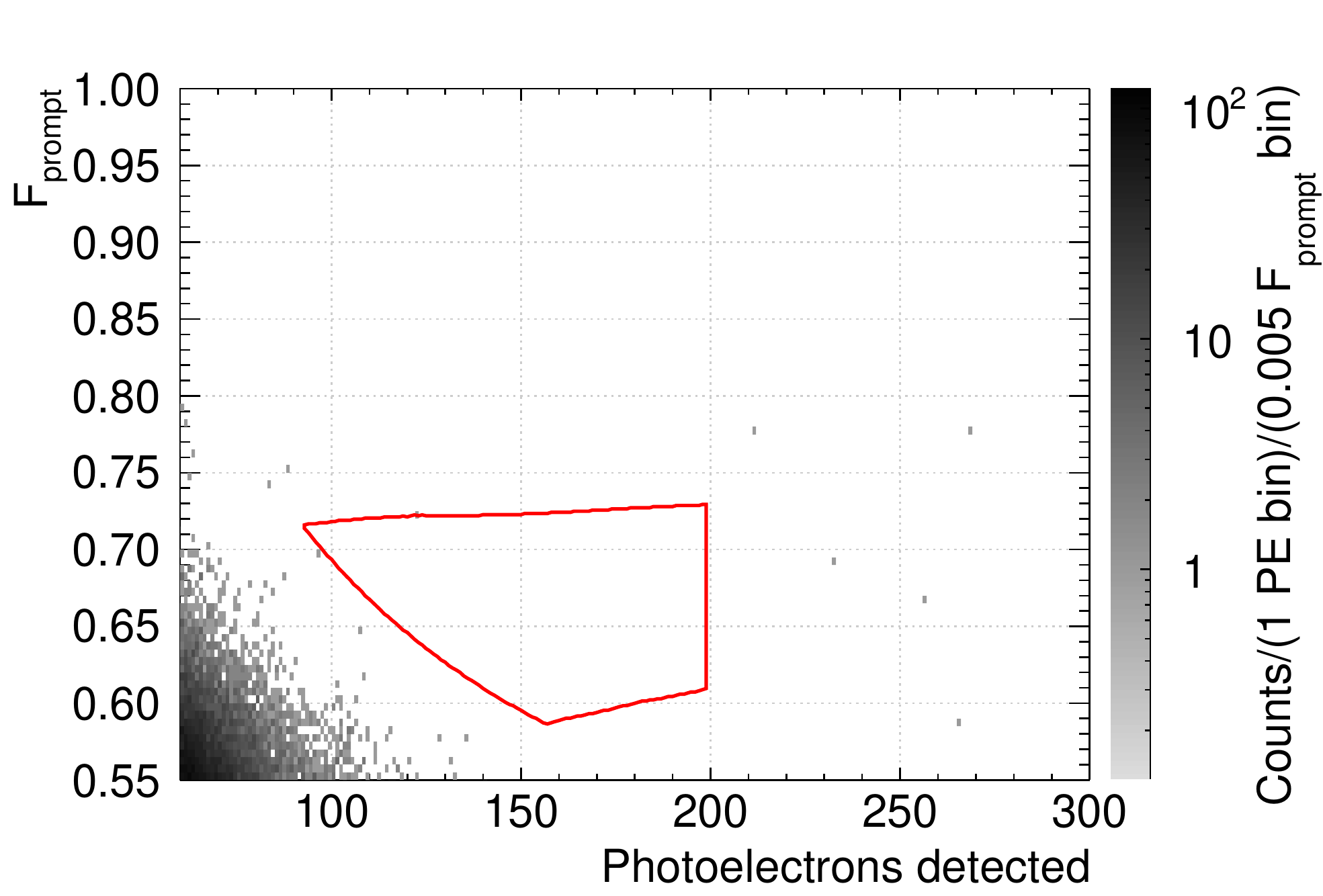}
 \caption{Observed \FPrompt\ vs. \PE\ distribution after all cuts. The region of interest is shown in red. 
 }
 \label{fig:roi_after_cuts}
\end{figure}

After applying all \WIMP\ search cuts described in Tables~\ref{tab:lowLevelCuts-fiducial-cuts} and~\ref{tab:cuts}, the events shown in Figure~\ref{fig:roi_after_cuts} remain. 
There are no events remaining in the region of interest. There is one event close to the \ROI\ border, with \FPrompt\SI{<0.75}{} and approximately \SI{125}{\PE} that is above the upper \FPrompt\ bound of the \ROI. 
There are also 5 events in the \SIrange{200}{300} \PE\ range with \FPrange{0.55}{1.0}. 
The background model discussed here is used to determine the probability that either of these two event populations are likely.

In the \WIMPPERange\ range, the background model predicts \PaperTwoTotalPredictionHighFP\ events with \FPrompt\ values between the top boundary of the \ROI\ and \FPrompt\SI{<0.75}{}.
The probability of observing one or more events in this region is \SI{36}{\percent}, and so the observed event is consistent with the model. 
Between \SIrange{200}{300}{\PE}, a total of \PaperTwoTotalPredictionHighPE\ background events are predicted with \FPrange{0.55}{1.0}. 
In this region, the number of predicted events from \alpds\ in the \AV\ neck depends most strongly on the uncertainty in modeling the light yield for events originating in the neck. 
In order for the background model to be consistent with the events observed in this region, the optical properties of the neck or the position resolution must change, and in the case of the latter by several times its uncertainty. Systematic uncertainties on optical properties of the neck relevant to events in this energy range are still being evaluated.
Varying the systematic uncertainties at the required levels does not significantly affect the \WIMP\ exclusion curve presented here. 
The observed excess over the nominal model extends above \SI{300}{\PE}.
Future analyses will explore adding new background sources to the model and further constrain the relevant response functions.

\begin{figure}
 \centering
 \includegraphics[width=1.01\linewidth]{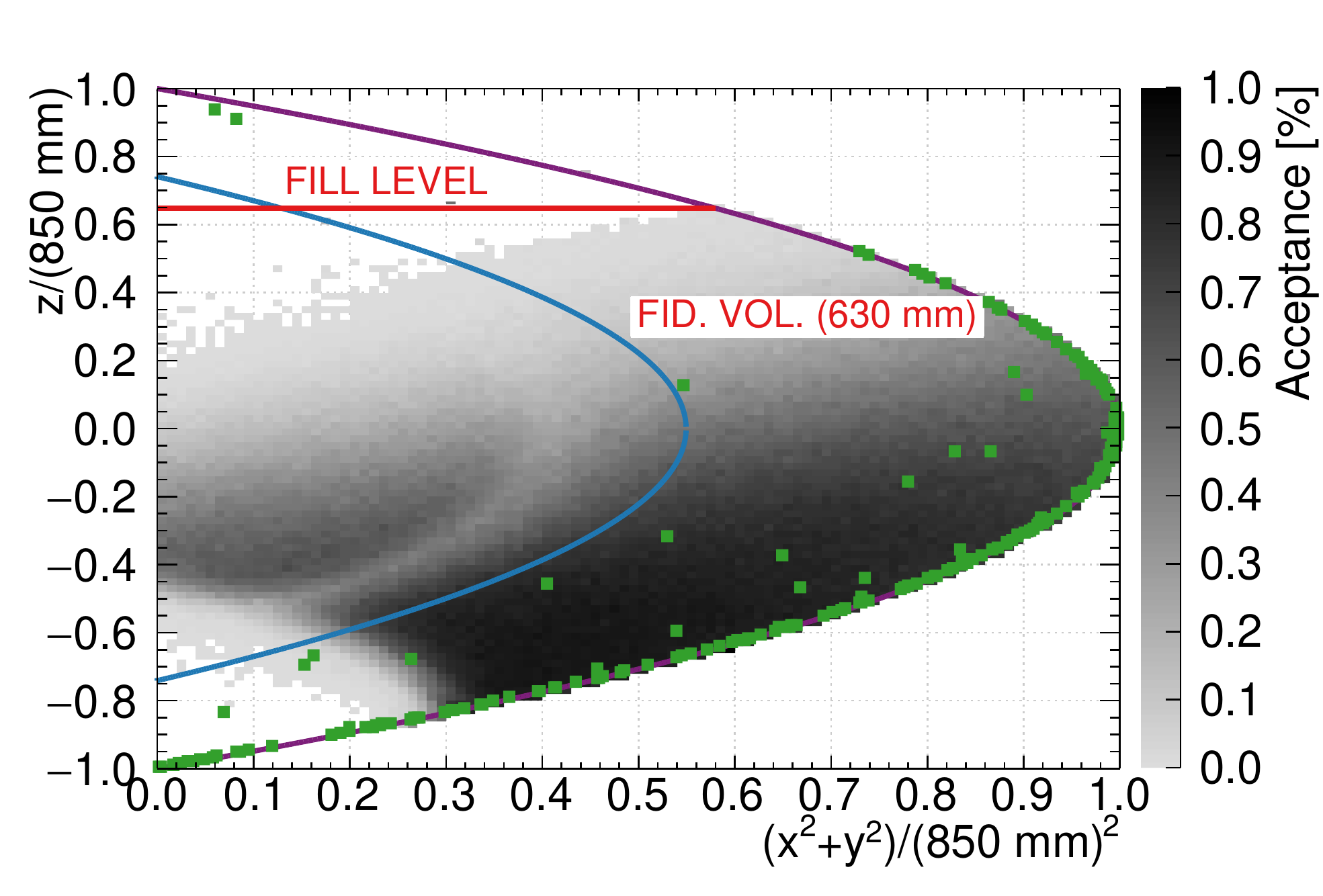}
  \caption{Observed spatial distribution for all events surviving all cuts other than the cut on reconstructed radius. The color scale in the background shows the acceptance for \arnine\ events measured as a function of position after all but the radial cut; green points represent events in the \ROI\ after all background rejection cuts. The fill level and radial fiducial cuts are drawn as well. 
  }
 \label{fig:pos_after_cuts}
\end{figure}

Figure~\ref{fig:pos_after_cuts} shows the spatial distribution of events within the \WIMP\ \ROI\ after all event selection cuts have been applied other than the fiducial radial cut. 
The fill level and the fiducial radius are both shown, and the acceptance as a function of position is illustrated in the background. 
The fiducializing effects of the cut on the fraction of observed charge in the 2 rows of \PMTs\ and bottom 3 rows of \PMTs, as summarized in Table~\ref{tab:lowLevelCuts-fiducial-cuts}, can be seen in this figure.

\begin{figure}[htb]
 \centering
 \includegraphics[width=\linewidth]{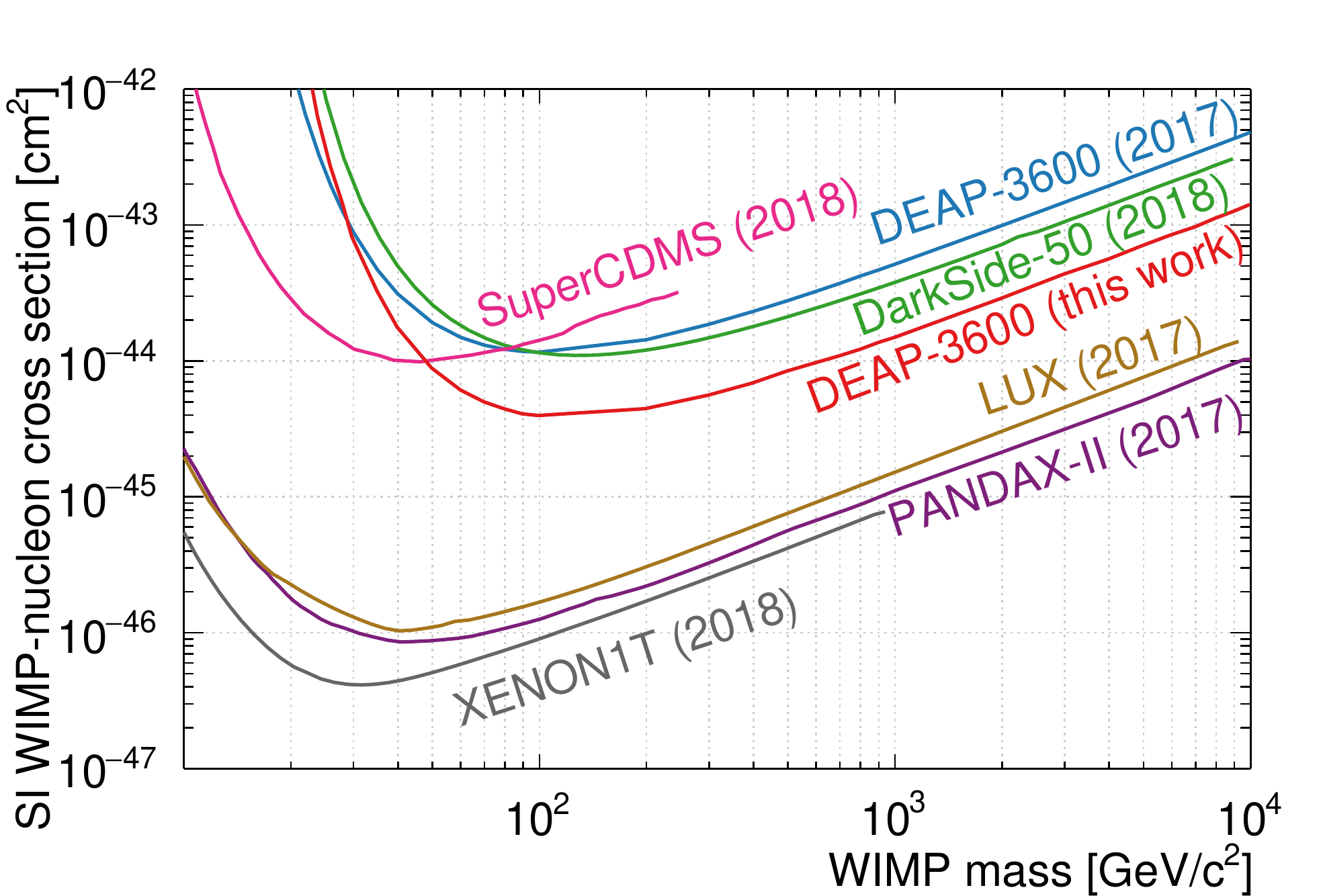}
 \caption{90\% confidence upper limit on the spin-independent \WIMP-nucleon cross sections based on the analysis presented in this paper (blue), compared to other published limits, including our previous limit~\cite{amaudruz_first_2017}, SuperCDMS~\cite{supercdms_collaboration_results_2018}, DarkSide-50~\cite{the_darkside_collaboration_darkside-50_2018}, LUX~\cite{lux_collaboration_results_2017}, PANDAX-II~\cite{pandax-ii_collaboration_dark_2017}, and XENON1T~\cite{aprile_dark_2018}.
 }
 \label{fig:exclusion}
\end{figure}

Figure~\ref{fig:exclusion} shows the 90\%~\CL\ upper limit on the spin-independent \WIMP-nucleon cross section as a function of \WIMP\ mass.
These upper limits are calculated accounting for the systematic uncertainties in the detector response function, following the prescription outlined by Highland and Cousins~\cite{cousins_incorporating_1992}. 
Uncertainties considered include those for the energy scale parameters in Table~\ref{tab:fitresults}, the \PSD\ model fit parameters in Equation~\ref{eq:psdnrmodel}, the \WIMP\ acceptance as shown in Figure~\ref{fig:acceptance}, the \NR\ quenching factors and mean \FPrompt\ values, as derived from~\cite{Cao:2015ks}, and a \PaperTwoExpoUncertainty\ uncertainty on the total exposure.

This analysis excludes spin-independent \WIMP-nucleon cross sections above \PaperTwoWIMPLimitOneHundredGeV\ (\PaperTwoWIMPLimitOneTeV) for \WIMPs\ with a mass of \SI{100}{GeV\per\square c} (\SI{1}{TeV\per\square c}), assuming the standard halo dark matter model described in~\cite{mccabe_astrophysical_2010}, with a Maxwell-Boltzmann velocity distribution below an escape velocity of \WIMPEscapeVelocity\ and $v_0=\WIMPVzero$, and a local density of \WIMPDensity.

\section{Conclusions}\label{sec:conclusions}
This work improves upon the result reported in \cite{amaudruz_first_2017}, setting the most sensitive limit for the spin-independent \WIMP-nucleon cross section achieved using a \LAr\ target for \WIMPs\ with mass greater than \SI{30}{\GeV}. 
These results are complementary to results reported by liquid xenon-based experiments, allowing for further constraints on the nature of the \WIMP-nucleon coupling~\cite{hoferichter_analysis_2016,fitzpatrick_effective_2013}. 

The use of \LAr\ here demonstrates the power of \PSD\ as a tool to achieve low backgrounds in \WIMP\ searches, emphasizing the future prospect of much larger \LAr-based detectors designed to achieve sensitivity to \WIMP\ interaction cross-sections at the level of the neutrino floor. 

Additionally, a detailed description of backgrounds in the detector has been presented alongside the analysis methods and simulation models which characterize them. 
Using these models, a total background expectation of \SI{<1} event has been achieved; this model is consistent with observations in data in the \ROI. 
Multivariate techniques are currently being explored to utilize these models to maximize the sensitivity to dark matter signals.
Since the end of the data collection period presented here (\PaperTwoEndDate) \DEAP\ has continued to collect data. 
Updated results including a blind analysis of additional data are planned for the near future.

\section*{Acknowledgments}
This work is supported by the Natural Sciences and Engineering Research Council of Canada, the Canadian Foundation for Innovation (CFI), the Ontario Ministry of Research and Innovation (MRI), and Alberta Advanced Education and Technology (ASRIP), Queen's University, University of Alberta, Carleton University, the Canada First Research Excellence Fund, the Arthur B. McDonald Canadian Astroparticle Physics Research Institute, DGAPA-UNAM (PAPIIT No. IA100118) and Consejo Nacional de Ciencia y Tecnolog\'{\i}a (CONACyT, Mexico, Grants No. 252167 and A1-S-8960), the European Research Council (ERC StG 279980), the UK Science \& Technology Facilities Council (STFC) (ST/K002570/1 and ST/R002908/1), the Leverhulme Trust (ECF-20130496). Studentship support by the Rutherford Appleton Laboratory Particle Physics Division, STFC and SEPNet PhD is acknowledged.  We thank SNOLAB and its staff for support through underground space, logistical and technical services. SNOLAB operations are supported by CFI and the Province of Ontario MRI, with underground access provided by Vale at the Creighton mine site.
We thank Compute Canada, Calcul Qu\'ebec, the Centre for Advanced Computing at Queen's University, and the Computational Centre for Particle and Astrophysics (C2PAP) at the Leibniz Supercomputer Centre (LRZ) for providing the computing resources required for this work.

\bibliographystyle{deap}
\bibliography{deap}
\end{document}